\documentclass[review]{elsarticle}
 \biboptions{comma,sort&compress}
\usepackage{graphicx}
\usepackage{amsmath}
\usepackage{here}
\usepackage{cuted}

\usepackage{here}
\usepackage[dvips]{epsfig}
\def\beq{\begin{equation}}
\def\eeq{\end{equation}}
\def\bea{\begin{eqnarray}}
\def\eea{\end{eqnarray}}
\def\bq{\begin{quote}}
\def\eq{\end{quote}}

\def\nnb{\nonumber}
\def\ga{\left(}
\def\dr{\right)}

\def\nnb{\nonumber}

\def\la{\langle}
\def\ra{\rangle}
\def\nin{\noindent}
\def\ba{\vspace*{-0.2cm}\begin{array}}
\def\ea{\end{array}\vspace*{-0.2cm}}

\def\b{$\bullet~$}
\def\d{$\diamond~$}

\def\als{\alpha_s}

\def\gg2{\la\alpha_s G^2 \ra}
\def\gg3{g^3f_{abc}\la G^aG^bG^c \ra}
\def\ggg4{\la\als^2G^4\ra}

\def\gg{\lag g^{2}_{s} G^2 \rag}
\def\ggg{\lag g^{3}_{s}G^3\rag}

\usepackage{amsmath}
\usepackage{slashed}
\usepackage{color}

\newcommand{\pipi}{\mbox{$\pi\pi$}}

\begin{document}
\begin{frontmatter}

\title{Scrutinizing the Light Scalar Quarkonia  from  LSR at Higher Orders }

\author{R. Albuquerque}
\address{Faculty of Technology,Rio de Janeiro State University (FAT,UERJ), Brazil}
\ead{raphael.albuquerque@uerj.br}

\author{S. Narison
}
\address{Laboratoire
Univers et Particules de Montpellier (LUPM), CNRS-IN2P3, \\
Case 070, Place Eug\`ene
Bataillon, 34095 - Montpellier, France\\
and\\
Institute of High-Energy Physics of Madagascar (iHEPMAD)\\
University of Ankatso, Antananarivo 101, Madagascar}
\ead{snarison@yahoo.fr}

\author{D. Rabetiarivony}
\ead{rd.bidds@gmail.com}

\address{Institute of High-Energy Physics of Madagascar (iHEPMAD)\\
University of Ankatso, Antananarivo 101, Madagascar}

\date{\today}
\begin{abstract}
\noindent
We scrutinize, improve some determinations of the masses and couplings of light scalar quarkonia ($\bar qq$ and four-quark states) and present new results for the $\pi^+\pi^-,K^+K^-,\dots$ molecules using QCD Laplace Sum Rule (LSR) truncated at the $D=6$ dimension vacuum condensates.  We pay a special attention on the  higher order  perturbative (PT)  corrections  up to the (estimated) ${\cal O}(\alpha_s^5)$ which improve  the quality of the analysis.   We request that the optimal results obey the rigorous constraint: {\it Resonance  $\geq$ QCD continuum contributions ($R_{P/C}\geq 1$)} in the LSR which  excludes a Breit-Wigner / on-shell (not to be confused with a complex pole) scalar meson mass around (500-600) MeV obtained for values [$t_c\leq (1\sim 1.5)$ GeV$^2$] of the QCD continuum threshold. Mass-splittings due to $SU(3)$ breakings are small. We discuss the different assignments of the observed scalar mesons below 2 GeV in the Conclusions where the $I=0$ states are compared with the scalar gluonia. The results are compiled in Table\,\ref{tab:res} to \ref{tab:res-rad}.  {\it We hope that the systematic analysis done in this paper  helps to clarifiy the complex spectra of the light scalar mesons. }  
   \end{abstract}
\end{frontmatter}
\pagestyle{plain}

\newpage
 \section{Introduction}
 Since several years, a large amount of efforts have been furnished for understanding the nature of the light scalar mesons both from theory and from experiments\,\cite{PDG,AMSLER}. In this paper, first, we shortly review the properties of these states from $\pi\pi,\,\bar KK$ and $\gamma\gamma$ scatterings data and some theoretical attempts to understand their nature. Secondly, we present an improved analysis of the determination of these states from QCD spectral sum rules (QSSR) within the ordinary $\bar qq$ and the exotic four-quark state configurations. We also present new results for the molecule configurations.
 
 \subsection*{\b The $\sigma/f_0(500)$ from $\pi\pi,~\bar KK$ and $\gamma\gamma$ scatterings} 
The nature of the $\sigma/f_0(500)$ meson and some other scalar mesons remains 
 still mysterious despite many theoretical efforts. In particular, the $\sigma/f_0(500)$ appeared and disappeared in the Particle Data Group. Since few years, its evidence has been settled\,\cite{PDG} but its nature still remains unclear. Taking the average of the $\sigma$ mass and $\pi\pi$ width compiled in Table\,\ref{tab:sigma}
 from $\pi\pi,\,KK$ scatterings, $J/\psi $ and $D$ decays data, one finds the complex pole:
  \beq
 M_\sigma^{pole}[\rm MeV]=(455\pm 12) -i\,(260\pm 12),
 \label{eq:msigma}
  \eeq
 with its  hadronic couplings from $Ke4\oplus \pi\pi$ scatterings within an improved analytic K-matrix model are\,\cite{WANG1}:
 \beq
 \vert g_{\sigma\pi^+\pi^-}\vert\simeq 2.65(10)~{\rm GeV},\,\,\,\,\,\,\,\,\,\,\,\,\,\,\,\,\,\,\,\,\,\,\,\,\,\,\,\,\,\,\,\,\,\,\,\,\,\,\,\,\,\,\,\,\,\,\,\,\,\,\,\,\,\,\,\,
r_{\sigma\pi K}\simeq \frac{\vert g_{\sigma K^+K^-}\vert} { \vert g_{\sigma\pi^+\pi^-}\vert}\simeq 0.37(6)~.
\label{eq:sig-coupling}
\eeq
Some other analysis quoted  in Table 6 of Ref.\,\cite{WANG1} and the one from $\pi\pi\to \pi\pi/\bar KK$ in Ref.\,\cite{KMN} give the range $r_{\sigma\pi K} \simeq (0.47\sim 1)$ indicating that the $\sigma$ coupling to $\bar KK$ is expected to be large (see however Ref.\,\cite{HANHART}\,\footnote{We thank C. Hanhart for bringing this reference to our attention and for some discussions on this point.}). 
The strong coupling of the $\sigma$ to $\bar KK$  does not (a priori) favour its molecule / four-quark interpretation\,\cite{JAFFE,ACHASOV,ISGUR} which is not expected to couple to $\bar KK$ unless there is a huge Zweig rule violation. 
{\scriptsize
\begin{table}[hbt]
\setlength{\tabcolsep}{2.4pc}
\begin{tabular}{lll}
&\\
\hline
Processes&$M_\sigma -i\Gamma_\sigma/2$&Refs.   \\
\hline
$\pi\pi\to \pi\pi\oplus$\,Roy$\,\oplus\,$ChPT&$(441^{+16}_{-8}) -{\rm i}~(272^{+9}_{-15})$&\cite{LEUT}\\
$\pi\pi\to \pi\pi/\bar KK\,\oplus\,$Roy&$(461 \pm 15 )-{\rm i} ~(255 \pm 16)$&\cite{YND}\\
$J/\psi\to \omega\pi\pi$&$(541\pm 39) -{\rm i}~(222\pm 42)$& \cite{BES05}\\
$D^+\to\pi^+\pi^-\pi^+$ &$(478 \pm 29)-{\rm i}~(162 \pm 46)$&\cite{E741}\\
$\pi\pi\to \pi\pi\,\oplus$\,Roy $\oplus$ 1 resonance&$(456\pm 19)-{\rm i}~(265\pm 18)$&\cite{WANG1,MNO}\\
$Ke4$ $\oplus$ $\pi\pi\to \pi\pi$&$(452\pm 13) -{\rm i}~(259\pm 16)$&\cite{WANG1}\\
$Ke4$ $\oplus$ $\pi\pi\to \pi\pi/K\bar K$&$(448\pm 43) -{\rm i}~(266\pm 43)$&\cite{WANG1}\\
\it Mean & \it $(455\pm 12) -{\rm i}~ (260\pm 12)$& {\it This paper} \\
\hline
\end{tabular}
 \caption{\scriptsize    Mass and 1/2 width in MeV of the $\sigma$ meson in the complex plane.  The error in the mean value comes from the most precise determination. For the assymetric errors, we have taken their mean value.}
\label{tab:sigma}
\end{table}
}
\nin

\subsection*{\b The $f_0(980)$ from $\pi\pi$ and $\bar KK$ scatterings} 
The nature of the $f_0(980)$ and $a_0(980)$ is also intriguing due to their vicinity of the $\bar KK$ threshold and to their anomalous strong coupling to $\bar KK$.  One finds from recent $\pi\pi$ and $ KK$ scattering data\,\cite{WANG1}:
\beq
M^{pole}_{f_0}[{\rm MeV}]\simeq 981(34)-i18(11) ~,~~~~\vert g_{f_0\pi^+\pi^-}\vert\simeq (1.17\pm 0.26)~{\rm GeV},~~~~r_{f_0\pi K}\equiv \frac{\vert g_{f_0 K^+K^-}\vert}{ \vert g_{f_0\pi^+\pi^-}\vert}\simeq (2.6\pm 1.3)~,
\eeq
where $r_{f_0\pi K}$  is in the range 1.2 to 1.8 of  the other determinations quoted in Table 6 of Ref.\,\cite{WANG1} and the value of about 1.8 in Ref.\,\cite{KMN}. 
 A  problem similar to the case of the $\sigma$ occurs for the non-vanishing of the coupling of $f_0(980)$ to $\pi\pi$ (not expected from the four-quark picture). 
 
\subsection*{\b Complex Pole and the On-shell / Breit-Wigner Mass and Width}
 However, in order to compare properly the experimental results with the QCD spectral sum rules ones where the spectral function is used in the real axis, one has to introduce the On-shell or Breit-Wigner (os) mass defined in Refs.\,\cite{KNIEHL,MNO} and width\,\cite{MNO}, where the amplitude is purely imaginary at the phase $90^0$\,\footnote{We should note that lattice calculations may have access directly to the complex pole mass.}:
\beq
Re{\cal D}((M_{S}^{os})^2)=0~,\,\,\,\,\,\,\,\,\,\,\,\,\,\,\,\,\,\,\,\,\,\,\,\,\,\,\,\,
M^{\rm os}_S \Gamma^{\rm os}_S\simeq \frac{{\rm Im}~{\cal D}} {-{\rm Re~} {\cal D}'}.: ~~~ S\equiv \sigma,f_0,
\eeq
where ${\cal D}$ is the propagator appearing in the unitary $\pi\pi$ amplitude.  
Then, one can deduce from the complex pole obtained from $\pi\pi, KK$ scattering data\,\cite{MNO,WANG2}:
\beq
(M^{os}_\sigma, \Gamma^{os}_\sigma) \simeq (920, 700)~{\rm MeV}~, \,\,\,\,\,\,\,\,\,\,\,\,\,\,\,\,\,\,\,\,\,\,\,\,\,\,\,\,(M^{os}_{f_0}, \Gamma^{os}_{f_0}) \simeq  (981, 36)~{\rm MeV}~,
\label{eq:sigdata}
\eeq
evaluated at $s=\ga M^{os}_S\dr^2~:~S\equiv \sigma, f(980)$. 
\subsection*{\b Some theoretical interpretations for  $\sigma/f_0(500)$ and $f_0(980)$} 
 The unexpected values of the $\bar\pi\pi$ and $\bar KK$ couplings from scatterings data can  disfavour the molecule and four-quark picture of the $\sigma/f_0(500)$  and $f_0(980)$ but may  instead support a large gluon component in their wave functions where the $\bar\pi\pi$ and $\bar KK$ couplings are expected to have about the same strength\,\cite{VENEZIA,SNG,SNGS}\,\footnote{More discussions and references  on glueball/gluonium works can be e.g. found in\,\cite{OCHS,GASTALDI,PDG,SNSCAL,SNPRD}.}. The observed states  may result from the mixing between gluonia / glueball and ordinary $\bar qq$ states where the size of the mixing angle is small\,\cite{SNGMIX} but maybe sizable in the presence of some eventual instanton effects\,\cite{STEELES}. 

Among different theoretical interpretations of the nature of these light scalar states\,\footnote{For reviews, see e.g.\,\cite{AMSLER,RICHARD,KLEMPT,OCHS,SNPRD}.},we have :

\d {\it Ordinary $\frac{1}{\sqrt{2}}\ga \bar uu +\bar dd\dr, \bar ud$ and gluonium/glueball states} which are related to the trace of the energy momentum tensor : 
\bea
\theta^\mu_\mu&=&\frac{1}{4} \beta(\alpha_s) G^{\mu\nu}_aG^a_{\mu\nu}+\ga1+\gamma_m(\alpha_s)\dr\sum_{u,d,s}m_i\bar\psi_i\psi_i ,
\label{eq:theta}
\eea
with :  $\gamma_m=2a_s+\cdots$ is the quark mass anomalous dimension and $a_s\equiv \alpha_s/\pi$.  $\beta(\alpha_s)$ is the $\beta$-function, where one should notice that   in the chiral limit $m_q=0$, the corresponding two-point correlator is given by the scalar gluonium one which has been recently revised in Refs.\,\cite{SNGS,SNG} (where earlier references can be found) 
for extracting the scalar glueball/gluonia masses and couplings from QCD spectral sum rules.  One should also mention that the mixing of the gluonium and the $\bar qq$ correlator has been also discussed in Refs.\,\cite{SNGMIX,SNG} within the ordinary Operator Product Expansion (OPE) where the mixing angle has been shown to be relatively small. 

\d {\it Ordinary $\bar uq$ state} which is related to the divergence of the vector current:
\beq
\partial_\mu V^\mu_{\bar uq}(x)=(m_u-m_q) \bar u q(x)~:~~~~~~~~~~~~~~~~~~~~q\equiv d,s.
\eeq
It has been used to estimate (for the first time) the running quark mass difference $(m_s-m_u)$ from $K\pi$ system in a $S$-wave and $I=1/2$ state and $(m_d-m_u)$ from the relation of the $a_0$ form factor at the origin to the tadpole difference $(M_{K^+}^2-M_{K^0}^2)$ where the electromagnetic contribution has been subtracted\,\cite{SCAL}. The mass of the scalar meson has been also estimated from the ratio of sum rules in\,\cite{RRY,SNB2,SNG,SNPRD} while their hadronic and $\gamma\gamma$ couplings have been estimated from vertex sum rules\,\cite{SNA0,SNG}. These results  have been used for interpreting the $\bar qq$ nature of the $f_0(980), a_0(980)$ and $K^*_0(700)$ states\,\cite{BN,SNPRD}.

\d {\it Exotic Molecule $\pi^+\pi^-,~K^+K^-,...$  and Four-quark states} which are based on the strong couplings of the $f_0(980)$ and $a_0(980)$ to  $\bar KK$ and on their almost degeneracy and on the anomalously light mass of the $\sigma$\,\cite{JAFFE,ACHASOV,ISGUR}\,\footnote{$1/N_c$ expansion has been also used in\,\cite{ROSSI} for interpreting baryonium molecule state.}:
\beq
\sigma = \bar u\bar d ud,~~~~~~~~~~~~~f_0=\frac{1}{\sqrt {2}}\ga \bar u\bar s us +\bar d\bar s ds\dr,~~~~~~~~~~~~~a_0=\frac{1}{\sqrt {2}}\ga \bar u\bar s us -\bar d\bar s ds\dr, ~~~~~~~~~~~~~K^*_0 =\bar d\bar s ud~.
\eeq
The analysis of these states has been pursued in Refs.\,\cite{LATORRE,SNA0,SNB2,MARINA,ZHU,STEELE} using QCD spectral sum rules (QSSR), in Ref.\,\cite{JAFFE2} using lattice calculations, in Ref.\,\cite{THOOFT} using a six-fermion effective Lagrangian induced by instantons, in Ref.\,\cite{BRODSKY} using light front holographic QCD  and in some other approaches\,\footnote{For reviews, see e.g.\cite{RICHARD,KLEMPT,OCHS}.}.

\section{The QCD Laplace sum rule (LSR) approach}
 In the following, we shall re-analyze the previous different results within  the  framework of QCD Laplace Sum Rule (LSR) which has been successfully applied to different light, heavy-light and heavy hadronic states for predicting their masses and decay constants/couplings and to a lesser accuracy their hadronic widths\,\cite{SVZa,SVZb,ZAKA,BELLa,BELLb,BERT,SNR}\,\footnote{For reviews and introductory books on QCD spectral sum rules, see e.g.\,\cite{SNB1,SNB2,SNB3}.}.  
\subsection*{\b Form of the LSR}
 For extracting the coupling/ decay constant and the mass of the scalar quarkonia mesons, we shall work with the Laplace Sum Rule (LSR):
 \bea
{\cal L}_0^c(\tau,\mu)&\equiv&\lim_ {\begin{tabular}{c}
$Q^2,n\to\infty$ \\ $n/Q^2\equiv\tau$
\end{tabular}}
\frac{(-Q^2)^n}{(n-1)!}\frac{\partial^n \psi}{ ( \partial Q^2)^n}\,\,\,\,\,\,\,\,
=\,\,\,\,\,\,\,\,\int_{t_>}^{t_c}\hspace*{-0.25cm}dt~e^{-t\tau}\frac{1}{\pi} \mbox{Im}\,\psi_{S}(t,\mu)~,
\nnb\\
 {\cal R}^c_{10}(\tau)&\equiv&\frac{{\cal L}^c_{1}} {{\cal L}^c_0}\,\,\,\,\,\,\,\,= \,\,\,\,\,\,\,\,\frac{\int_{t_>}^{t_c}dt~e^{-t\tau}t\, \mbox{Im}\,\psi_{S}(t,\mu)}{\int_{t_>}^{t_c}dt~e^{-t\tau} \mbox{Im}\,\psi_{S}(t,\mu)},~~~~
\label{eq:lsr}
\eea
associated to the generic two-point function:
\beq
\psi_S(q^2)=i\int d^4x\,e^{iqx}\,\la 0\vert {\cal T} {\cal O}_S(x){\cal O}_S^\dagger(0)\vert 0\ra~,
\eeq 
where : 
${\cal O}_S(x)\equiv \bar qq', ~(\bar qq')(\bar q'q)$ or $(\bar q\bar q')(qq')$ are the quark operators which describe respectively the ordinary mesons, molecules or four-quark states; $Q^2\equiv -q^2>0$ is the momentum transfer squared; $\tau$ is the LSR variable and $t_>$   is the quark threshold.  Here $t_c$ is  the threshold of the ``QCD continuum" which parametrizes, from the discontinuity of the Feynman diagrams, the spectral function  ${\rm Im}\,\psi_{S}(t,m_Q^2,\mu^2)$.


\subsection*{\b Parametrization of the spectral function}
 The contribution of the $\sigma$ to the spectral function can be introduced within
 the Minimal Duality Ansatz (MDA):
\beq
\frac{1}{\pi} \mbox{Im}\,\psi_{S}(t)=2f_S^2 M_S^{2(d-2)}\,\,\delta\ga t- M^2_{S}\dr \,\oplus \, \theta(t-t_c)\,``{\rm QCD \,\,  Continuum} "~: \,\,\,\,\,\,\,\,\ \la 0\vert {\cal O}_ S(x)\vert S\ra= \sqrt 2 f_{S} M_{S}^{d-2}.
\eeq
where $d$ is the dimension of the current.  The QCD continuum smears all higher radial excitations from the  threshold $t_c$ which is expected to be above the lowest resonance mass squared :
\beq
t_c\geq M_S^2.
\eeq
However, contrary to some intuitive claims in the literature, $\sqrt{t_c}$ does not necessarily coincide with the mass of the first radial excitation but can be higher as found from FESR in the well-known example of the $\rho$-meson channel\,\cite{FESR}.  We expect that this MDA gives a good description of the lowest resonance like in the case of the $\rho,D,B,B_c,J/\psi$ ordinary mesons and heavy molecules / four-quark states studied in the current SVZ sum rules literature\,\cite{SNB1,SNB2} thanks to the exponential weight enhancement of their contributions in the Laplace sum rules. 
At the first step, we shall  use a Narrow Width Approximation (NWA). The finite width correction will be estimated at the second step  of the analysis. 
\subsection*{\b Optimization procedure}
We shall base our analysis on the stability of the results versus the change of the external input parameters (sum rule variable $\tau$ and continuum threshold $t_c$\,\cite{SNB1,SNB2} to which we implement the (rigorous) condition\,\cite{TQQ} :
\beq
R_{P/C} \equiv \frac {\rm Lowest\,Pole} {\rm QCD~ continuum}=\frac{\int_{t_>}^{t_c}dt~e^{-t\tau}\, \mbox{Im}\,\psi_{S}(t,\mu)}{\int_{t_c}^{\infty}dt~e^{-t\tau} \mbox{Im}\,\psi_{S}(t,\mu)} \geq 1.
\label{eq:rpc}
\eeq
This condition is necessary for a reliable prediction of the resonance contribution which is disentangled from the QCD continuum one. 
 \section{The QCD input parameters}
 \subsection*{\b Definitions and Notations}
The QCD parameters which shall appear in the following analysis will be the QCD coupling $\alpha_s$, the light quark masses $m_{q}$, the light quark condensates $\la\bar qq\ra$, 
$\la\bar qq\ra^2$,
 the gluon condensates $ \la\alpha_sG^2\ra$,  $ \la g^3G^3\ra$, the quark-gluon mixed condensate $g\la \bar qGq\ra\equiv \la \bar qq\ra M^2_0$
 and the four-quark condensate $\rho\la\bar qq\ra^2$. 
 $q\equiv u,d,~G^2\equiv G^{\mu\nu}_aG^a_{\mu\nu},~  \bar qGq\equiv \bar q (\lambda_a/2)\sigma^{\mu\nu} G^a_{\mu\nu}q,~ G^3\equiv g^3f_{abc}G^{\mu\nu}_aG^\sigma_{\nu,b}G_{\sigma\mu,c}$ where $u,d$ are quark fields and $G_{\mu\nu}$ the gluon field strengths. $\la ...\ra\equiv \la 0\vert  ...\vert 0\ra$ is a compact notation for the QCD vacuum condensates and $\rho$ measures the deviation from the vacuum saturation of the four-quark condensates. 
$\alpha_s$ is the running QCD coupling, $\bar m_q,\, \la\overline{\bar qq}\ra, \la\overline{\bar qGq}\ra$ are running quark mass and condensates. 

For a guidance, we shall give the expressions of the running parameters to leading order:
\bea
 a_s&\equiv& \ga \frac{\alpha_s}{\pi}\dr=\frac{2}{-\beta_1\log{ Q^2}/{\Lambda^2}},\,\,\,\,\,\,\,\,\,\,\,\,\,\,\,\,\,\,\,\,\,\,\,\,\,\,\,\,
\bar m_q = {\hat m_q} \ga -\beta_1 a_s\dr^{\gamma_1/-\beta_1},\nnb\\
 \la\overline{\bar qq}\ra &=& -\hat\mu_q^3 \ga -\beta_1 a_s\dr^{\gamma_1/\beta_1},\,\,\,\,\,\,\,\,\,\,\,\,\,\,\,\,\,\,\,\,\,\,\,\,\,\,\,\,\,\,\,\,\,\,\,\,\,\,\,\,\,\,
\la\overline{\bar qGq}\ra= \ga -\beta_1 a_s\dr^{\gamma_M/-\beta_1}M_0^2(-\hat\mu_q)^3,
\label{eq:rgi}
\eea
 where: 
$\gamma_1=2$ is the first coeeficient of the quark mass anomalous dimensions, $\gamma_M=1/3$  is the first coefficient of the mixed quark gluon condensate anomalous dimensions and  $\beta_1=-(1/2)(11-2n_f/3)$ is the first coefficients of the $\beta$-function: \,\cite{BECCHI,SNB1,SNB2,SNB3} :
\beq
\gamma = \gamma_1\,a_s+\gamma_2\,a^2_s+\cdots,\,\,\,\,\,\,\,\,\,\,\,\,\,\,\,\,\,\,\,\,\,\,\,\,\beta=\beta_1\,a_s+\beta_2\,a_s^2+\cdots
\eeq
\subsection*{\b Values of the QCD input parameters}

 In this paper, their  expressions are used to order ${\cal O}(\alpha_s^3)$ and their values are given in Table\,\ref{tab:param} together with the ones of the gluon condensates. We shall also use for 3 flavours:
 \beq
 \Lambda= 332(8)~{\rm MeV}
 \eeq
 using the central value of $\alpha_s(M_Z)=0.1181$ in the Table but taking the precise error from the world average given by PDG\,\cite{PDG}. The values of the other QCD parameters used in our analysis are given in Table\,\ref{tab:param}. 
\vspace*{-0.25cm}

{\scriptsize
\begin{table}[hbt]
\setlength{\tabcolsep}{2.2pc}
\begin{center}
    {\small
  \begin{tabular}{llcc}
&\\
\hline
Parameters&Values&Sources& Refs.    \\
\hline
\it Heavy \\
$\alpha_s(M_Z)$& $0.1181(16)(3)$&$M_{\chi_{0c,b}-M_{\eta_{c,b}}}$&
\cite{SNparam} \\
$\la\alpha_s G^2\ra$ [GeV$^4$]& $6.49(35) 10^{-2}$&Light, Heavy &
 \cite{SNparam,SNREV1}\\
${\la g^3  G^3\ra}/{\la\alpha_s G^2\ra}$& $8.2(1.0)$[GeV$^2$]&${J/\psi}$&
\cite{SNcb1}\\
 \it{Light} \\
$\hat \mu_\psi$ [MeV]&$253(6)$ &Light &\,\cite{SNB1,SNB2,SNp15,SNLIGHT} \\ 
$\la\overline{\bar \psi\psi}\ra(2)$ [MeV]$^3$&$-(276\pm 7)^3$ &--&\,\cite{SNB1,SNB2,SNp15,SNLIGHT,DOSCHSN}  \\
$\kappa\equiv\la \bar ss\ra/\la\bar dd\ra$& $0.74(6)$&Light, Heavy&\cite{SNB1,SNB2,SNp15,SNLIGHT,HBARYON2}\\
$\hat m_u$ [MeV]&$3.05\pm 0.32$&Light &\,\cite{SNB1,SNB2,SNp15,SNLIGHT} \\
$\hat m_d$ [MeV]&$6.10\pm0.57$ &-- &-- \\
$\hat m_s$ [MeV]&$114(6)$ &-- & -- \\
$\overline{ m}_u$ (2) [MeV]&$2.64\pm 0.28$ &-- &-- \\
$\overline{ m}_d$ (2) [MeV]&$5.27\pm 0.49$ &-- & -- \\
$\overline{ m}_s$ (2) [MeV]&$98.5\pm 5.5$ &-- & --\\
$M_0^2$ [GeV$^2$]&$0.8(2)$ &Light, Heavy&\,\cite{SNB1,SNB2,IOFFE,DOSCH,PIVOm,SNhl}\\

$\rho \alpha_s\la \bar \psi\psi\ra^2\times 10^{4}$ &$5.8(9) $[GeV$^6$] &Light, $\tau$&\cite{FESR2,SNTAU,LNT,DOSCH,SN95,SOLA,SNe23}\\
\hline
\end{tabular}}
 \caption{QCD input  parameters from light  and  heavy quarks QSSR (Moments, LSR and ratios of sum rules) within stability criteria (compilation  from\,\cite{SN22}). The running light quark masses and condensates have been evaluated at 2 GeV within the SVZ expansion without instantons contributions. }
\label{tab:param}
\vspace*{-0.25cm}
\end{center}
\end{table}
} 
\section{The light scalar-isoscalar  $(\bar uu+\bar dd)$ state}
In QCD, the isoscalar states are associated to the trace of the energy-momentum tensor defined in the introduction. 
We shall separate the quark part of $\theta_\mu^\mu$, into two currents:
\beq
J_2=\frac{m_{q}}{\sqrt{2}}\ga \bar uu+\bar dd \dr \,\,\,\,\,\,\,\,\,\,\,\,\,\,\,\,\,\,\,\,\,\,\,\,\,\,\,\,\,\,\,\,\,\,\,\,
{\rm and} \,\,\,\,\,\,\,\,\,\,\,\,\,\,\,\,\,\,\,\,\,\,\,\,\,\,\,\,\,\,\,\,\,\,\,\, J_3=m_s\bar ss
\eeq
where : $m_q=(m_u+m_d)/2$. In this section, we shall first analyze the correlator:
\beq
\psi_{\bar qq}(q^2)=i\int\,d^4x\,e^{iqx}\,\la 0\vert {\cal T} J_2(x)J_2^\dagger(0)\vert 0\ra.
\eeq 
In this paper, we shall improve the estimate  done in Refs.\,\cite{ SCAL,RRY,SNB2,SNA0,BN,SNPRD} by adding higher order PT corrections.  The mass corrections and non-pertubative contributions up to D=6  are listed in Eqs.\,53.24 to 53.31 of Ref.\,\cite{SNB1} and in\,\cite{SCAL,JM95,SNp15}. 
\subsection*{\b Estimate of the Mass and Coupling}
\begin{figure}[hbt]
\begin{center}
\hspace*{-7cm} {\bf a) \hspace*{8.cm} \bf b)} \\
\includegraphics[width=7.9cm]{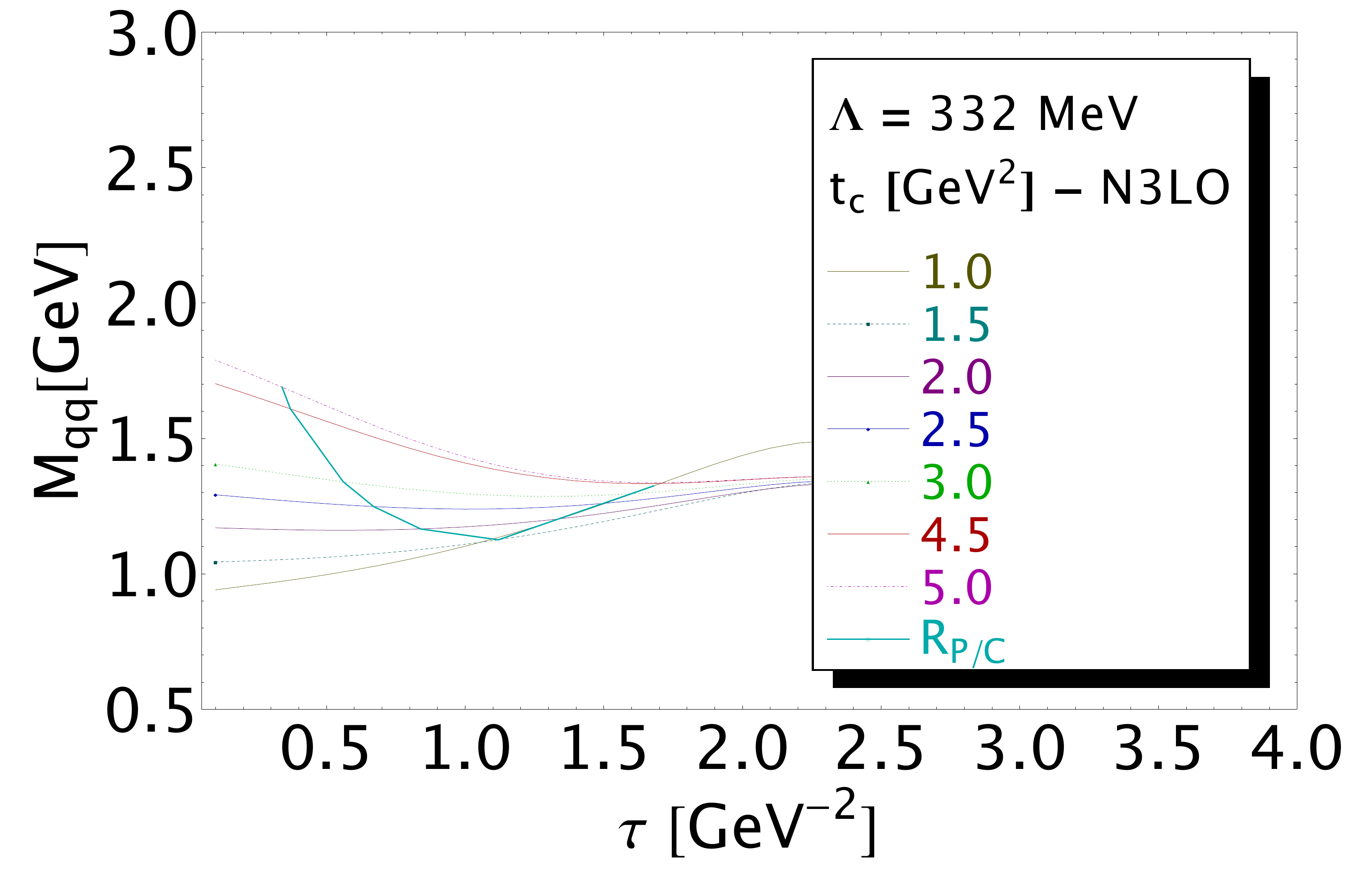}
\includegraphics[width=8.4cm]{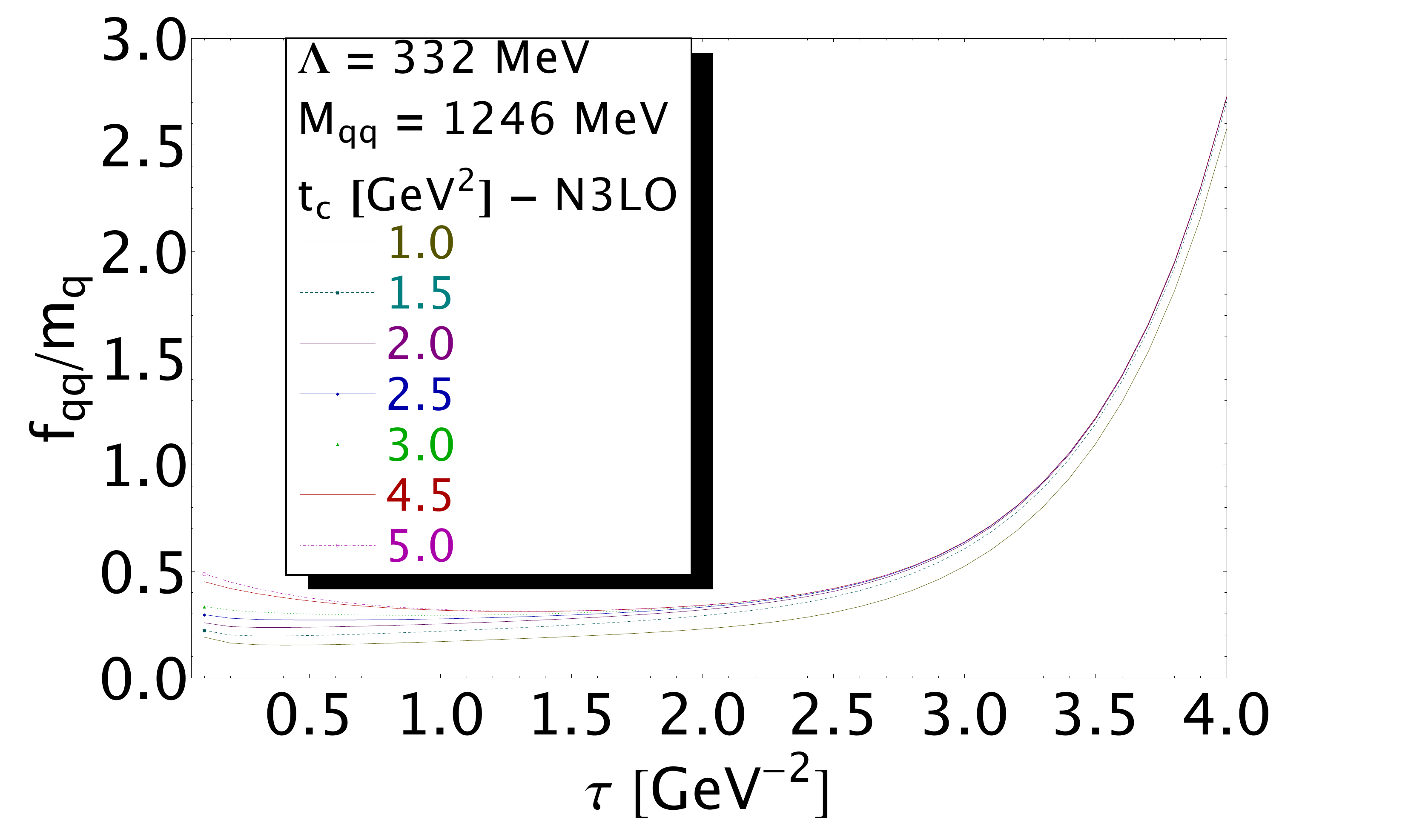} 
\vspace*{-0.5cm}
\caption{\footnotesize   Behaviour of a) mass and  b) coupling of the $\bar qq\equiv\frac{1}{\sqrt{2}}\ga \bar u u \pm \bar dd\dr $ meson states versus $\tau$ for different values of $t_c$. } 
\label{fig:mass-qq}
\end{center}
\vspace*{-0.5cm}
\end{figure}

One can estimate the coupling from the low moment ${\cal L}_0$ and the mass from the ratio of moments ${\cal R}_{10}$.  
The LSR expression of the moment reads to N3LO of PT series and to leading order of the NPT contributions truncated at $D=6$:
\bea
{\cal L}_0^c(\tau)\equiv \int_{(m_u+m_d)^2}^{t_c}\hspace*{-1.1cm}dt\,e^{-t\tau}\,\frac{1}{\pi}\,{\rm Im}\,\psi_S(t)
=
 \overline{m}_{q}^2
\Bigg{\{} \int_{(m_u+m_d)^2}^{t_c}\hspace*{-1.1cm}dt\,e^{-t\tau}\,\frac{1}{\pi}\,{\rm Im}\,\psi_{\rm Pert}(t)  +
  \frac{3}{8\pi^2}\tau^{-2}  
  \sum_{n=1}^3\delta^{(2n)}\tau^n\Bigg{\}} ~,
\label{eq:lsrqcd}
\eea
where ${\rm Im}\,\psi_{\rm Pert}(t)$ is known up to order $\alpha_s^4$\,\cite{BECCHI,BROAD1,LARIN2,CHET3,CHET4}.  For 3 flavours, it reads including the $\pi^2$-term from the analytic continuation :
\beq
 \frac{1}{\pi}\,{\rm Im}\psi_\pi(t)\vert_{\rm Pert}= \frac{3}{8\pi^2}\,t\,\ga 1+\frac{17}{3}a_s+31.863a_s^2+89.157a_s^3-536.749a_s^4+{\cal O}(a_s^5)\dr.
 \label{eq:pert}
 \eeq
The mass corrections to order $\alpha_s$ and non-perturbative corrections to lowest order read:
\bea
\delta^{(2)}&=&-2\Big{[}(1+7.64\,a_s)(\bar m_u^2+\bar m_d^2)+(1+11.64\,a_s)\,\bar m_u\bar m_d\Big{]},\nnb\\
\delta^{(4)}&=& \frac{8\pi^2}{3} \Big{[} (m_u+\frac{m_d}{2})\la \bar uu\ra +(m_d+\frac{m_u}{2})\la \bar dd\ra \Big{]} +\frac{\pi}{3}\la \alpha_s G^2\ra,\nnb\\
\delta^{(6)}&=& \frac{4\pi^2}{3}\Big{[}\ga m_u\la\bar dGd\ra +m_d\la \bar u Gu\ra\dr -\frac{32}{27}\pi\alpha_s\ga \la \bar uu\ra^2
+ \la \bar dd\ra^2+9 \la \bar uu\ra \la \bar dd\ra\dr\Big{]},
\label{eq:npert}
\eea
We extract the meson mass from the ratio ${\cal R}_{10}(\tau,t_c)$ of moments defined in Eq.\,\ref{eq:lsr}. The analysis for the N3LO PT series approximation is shown in Fig.\,\ref{fig:mass-qq}. 
\subsection*{\b On the truncation of the PT series}
\begin{figure}[hbt]
\begin{center}
\hspace*{-7cm} {\bf a) \hspace*{8.cm} \bf b)} \\
\includegraphics[width=8.cm]{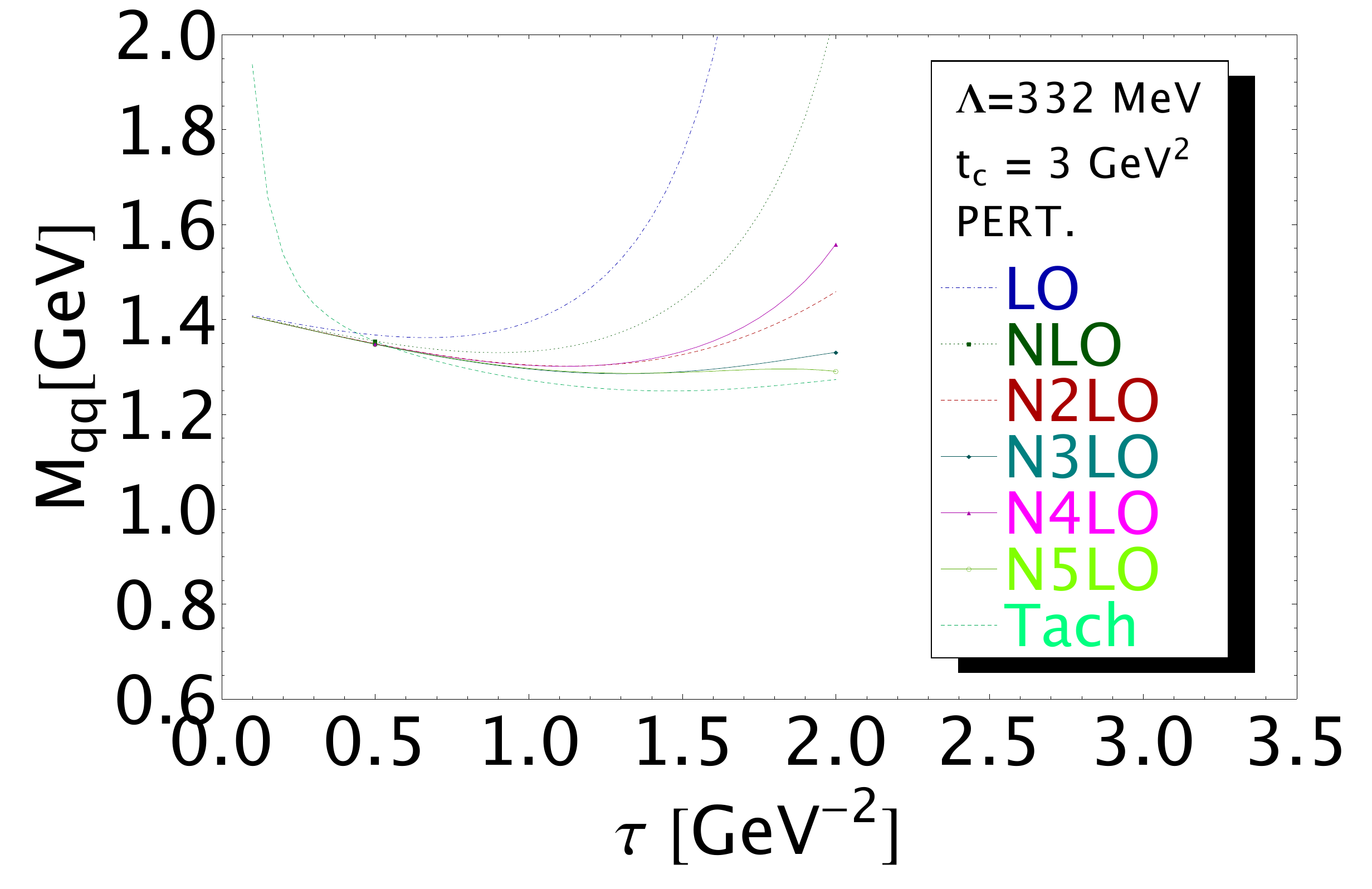}
\includegraphics[width=8.cm]{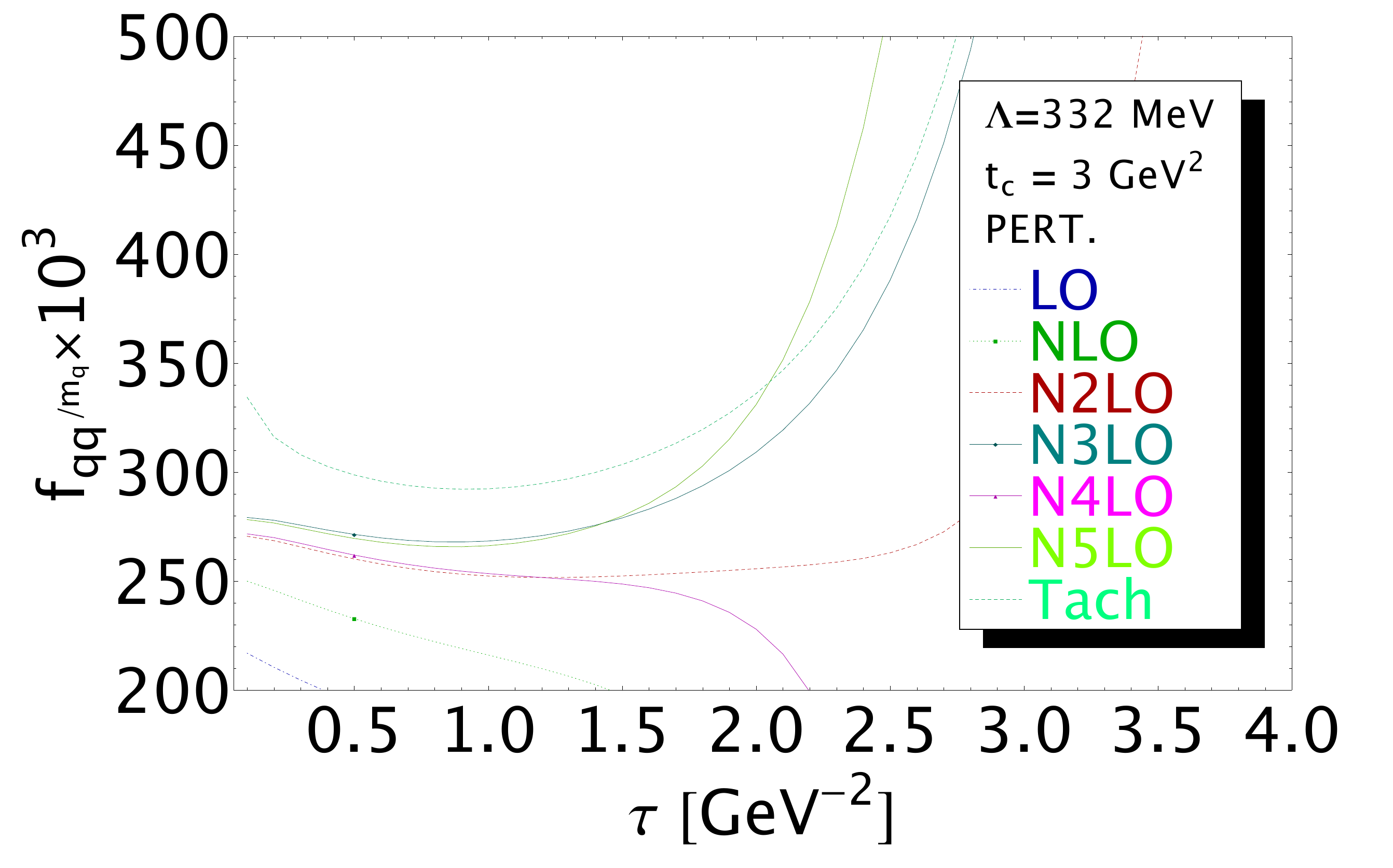} 
\vspace*{-0.5cm}
\caption{\footnotesize   Behaviour of a) mass and  b) coupling of the $\bar qq\equiv\frac{1}{\sqrt{2}}\ga \bar u u \pm \bar dd\dr $  meson states for different truncation of the PT series for fixed value of $t_c=3$ GeV$^2$ versus $\tau$. } 
\label{fig:pert-qq}
\end{center}
\vspace*{-0.5cm}
\end{figure}

\begin{figure}[hbt]
\begin{center}
\hspace*{-7cm} {\bf a) \hspace*{8.cm} \bf b)} \\
\includegraphics[width=8.cm]{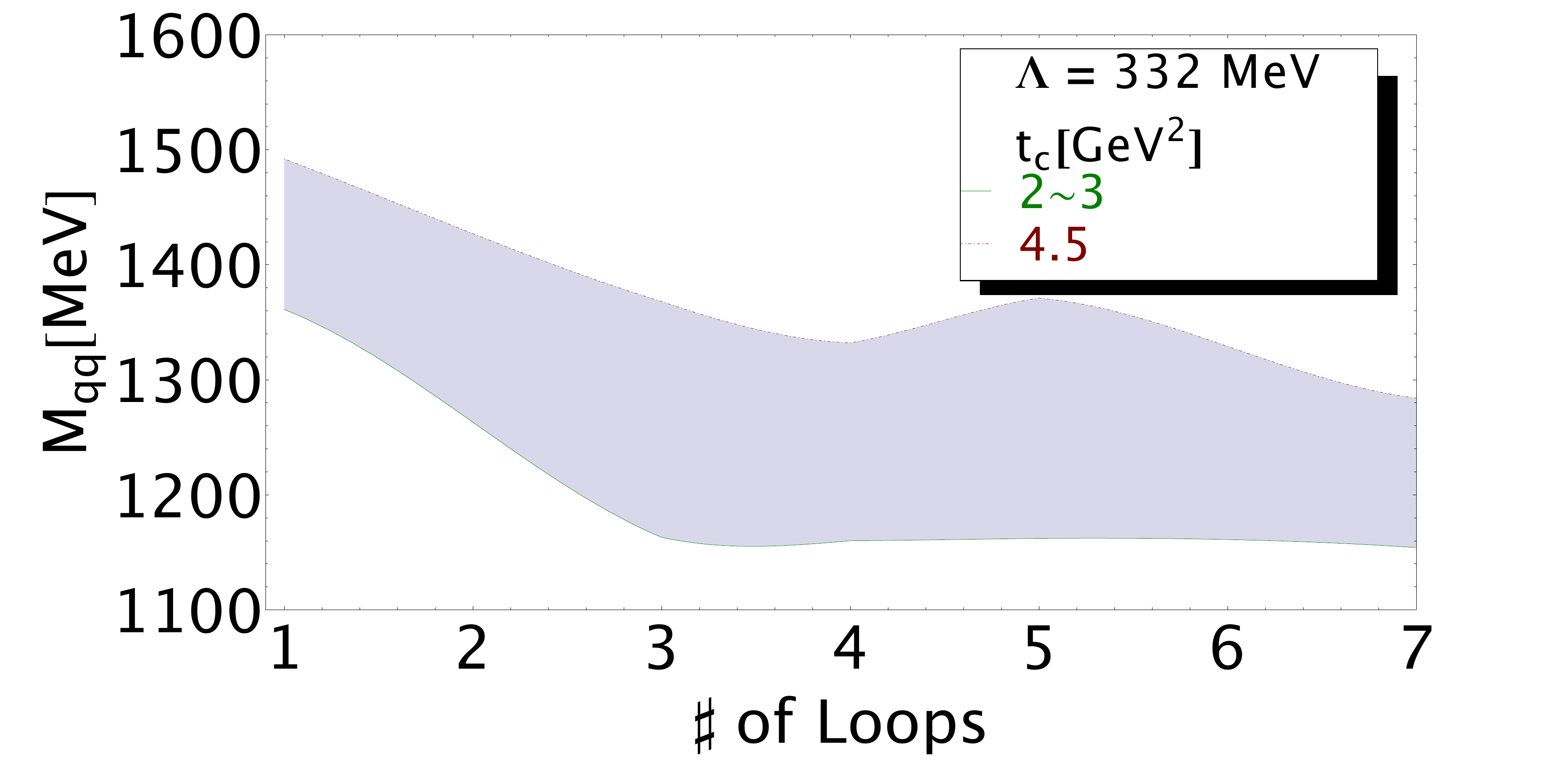}
\includegraphics[width=8.cm]{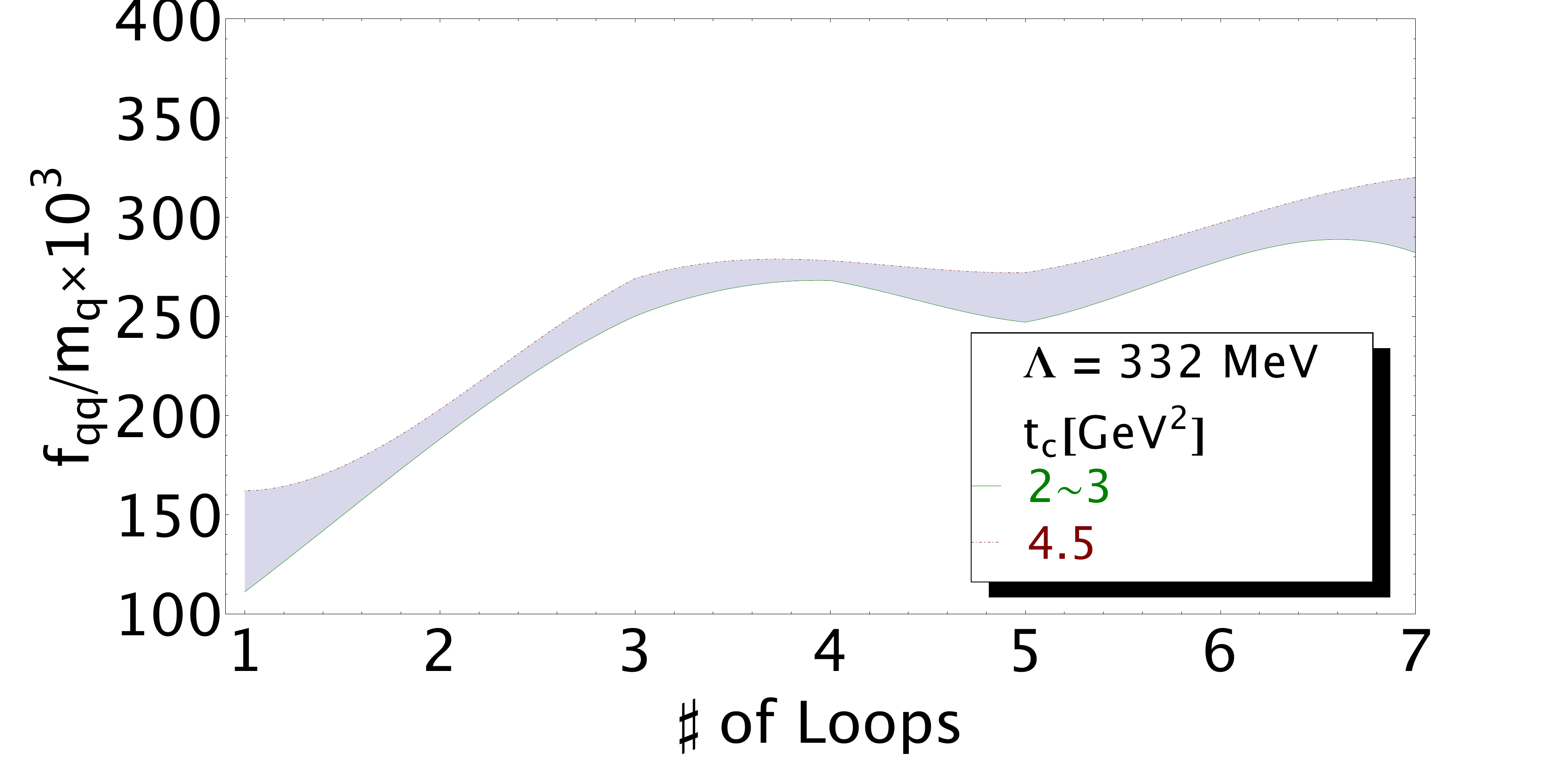} 
\vspace*{-0.5cm}
\caption{\footnotesize   Behaviour of a) mass and  b) coupling of the $\bar qq\equiv\frac{1}{\sqrt{2}}\ga \bar u u \pm \bar dd\dr $  meson states for different truncation of the PT series and for two extremal values of $t_c$ where the $\tau$-stability is reached. } 
\label{fig:trunc-qq}
\end{center}
\vspace*{-0.5cm}
\end{figure}

We show in Fig.\,\ref{fig:pert-qq} the behaviour of $M_{S}$ and $f_{S}$ versus $\tau$ for a given value of $t_c=3$ GeV$^2$ (which reproduces their central values) for different truncation of the PT series. We also show in Fig.\,\ref{fig:trunc-qq} the behaviour of the mass and coupling for different truncation of the optimal results in $\tau$ for given two extremal values of $t_c$. The last point n=7  in the loop coordinate corresponds to the effect of tachyonic gluon  mass which we consider as an alternative source of the errors for the truncation of the PT series. 

For the mass estimate, one can notice that higher order corrections shift the position of the minima to larger values of $\tau$ from 0.65 GeV$^{-2}$ for LO to 1.35 GeV$^{-2}$ for N5LO. However, the $\tau\simeq (0.9-1.0)$ GeV$^{-2}$ value for the coupling is almost stable.  We also notice that the inclusion of the NLO to N3LO corrections improve its estimate   as the $\tau$-stability appears from N2LO. 

Due to its negative sign, the N4LO contribution tends to increase the value of the mass but decrease the one of the coupling. We interpret this change of the N4LO sign as a signal of the beginning of the alternate signs of the QCD series where the asymptotic form of the series may already be reached at N3LO. To check our argument, a non-trivial evaluation by experts of the N5LO term is needed.  

Therefore, in order to estimate the error due to the truncation of the PT series, we first assume that the next term of the series has a positive sign (alternate sign). We estimate its absolute size by observing that the coefficients of the series have a geometrical growth behaviour similar to the one found in Ref.\cite{CNZb} for the complete two-point correlator:
\beq
 \frac{1}{\pi}\,{\rm Im}\psi_{\bar qq}(t)\vert_{\rm Pert}= \frac{3}{8\pi^2}\,t\times \Big{\{} 1+c\,a_s+(c\,a_s)^2 +\frac{1}{2}\big{[} (c\,a_s)^3-(c\,a_s)^4+(c\,a_s)^5 \big{]}\Big{\}},
\eeq
where $c=17/3$ and the last term is our prediction. Then, the estimated coefficients lead to:
\beq
 \frac{1}{\pi}\,{\rm Im}\psi_{\bar qq}(t)\vert^{\rm estimate}_{\rm Pert}=  \frac{3}{8\pi^2}\,t\,\ga 1+\frac{17}{3}a_s+32.1a_s^2+90.98a_s^3-515.56a_s^4+2921.52a_s^5 \dr,
\eeq
where the coefficients up to order $\alpha_s^4$ agree remarkably with the analytic ones given in Eq.\,\ref{eq:pert}.  The estimated  N5LO contribution is:
\beq
\frac{1}{ \pi}{\rm Im}\, \psi_{\bar qq}(t)\vert_{\rm N5LO}=\ga \frac{3}{8\pi^2}\dr\,t \times 2921.51\,a^5_s~,
\eeq
which should be checked from a direct calculation.  We truncate the PT series at N3LO for extracting the optimal result and consider that the error due to the truncation of the PT series comes from the sum N4LO\,$\oplus$\,N5LO. Then,  we obtain at the optimal value of $\tau \simeq 1$ GeV$^{-2}$:
\beq
\Delta M_{\bar qq}\vert_{\rm Geom}= \pm 1\,{\rm MeV},\,\,\,\,\,\,\,\,\,\,\,\,\,\,\,\,\,\,\,\,\,\,\,\,\,\,\,\,\,\,\,\,\,\,\,\,
\Delta f_{\bar qq}/m_q\vert_{\rm Geom}= \pm 15\times 10^{-3}.
\eeq
An alternative way to estimate the error is to parametrize the contribution of the non-calculated order term by the tachyonic gluon mass $\lambda$ proposed in \,\cite{CNZa,ZAKa,ZAKb,ADS1,ADS2,ADS3} which is a phenomenological alternative to the large $\beta$  approach on UV renormalon:
\beq
\frac{1}{\pi}\,{\rm Im}\, \psi(t)\vert_{\rm Tach}=  - \ga\frac{3}{8\pi^2} \dr4\, a_s\,\lambda^2~,
\eeq
with\,\cite{SND2,CNZa,TERAYEV} :
\beq
a_s\,\lambda^2= -(7\pm 3)\times 10^{-2}~{\rm GeV}^2.
\eeq
This leads to :
\beq
\Delta M_{\bar qq}\vert_{\rm Tach}= \pm 27\,{\rm MeV},\,\,\,\,\,\,\,\,\,\,\,\,\,\,\,\,\,\,\,\,\,\,\,\,\,\,\,\,\,\,\,\,\,\,\,\,
\Delta f_{\bar qq}/m_q\vert_{\rm Tach}= \pm 24\times 10^{-3}.
\eeq
 Though it has been shown in Ref.\,\cite{CNZb} that the two approaches are equivalent, we shall see in the next sections that the tachyonic gluon mass tends to overestimate the error in some cases. For definiteness, we shall keep, here and in the following sections, the most optimistic errors from the geometric growth of the $\alpha_s$-coefficient. 

 One can notice from Figs.\,\ref{fig:pert-qq} and \ref{fig:trunc-qq} that the size of the coupling is strongly affected by the PT radiative corrections and starts to be stable from N2LO corrections. 
\subsection*{\b On the truncation of the OPE}

We  truncate the OPE by assuming that the next non-calculated term is of the form:
\beq
\Delta {\rm OPE} =(\Lambda^2\tau)\times (D=6\,\rm contributions).
\label{eq:ope}
\eeq
Adding these estimated contributions as some other sources of the errors in Table\,\ref{tab:res},  we obtain the final result within a NWA:
\beq
M_{\bar qq}=1246(95)~{\rm MeV}, \,\,\,\,\,\,\,\,\,\,\,\,\,\,\,\,\,\,\,\,\,\,\,\,\,\,\,\,\,\,\,\,\,\,\,\,f_{\bar qq}/\bar m_{q}(\tau) = 274(43)\times 10^{-3}.
\label{eq:nwa}
\eeq
where the mass $\bar m_q(\tau)$ is evaluated at $\tau\simeq 0.9$ GeV$^{-2}$ at which the optimal result has been extracted. 
\subsection*{\b Finite width correction}
\begin{figure}[hbt]
\begin{center}
\includegraphics[width=9.cm]{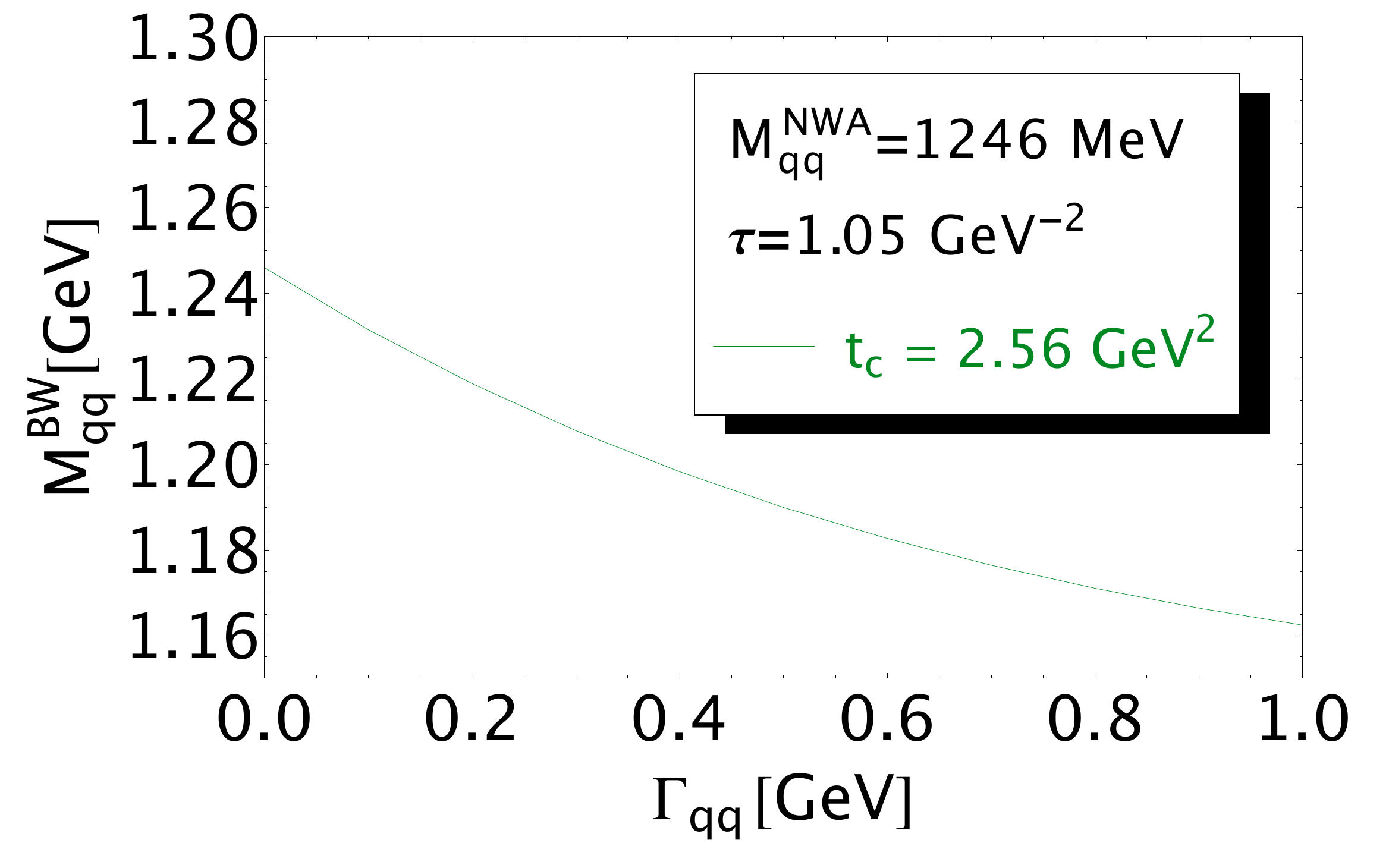}
\vspace*{-0.5cm}
\caption{\footnotesize   Analysis of the finite width effect on $M_{\bar qq}$  for $t_c=2.56$ GeV$^2$ corresponding to the central value of $M_{\bar qq}=1246 MeV $ in a NWA at  $\tau=1.05$ GeV$^{-2}$.} 
\label{fig:bw}
\end{center}
\vspace*{-0.5cm}
\end{figure}

In the literature,  one often identifies the lightest scalar meson with the broad $\sigma$ found from $\pi\pi$ and $\gamma\gamma$ scattering (see Eq.\,\ref{eq:msigma}). In order to take into account the finite width correction to the previous result obtained using a narrow width approximation (NWA), we make the replacement:
\beq
\pi\delta(t-M_{\sigma}^2)\to BW(t)= \frac{ M_{\sigma}\Gamma_{\sigma}}{(t- M^2_{\sigma})^2+ M^2_{\sigma}\Gamma^2_{\sigma}}.
\label{eq:bw}
\eeq
in the parametrization of the spectral function. Then, we study the effect of the width to the ratio:
\beq
\ga M_{\bar qq}^{\rm BW}\dr^2 =  \frac{\int_0^{t_c} dt\,t^2\,e^{-t\tau}\, BW(t)}{\int_0^{t_c} dt\,t\,e^{-t\tau}BW(t)}.
\label{eq:bwsr}
\eeq
We use in the integral the mass from the NWA and the value $\tau$=1.05 GeV$^-2$ where the optimal value of $M_{\bar qq}\vert_{NWA}$ has been obtained for $t_c$=2.56  GeV$^2$ corresponding to the mean value of the mass from the extremal values of $t_c$.   We show the analysis in Fig.\,\ref{fig:bw}. From the range $\Gamma_{\pipi}= 120$ MeV predicted from vertex sum rule\,\cite{BN,SNB2,SNG} to 520  MeV for the complex pole (see Table\,\ref{tab:sigma}) and 700 MeV for the On-shell / Breit-Wigner mass (see Eq.\,\ref{eq:sigdata}), one can see that the width decreases respectively slightly the mass by (in units of MeV):
\beq
\Delta M_{\bar qq}^{\rm BW}=-22\vert_{\rm Vertex \,SR}~, ~~~~~ - 60\vert_{\rm Pole}~, ~~~~~ -70\vert_{\rm On-shell}~,
\label{eq:width}
\eeq
 which is within the errors of the LSR determinations. 
Using the width predicted from vertex sum rules, we deduce:
\beq
M_{\bar qq}\vert_{\rm Vertex\, SR}=1229(95)~{\rm MeV},
\label{eq:respi-vertex}
\eeq
which can be compared with the on-shell mass in Eq.\,\ref{eq:sigdata}. 
\subsection*{\b Factorization of the four-quark condensates}
\begin{figure}[hbt]
\begin{center}
\hspace*{-7cm} {\bf a) \hspace*{8.cm} \bf b)} \\
\includegraphics[width=8.cm]{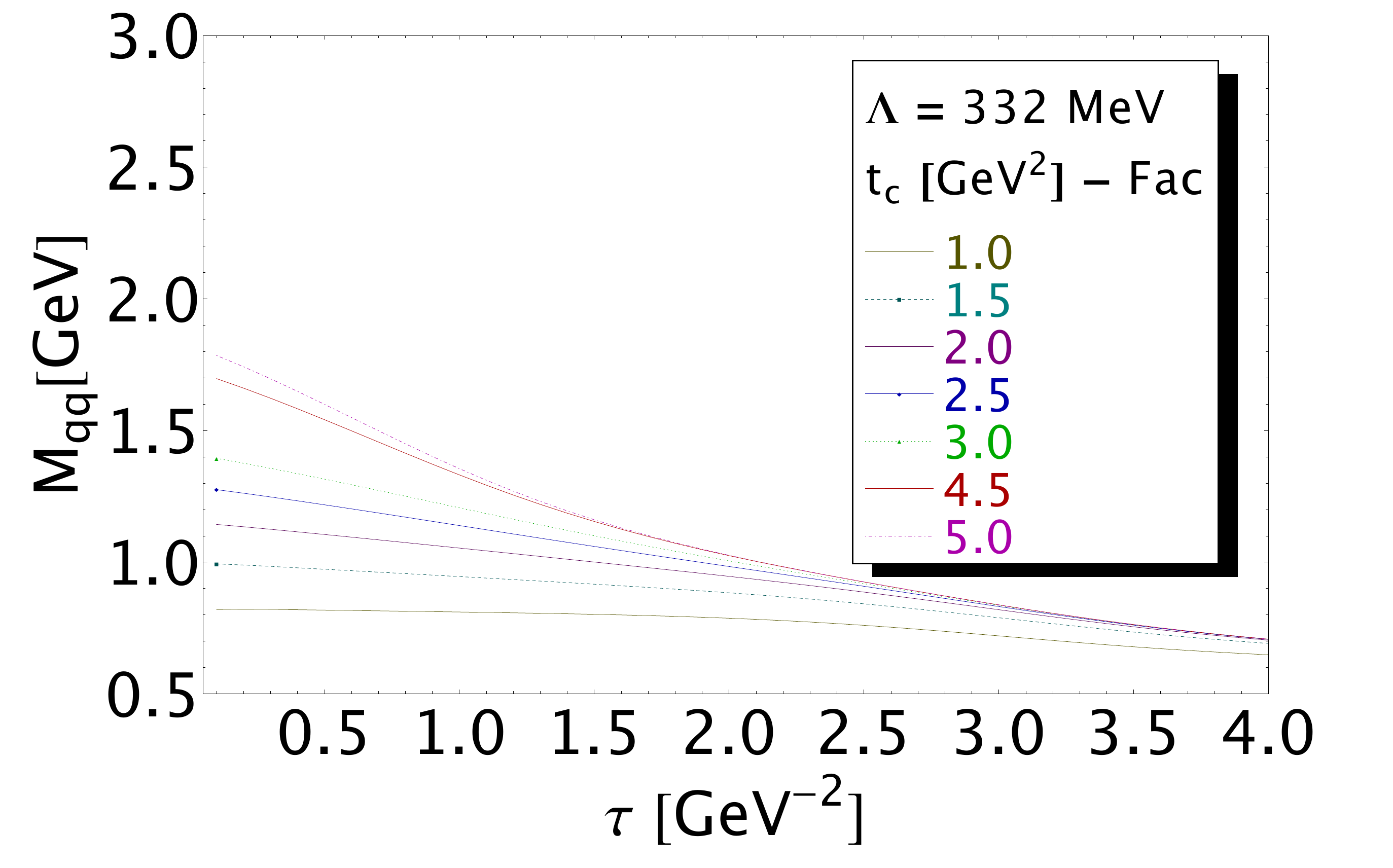}
\includegraphics[width=8.cm]{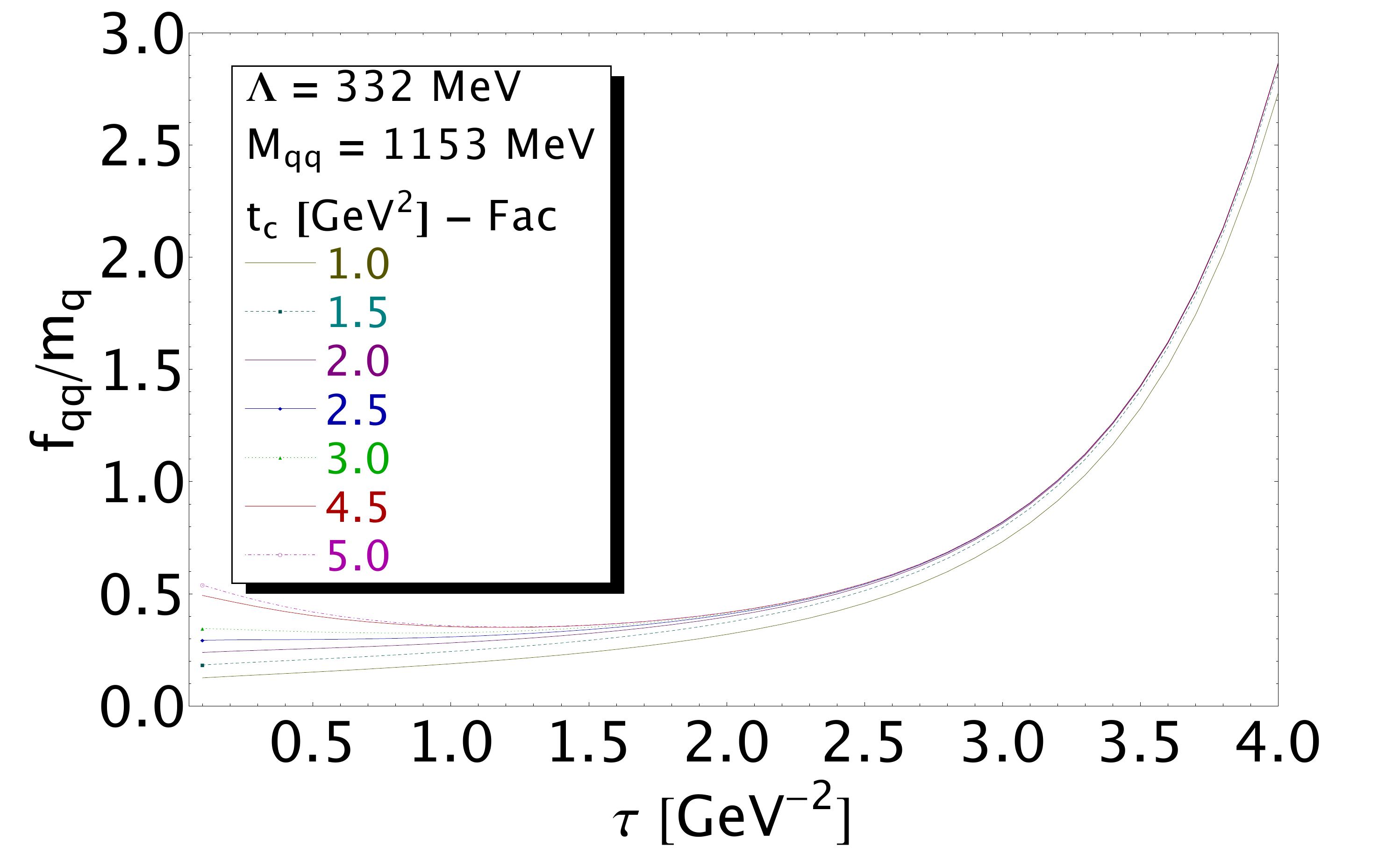} 
\vspace*{-0.5cm}
\caption{\footnotesize   Behaviour of a) mass and  b) coupling of the $\bar qq\equiv\frac{1}{\sqrt{2}}\ga \bar u u \pm \bar dd\dr $ meson states versus $\tau$ for different values of $t_c$ using a factorization of the four-quark condensate.} 
\label{fig:fac}
\end{center}
\vspace*{-0.5cm}
\end{figure}

Another point which does not make a consensus in the literature is the vacuum saturation or factorization in order to estimate the size of the four-quark condensate:
\beq
\la 0\vert \bar qq \bar q'q'\vert 0\ra \simeq  \rho \la 0\vert \bar qq \vert 0\ra\la 0\vert \bar q'q'\vert 0\ra,
\eeq
where $\rho$ quantifies the deviation from the vacuum saturation. Several analysis of $e^+e^-\to hadrons$ and $\tau$-decay data have shown that $\rho\simeq (3-6)$ \,\cite{FESR2,SNTAU,LNT,DOSCH,SN95,SOLA,SNe23} while the renormalization of the four-quark operators indicates that vacuum saturation is inconsistent with a renormalization group invariance of the four-quark operators\,\cite{SNT}.  In the following, we shall test the effect of this assumption on the mass and coupling predictions. The analysis is shown in Fig.\,\ref{fig:fac}.  One can notice that the $\tau$-stability of the mass is less good than in the case of a violation of factorization where it is an inflexion point here.  It makes its localization less precise while the range of $t_c$-values is more restricted for $t_c=(3-4.5)$ GeV$^2$. For the mass, this leads to an error of 33 MeV from $t_c$ and of 77 MeV  taking $\tau=1.5$ GeV$^{-2}$ and $\Delta\tau=0.2$ GeV$^{-2}$. For the coupling, the minimum appears at $\tau\simeq$ 0.9 and 1.2 GeV$^{-2}$. The errors due to $t_c$  and $\tau$ on the coupling are 12$\times 10^{-3}$ and 3$\times 10^{-3}$. Adding the other sources of errors from Table\,\ref{tab:res}, one obtains:
\beq
M_{\bar qq} = 1131(90)~{\rm MeV}, \,\,\,\,\,\,\,\,\,\,\,\,\,\,\,\,\,\,\,\,\,\,\,\,\,\,\,\,\,\,\,\,\,\,\,\,f_{\bar qq}/\bar m_{q}(\tau) = 338(24)\times 10^{-3},
\eeq
where we have subtracted to the mass the width correction of --22 MeV.

\section{The $\bar us$ and $\bar ss$ states}
One extends the analysis to the current:
\beq
J_{\bar us}\equiv \partial_\mu V^\mu_{\bar us}=(m_u-m_s)\bar us, \,\,\,\,\,\,\,\,\,\,\,\,\,\,\,\,\,{\rm and}\,\,\,\,\,\,\,\,\,\,\,\,\,\,\,\,\,\,\,J_3=m_s\,\bar ss.
\eeq
We use the previous expressions in Eqs.\,\ref{eq:pert} and \,\ref{eq:npert} by replacing $d$ by $s$ for $\bar us$ state and  $u,d$ by $s$ for the $\bar ss$ states. The behaviours of the different curves are similar to the previous case of $
\bar qq$ state and will not be repeated here. We just quote the results from Table\,\ref{tab:res}. 
\bea
M_{\bar us}&=&1276(61)~{\rm MeV}, \,\,\,\,\,\,\,\,\,\,\,\,\,\,\,\,\,\,\,\,\,\,\,\,\,\,\,\,\,\,\,\,\,\,\,\,f_{\bar us}/{(\bar m_{u}-\bar m_s)(\tau)} = 264(25)\times 10^{-3}\nnb\\
M_{\bar ss}&=&1288(65)~{\rm MeV}, \,\,\,\,\,\,\,\,\,\,\,\,\,\,\,\,\,\,\,\,\,\,\,\,\,\,\,\,\,\,\,\,\,\,\,\,f_{\bar ss}/{(\bar m_s)(\tau)}= 256(19)\times 10^{-3}.
\eea
Comparing these values with the one for the $\bar uu\pm\bar dd$ state, one can notice that the SU(3) breakings are small. They shift the mass by about 28 MeV for $\bar us$ and 40 MeV for $\bar ss$ states. The corresponding shift for the coupling is --$10\times 10^{-3}$ (resp. --$18\times 10^{-3}$) for the $\bar us$ (resp. $\bar ss$ state). These shifts are relatively tiny. The ratio of masses:
\beq
\frac{M_{\bar ss}}{M_{\bar us}}\simeq 1.01
\eeq
is consistent with the direct determination $(1.03\pm 0.02)$ in Ref.\,\cite{SNG}. 

\section{Comments on ordinary $\bar qq$ mesons}
\d We consider the results  in our previous analysis as an update of the ones obtained earlier in Refs.\,\cite{SNB2,SNG,SNPRD}. 

\d  A comparison of these results with the on-shell mass for the $\sigma$ and $f_0/a_0(980)$ are more appropriate than with the residue at the complex plane. These results indicate that the lightest $\bar qq$ mesons are in the range:
 \beq
 M_{\bar qq'}\simeq (1040 \sim 1353)~{\rm   MeV}~~~~for~~~~~~~~q,q'\equiv u,d,s~
  \eeq
   within the accuracy of the LSR approach. 
   
   \d This range of values is consistent with the on-shell masses of the observed mesons $\sigma/f_0(500), f_0/a_0(980)$ and $K^*_0(1430)$. However , the estimated hadronic width  of about 120 MeV does not favour a pure $\bar qq$ interpretation of the broad $\sigma/f_0(500)$ but may favour the  meson-gluonium mixing scenario proposed in\,\cite{BN,SNG,SNPRD}. 
   
\d The predicted value  $M_{\bar us}\simeq 1276(58)$ MeV is comparable with the one of the $K^*_0(1350)$ but its predicted total width from vertex sum rule is expected to be narrower than the experimental data as can be deduced from\,\cite{SNG}.   

\d   The predicted value of the $\bar ss$ state mass is too low compared to the candidate $f_0(1710)$ while the predicted $K^+K^-$ width\,\cite{SNG} is about 1/2 of the observed one. 


 

\section{The \boldmath$\sigma/f_0(500)$ as a dipion molecule}
 \subsection*{\b The $\pi^+\pi^-$ dipion field and its two-point function}
 The interpolating current of the dipion state:
 \beq
 {\cal O}_{\pi^+\pi^-}= J_{\pi^+}\otimes J_{\pi^-} (x)\equiv (m_u+m_d)^2\tilde{\cal O}_{\pi^+\pi^-}(x)
 \eeq
 is the convolution of the two renormalization group invariant divergences of  axial current (pion current):
 \beq
J_{\pi^+}(x)=(m_u+m_d):\bar d(i\gamma_5) u (x):.
 \eeq
 We do not consider the scalar $\bar d u$ current which cannot participate to leading order to the decay $\sigma\to\pi^+\pi^-$. A similar choice will be done for the $K^+K^-$ and $\eta\pi^0$ molecules.
For convenience, we shall omit the global factor $(m_u+m_d)$ of the pion current such that the corresponding two-point function has an anomalous dimension. We shall see that this procedure will not affect the mass but the coupling. 
Then, the $\sigma$ two-point correlator is given by :
  \beq
  \psi_{\pi^+\pi^-}(q^2)=i\int d^4x\,e^{iqx}{\cal T}\la 0\vert  \tilde{\cal O}_{\pi^+\pi^-}(x)\ga \tilde{\cal O}_{\pi^+\pi^-}(0)\dr^\dagger \vert 0\ra,
  \label{eq:psi}
  \eeq
where $\ga 1/\pi\dr$ Im\,$\psi_{\pi^+\pi^-}(t) \equiv \rho_{\pi^+\pi^-}(t)$ is :
\bea
\rho^{pert}_{\pi^+\pi^-}&=&\frac{t^4}{5\times 2^{14}\pi^6},\,\,\,\,\,\,\,\,\,\,\,\,\,\,\,\,\,\,\,\,\,\,\,\,\,\,\,\,\,
\rho^{\la \bar qq\ra}_{\pi^+\pi^-}=\frac{\ga m_d-2m_u\dr\la \bar dd\ra + 
\ga m_u-2m_d\dr\la \bar uu\ra }{ 2^{8}\pi^4}t^2\nnb\\
\rho^{\la G^2\ra }_{\pi^+\pi^-}&=&\frac{\la \alpha_s G^2\ra }{ 2^{10}\pi^5}\,t^2,\,\,\,\,\,\,\,\,\,\,\,\,\,\,\,\,\,\,\,\,\,\,\,\,\,\,\,\,\,
\rho^{\la \bar q Gq\ra}_{\pi^+\pi^-}=\frac{(2m_d+3m_u)\la \bar dGd\ra + (2m_u+3m_d)\la \bar uGu\ra }{2^{8}\pi^4}\,t\nnb\\
\rho^{\la \bar qq\ra^2}_{\pi^+\pi^-}&=& \frac{\rho\la \bar uu\ra\la \bar dd\ra }{2^{4}\pi^2}\,t,\,\,\,\,\,\,\,\,\,\,\,\,\,\,\,\,\,\,\,\,\,\,\,\,
\rho^{\la G^3\ra}_{\pi^+\pi^-}={\cal O}(m_q^2\,\la   G^3\ra),
\label{eq:pipi}
\eea
where:
\beq
\ga m_u+m_d\dr\la \bar uu +\bar dd\ra = -2f_\pi^2m_\pi^2,
\eeq
and $\rho$ indicates the deviation from the factorization of the four-quark condensate.

\subsection*{\b Higher order PT QCD corrections}
 For this purpose, we use a factorization of the molecule spectral function which is given 
 by the product of the two pseudoscalar ones. 
 In this way, we obtain the convolution integral\,\cite{PICH,SNPIVO}:
\beq
\hspace*{-0.1cm}\frac{1}{ \pi}{\rm Im}\, \psi_{\pi^+\pi^-}(t)= k_\pi\,\int_{0}^tdt_1\,
\int_{0}^{(\sqrt{t}-\sqrt{t_1})^2}\hspace*{-0.8cm}dt_2~\lambda^{1/2}
\ga \frac{t_1}{ t}+ \frac{t_2}{ t}-1\dr^2\times \frac{1}{\pi}{\rm Im} \,\psi_{\pi^+}(t_1) \frac{1}{\pi} {\rm Im}\,\psi_{\pi^-}(t_2),
\eeq
with the phase space factor:
\beq
\lambda=\ga 1-\frac{\ga \sqrt{t_1}- \sqrt{t_2}\dr^2}{ t}\dr \ga 1-\frac{\ga \sqrt{t_1}+ \sqrt{t_2}\dr^2}{ t}\dr~,
\eeq
where :
\beq
k_{\pi}=\frac{35}{208\pi^2}
\eeq
is an appropriate normalisation factor. 
The convolution  representation is expected to be valid for large $N_c$ while the non-factorized contribution is found to be small in the example of the $\bar BB$ system\,\cite{SNPIVO}. The expression of the pion spectral function is known up to order  $\alpha_s^4$ as defined in Eq.\ref{eq:pert}.  

\subsection*{\b Parametrization of the spectral function}
 The contribution of the $\sigma$ to the spectral function can be introduced within
 the minimal duality ansatz:
\beq
\frac{1}{\pi} \mbox{Im}\,\psi_{\pi^+\pi^-}(t)=2f^2_{\pi^+\pi-} M_{\pi^+\pi-}^8\delta\ga t- M^2_{\pi^+\pi-}\dr\,\oplus \, \theta(t-t_c)``{\rm QCD \,\, Continuum }"\,
\eeq
with the normalization:
\beq
\la 0\vert J_{\pi^+\pi-} \vert {\pi^+\pi-}\ra= \sqrt 2 f_{\pi^+\pi-} M_{\pi^+\pi-}^4.
\label{eq:coupling}
\eeq
At the first step, we shall use a NWA for the $\sigma$ and shall estimate later on the finite width correction. 
\subsection*{\b Mass and Coupling of the $\pi^+\pi^-$ Molecule}
\begin{figure}[hbt]
\begin{center}
\hspace*{-7cm} {\bf a) \hspace*{8.cm} \bf b)} \\
\includegraphics[width=8.cm]{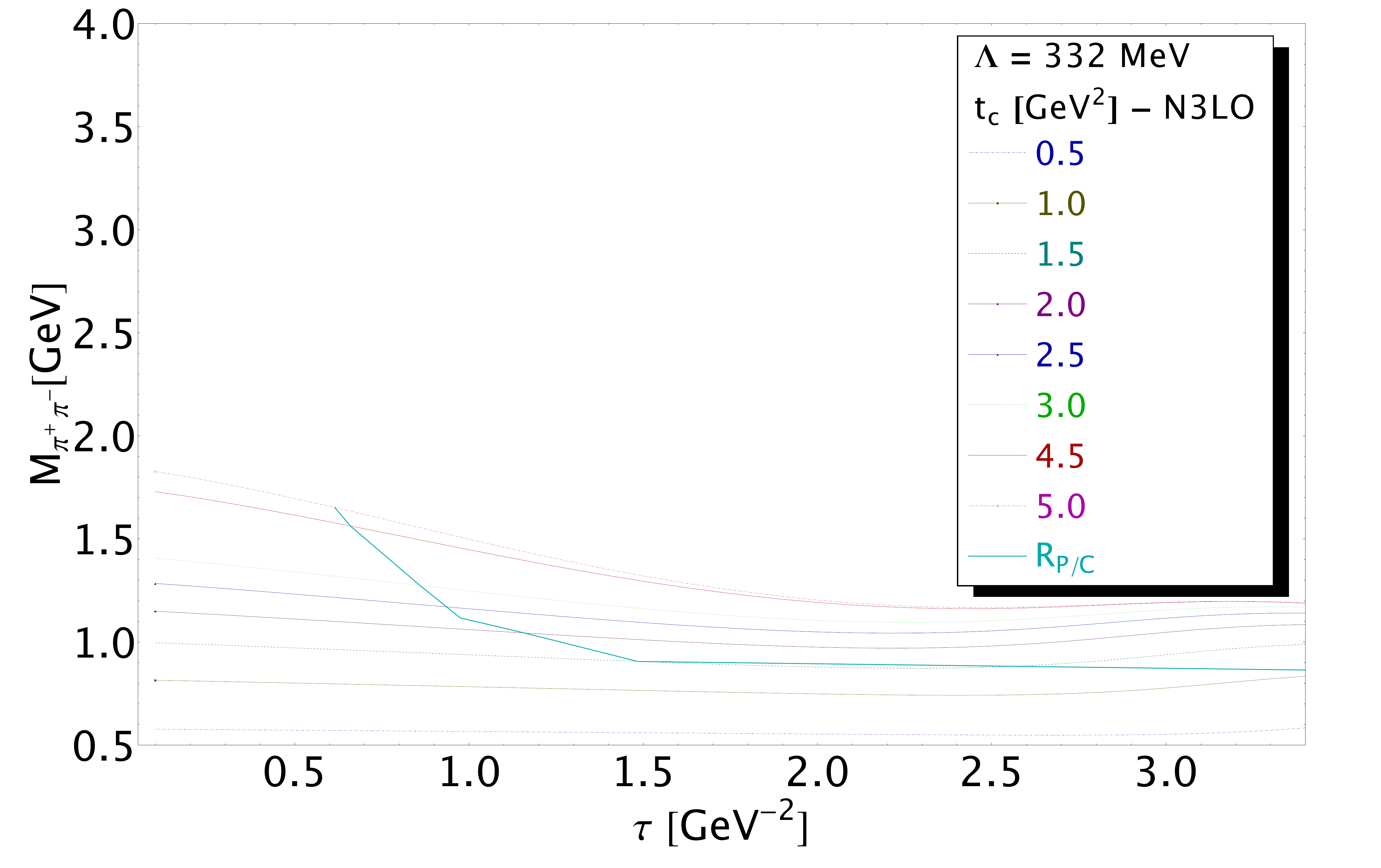}
\includegraphics[width=8cm]{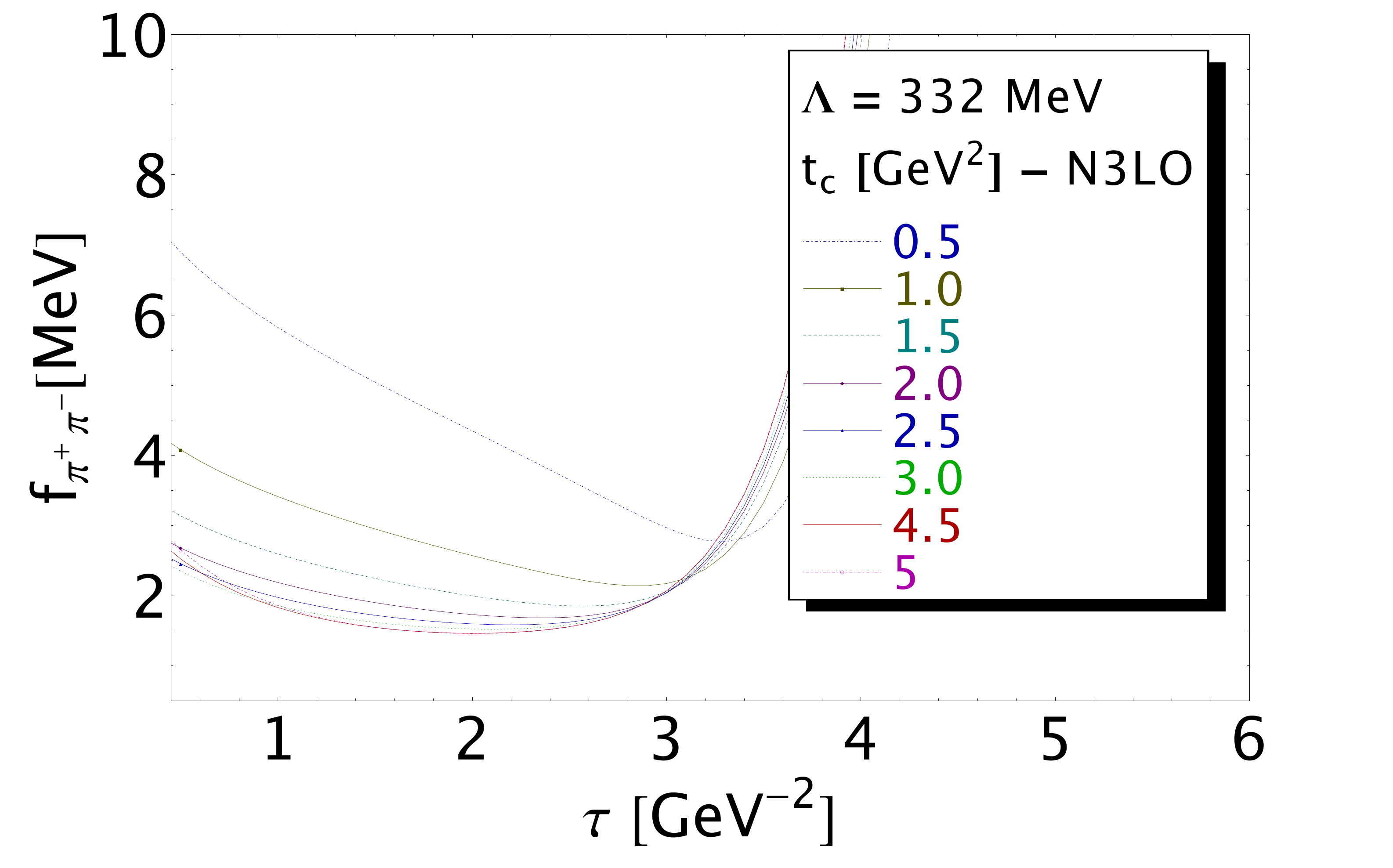} 
\vspace*{-0.5cm}
\caption{\footnotesize   $\pi^+\pi^-$ molecule: a) mass; b) coupling versus $\tau$ for different values of $t_c$. } 
\label{fig:pipi}
\end{center}
\vspace*{-0.5cm}
\end{figure}

 Using the previous QCD expressions, one can estimate the mass from the ratio of moments ${\cal R}_{10}$ and the coupling from the low moment ${\cal L}_0$.  We show the analysis of the mass and the one of the coupling in Fig.\,\ref{fig:pipi} at N3LO of PT series and retaining the $D=6$ condensate contributions. 
\begin{figure}[hbt]
\begin{center}
\includegraphics[width=9cm]{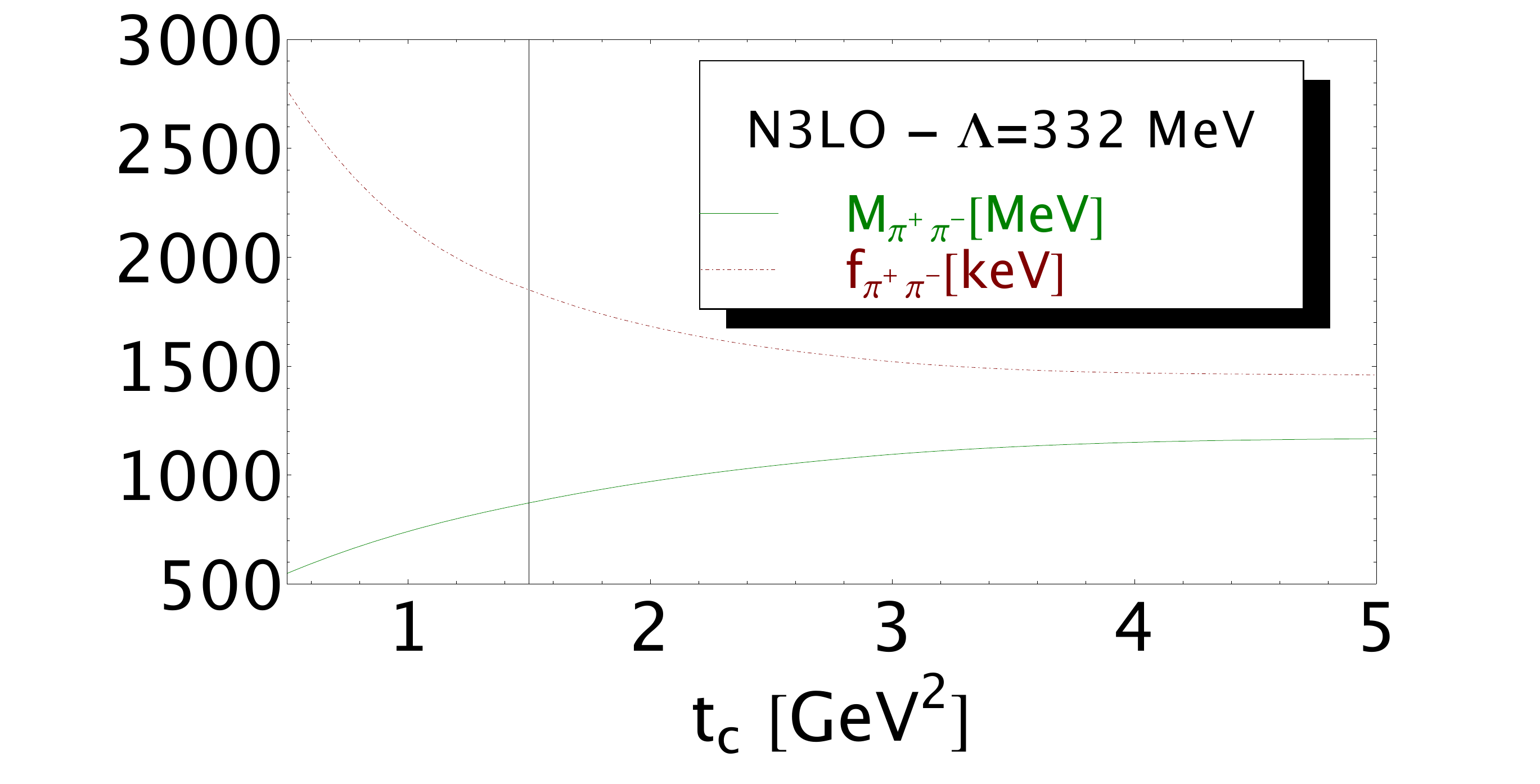}
\vspace*{-0.5cm}
\caption{\footnotesize   Behaviour of the optimal results versus $t_c$ for the $\pi^+\pi^-$ molecule.} 
\label{fig:tcpi}
\end{center}
\vspace*{-0.5cm}
\end{figure} 

The $\tau$-stability of the mass is obtained for  $t_c\geq 0.5$ GeV$^2$ while the result reaches a $t_c$-stability  above $t_c=4.5$ GeV$^2$ (see Fig.\,\ref{fig:pipi}). The $R_{P/C}$ condition does not allow the region on the left of the $R_{P/C}$ curve.

We show in Fig.\,\ref{fig:tcpi} the $t_c$-behaviour of the optimal results (minimum in $\tau$). The vertical line $t_c=1.5$ GeV$^2$ is the minimum value of $t_c$ allowed by the $R_{P/C}\geq 1$ condition.  Then, we obtain the optimal result for $t_c=(1.5-4.5)$ GeV$^2$ at N3LO within a NWA:
\beq
M_{\pi^+\pi^-}=1017(144)_{t_c}~{\rm MeV},\,\,\,\,\,\,\,\,\,\,\,\,\,\,\,\,\,\,\,\,\,\,\,\,\,\,\,\,\,\,\,\,\,\,\,\,
f_{\pi^+\pi^-}=1657(193)_{t_c}~{\rm keV}
\eeq
 where only the errors from $t_c$ are quoted. 
\subsection*{\b On the truncation of the PT series}
\begin{figure}[hbt]
\begin{center}
\hspace*{-7cm} {\bf a) \hspace*{8.cm} \bf b)} \\
\includegraphics[width=8.2cm]{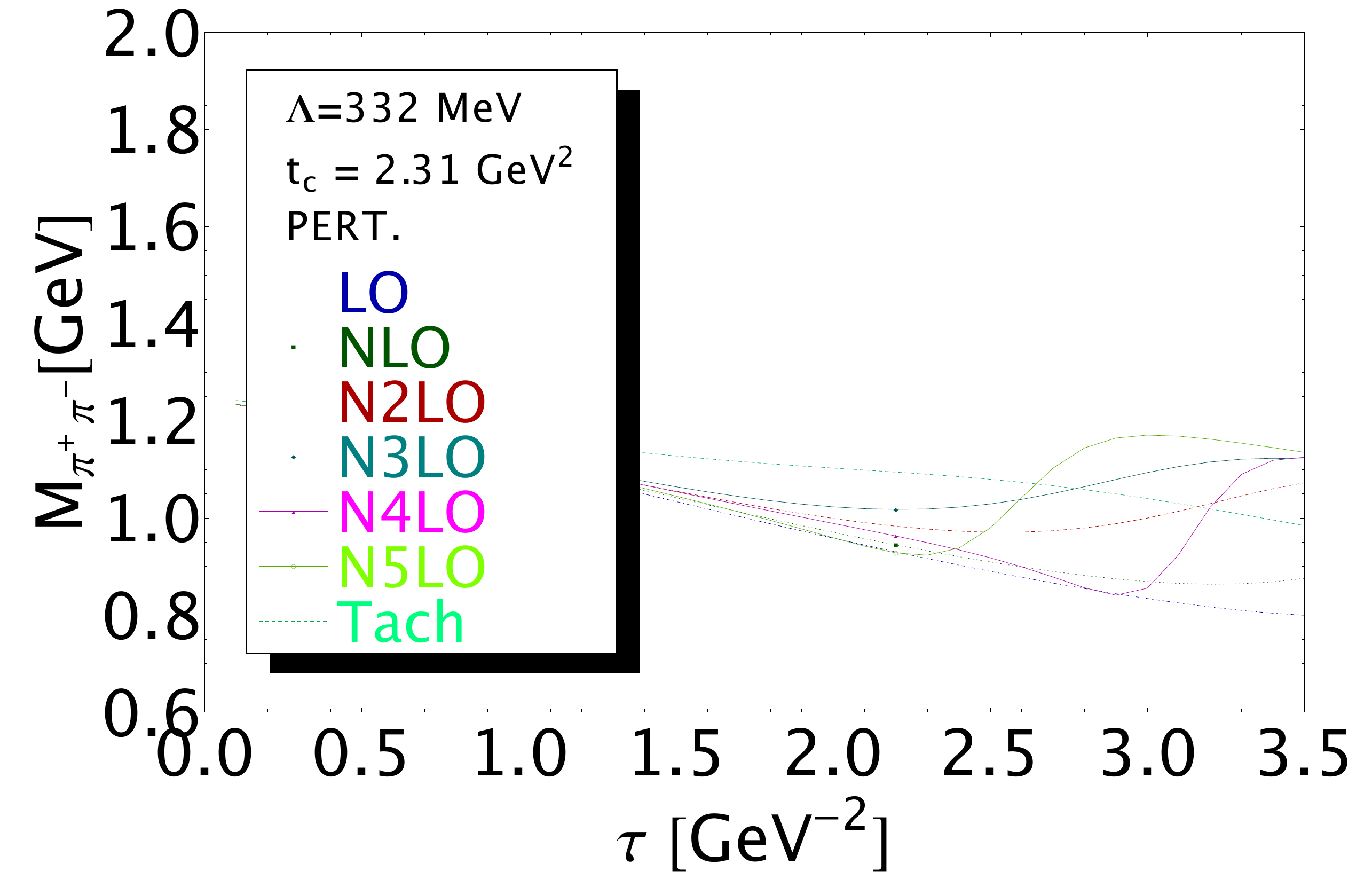}
\includegraphics[width=7.9cm]{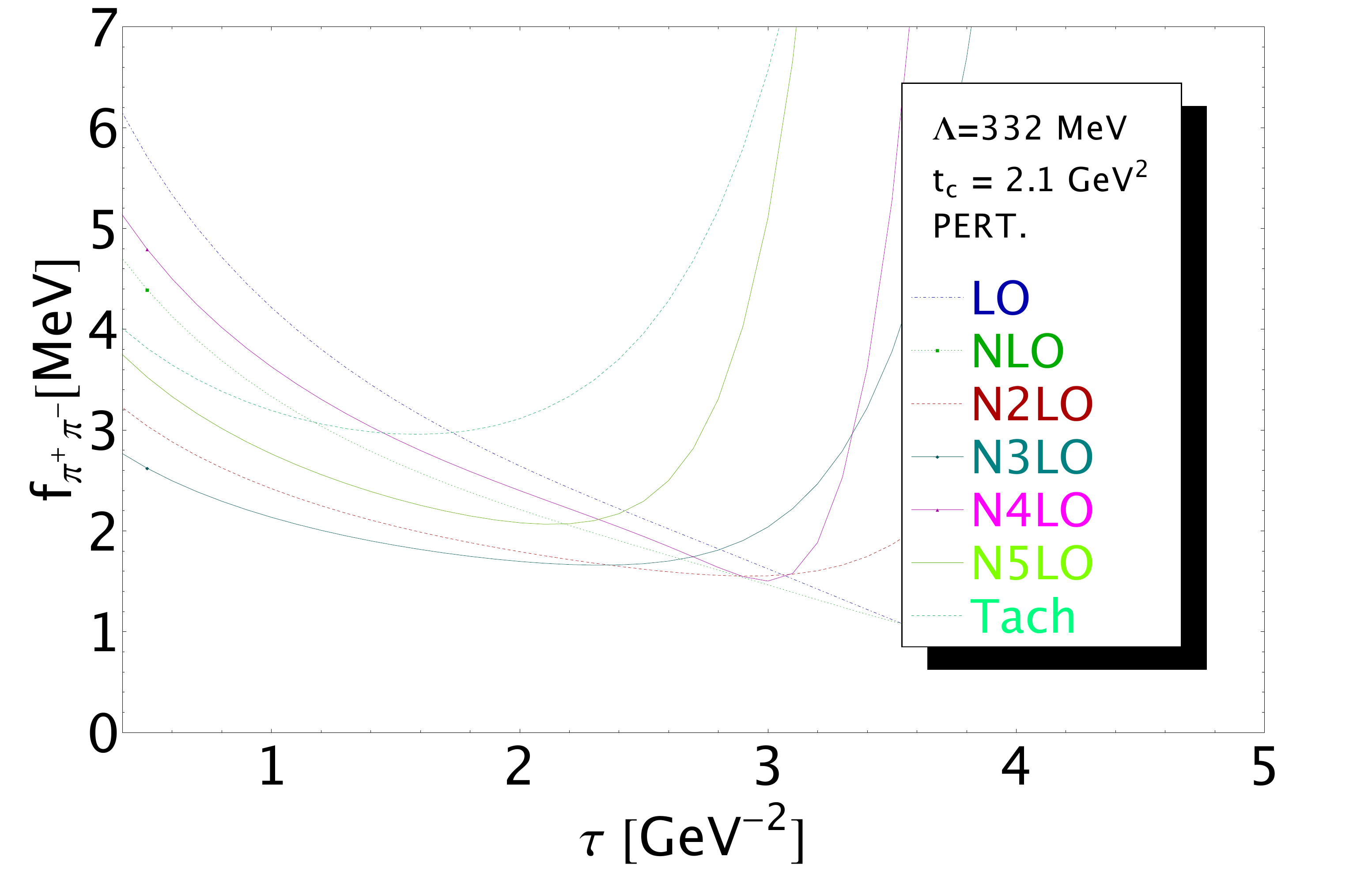} 
\vspace*{-0.5cm}
\caption{\footnotesize   Truncation of the PT series for the $\pi^+\pi^-$ molecule: a) mass; b) coupling versus $\tau$. The values of $t_c$ correspond to the central values of the mass and of the coupling.} 
\label{fig:pipi-pert}
\end{center}
\vspace*{-0.5cm}
\end{figure}

\begin{figure}[hbt]
\begin{center}
\includegraphics[width=8.2cm]{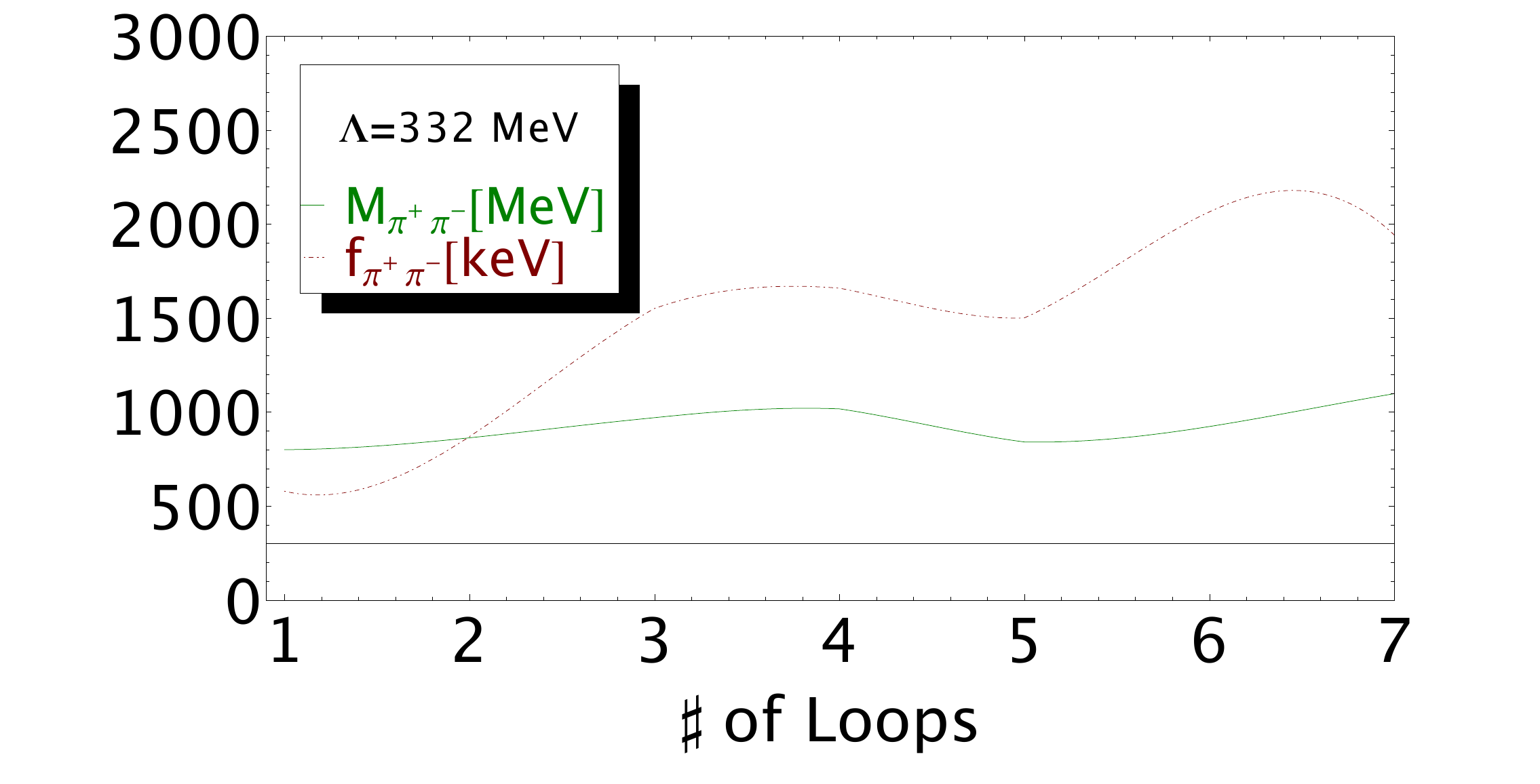}
\vspace*{-0.5cm}
\caption{\footnotesize   Behaviour of the optimal results versus the truncation of the PT series for $M_{\pi^+\pi^-}$ molecule. \# 7 corresponds to the tachyon gluon mass contribution. We take $t_c=$ 2.31 (resp. 2.10) GeV$^2$ which corresponds to the central value of the mass (resp. coupling). } 
\label{fig:pipi-trunc}
\end{center}
\vspace*{-0.5cm}
\end{figure}

\d We show in Fig.\,\ref{fig:pipi-pert} the behaviour of $M_{\pi^+\pi^-}$ and $f_{\pi^+\pi^-}$ versus $\tau$ for a given value of $t_c=2.31$ (resp. 2.1)  GeV$^2$ (which reproduces their central values) for different truncation of the PT series. We notice that the inclusion of the NLO to N3LO corrections improve the analysis as the optimal results shift to lower values of $\tau$. However, due to its negative sign the N4LO contribution tends to increase the $\tau$ minimum value and decrease the value of the mass and coupling. Like in the case of the $\bar qq$ state, we interpret this change of N4LO sign as a signal of the appearance of alternate signs of the QCD PT series where the asymptotic form of the series maybe reached at the N3LO.  We study the behaviour of the results versus the truncation of the PT series in Fig.\,\ref{fig:pipi-trunc} where a stability is obtained for N2LO-N3LO which we consider as our optimal result.

\d In order to estimate the error due to the truncation of the PT series, we again proceed like in the case of $\bar qq$ meson 
where the error comes from the sum of N4LO and N5LO contributions. 

\subsection*{\b On the truncation of the OPE}

We  truncate the OPE by assuming that the next non-calculated term is of the form in Eq.\,\ref{eq:ope}. 

\subsection*{\b Estimate of the errors}

One should also notice that the relative large value of the error coming  from the four-quark condensate compared to the one for ordinary $\bar qq$ meson in the previous section comes from the fact that here we parametrize the four-quark condensate as\,:
\beq
\rho\la\bar\psi\psi\ra^2= (3.4\pm 0.5)\la\overline{\bar\psi\psi}\ra^2(\tau)
\eeq
where we take into account the log-dependence of $\la\bar\psi\psi\ra$ as the four-quark condensate  contributes without $\alpha_s$ in the molecules and (as we shall see) in the four-quark states.  In the case of $\bar qq$ state we have neglected this log-dependence and use directly the value given in Table\,\ref{tab:param} as the four-quark contributes as $\alpha_s\la\bar\psi\psi\ra^2$.  Adding in Table\,\ref{tab:res} these previous estimated contributions as another sources of the errors,  we obtain the final result within a NWA:
\beq
M_{\pi^+\pi^-}=1017(159) ~{\rm MeV},\,\,\,\,\,\,\,\,\,\,\,\,\,\,\,\,\,\,\,\,\,\,\,\,\,\,\,\,\,\,\,\,\,\,\,\,
f_{\pi^+\pi^-}=1657(277)~{\rm keV}.
\label{eq:nwapi}
\eeq

\subsection*{\b Finite width corrections}
We proceed as in the previous section for estimating the finite width corrections by doing the replacement in Eq.\,\ref{eq:bw}.
Then, one obtains:
\beq
 \ga M_{\pi+\pi^-}^{BW}\dr^2\simeq \frac{\int_0^{t_c} dt\,t^5\,e^{-t\tau}\, BW(t)}  {\int_0^{t_c} dt\,t^4\,e^{-t\tau}BW(t)}
\label{eq:bwsrpi}
\eeq
One can remark that the shape of the width correction differs completly from the one of the $\bar qq$ meson. This is due to the  different powers of $t$ entering in the spectral  integral.  
From $\Gamma_{\pipi}= $ 0 to 700 MeV (width of the on-shell mass), the mass increases by 154 MeV leading to the
final prediction:
\beq
M_{\pi^+\pi^-}^{BW}=1171(159)~{\rm MeV}
\label{eq:respi}
\eeq

\begin{figure}[hbt]
\begin{center}
\includegraphics[width=9.cm]{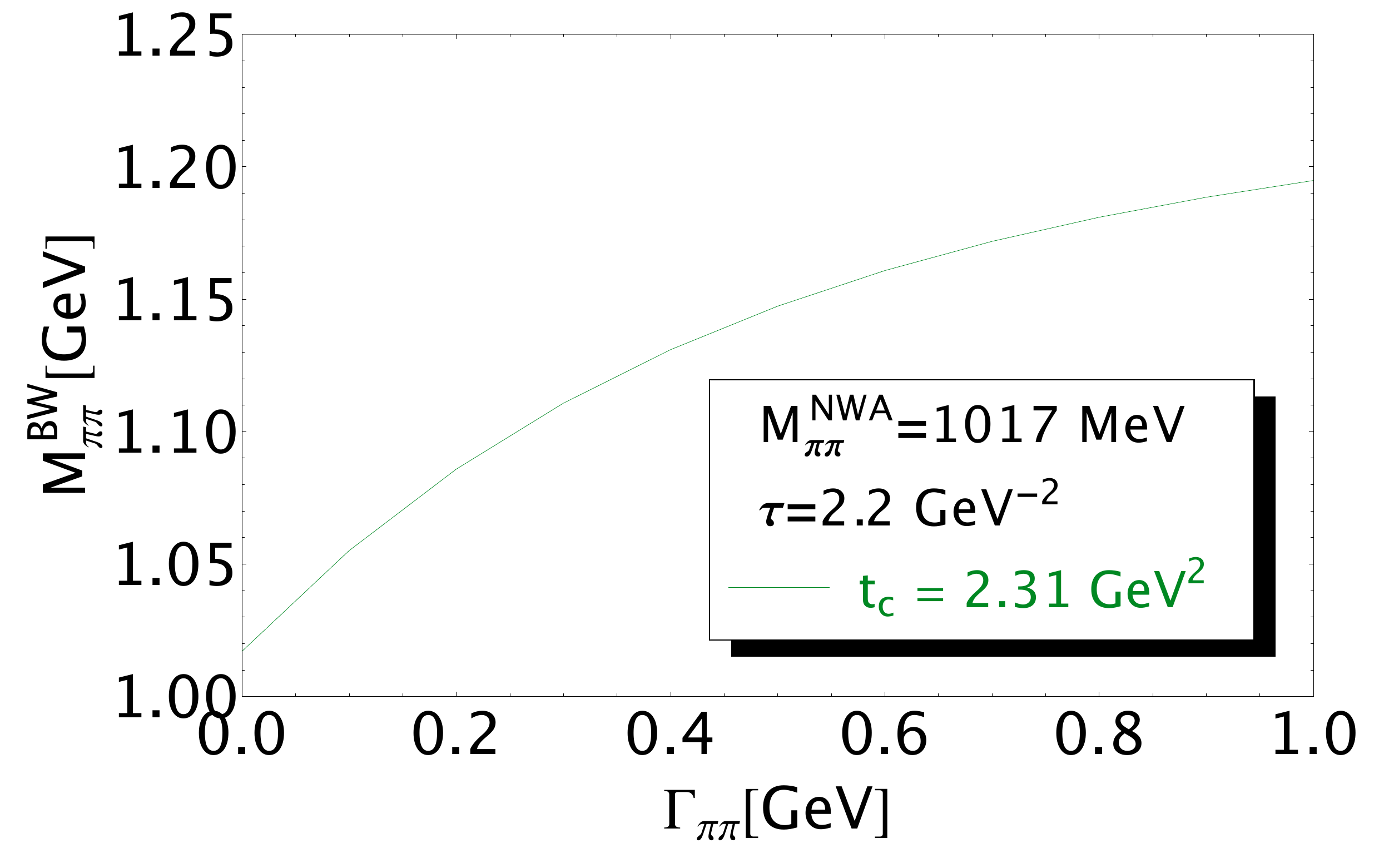}
\vspace*{-0.5cm}
\caption{\footnotesize   Analysis of the finite width effect on $M_{\bar qq}$  for $t_c=2.38$ GeV$^2$ corresponding to the central value of $M_{\bar \pi^+\pi^-}=1042 MeV $ in a NWA at  $\tau=2 $ GeV$^{-2}$.} 
\label{fig:bwpi}
\end{center}
\vspace*{-0.5cm}
\end{figure} 

\section{The $K^+K^-$ molecule}
We shall work with the current:
\beq
{\cal O}_{K^+K^-}(x)=  (\bar s\, i\,\gamma_5\,u)\otimes (\bar u\,i\,\gamma_5\, s)(x).
\eeq
The expression of the corresponding spectral function can be deduced from the one in Eq.\,\ref{eq:pipi} by replacing the $d$ by the $s$ quark to which 
we add the LO $m_s^2$ correction:
\beq
\frac{1}{ \pi}{\rm Im}\, \psi_{K^+K^-}(t)\vert_{m_s^2}=-\frac{m_s^2}{2^{11} \pi^6}t^3;
\eeq
The analysis and the shape of the different curves are very similar to the case of $\pi^+\pi^-$ and will not be shown. The different sources of the errors are given
in Table\,\ref{tab:res}. We obtain in the NWA:
\beq
M_{K^+K^-}=1056(214) ~{\rm MeV},\,\,\,\,\,\,\,\,\,\,\,\,\,\,\,\,\,\,\,\,\,\,\,\,\,\,\,\,\,\,\,\,\,\,\,\,
f_{K^+K^-}=1380(255)~{\rm keV}.
\label{eq:nwa-kk}
\eeq
\section{The $K^+\pi^-$ molecule}
We shall work with the current:
\beq
{\cal O}_{K^+\pi^-}(x)= (\bar s\, i\,\gamma_5\,u)\otimes (\bar u\,i\,\gamma_5\, d)(x).
\eeq
The expression of the corresponding spectral function reads to LO and up to dimension-six:
\bea
\rho^{pert}_{K^+\pi^-}&=&\frac{t^4}{5\times 2^{14}\pi^6} - \frac{m_s^2}{2^{12} \pi^6} t^3,\,\,\,\,\,\,\,\,\,\,\,\,\,\,\,\,\,\,\,\,\,\,\,\,\,\,\,\,\,\,\,\,\,\,\,\,\,\,\,\,\,\,\,\,\,\,
\rho^{\la G^2\ra }_{K^+\pi^-}=\frac{\la \alpha_s G^2\ra }{ 2^{10}\pi^5}\,t^2,\nnb\\
\rho^{\la \bar qq\ra}_{K^+\pi^-}&=&\frac{\ga m_d-2m_u\dr\la \bar dd\ra 
-2(m_d+m_s-m_u)\la \bar uu\ra 
+\ga m_s-2m_u\dr\la \bar ss\ra }{ 2^{9}\pi^4}t^2\nnb\\
\rho^{\la \bar q Gq\ra}_{K^+\pi^-}&=&\frac{(2m_d+3m_u)\la \bar dGd\ra + (4m_u+3(m_d+m_s))\la \bar uGu\ra 
+(2m_s+3m_u)\la\bar sGs\ra}{2^{9}\pi^4}\,t\nnb\\
\rho^{\la \bar qq\ra^2}_{K^+\pi^-}&=& \frac{\rho\la \bar uu\ra(\la \bar dd\ra +\la \bar ss\ra)}{2^{5}\pi^2}\,t, \,\,\,\,\,\,\,\,\,\,\,\,\,\,\,\,\,\,\,\,\,\,\,\,\,\,\,\,\,\,\,\,\,\,\,\,\,\,\,\,\,\,\,\,\,\,
\rho^{\la G^3\ra}_{\pi^+\pi^-}= {\cal O}(m_q^2\,\la   G^3\ra).
\label{eq:kpi}
\eea
We include the PT corrections like done in the preceeding sections. The analysis and the shape of different curves are similar to the previous case and will not be shown. The different sources of the errors are given in Table\,\ref{tab:res}. We obtain within a NWA  for $\tau\simeq 2.5$ GeV$^{-2}$ and $t_c\simeq (1.5-4.5)$ GeV$^2$:
\beq
M_{K^+\pi^-}=1035(134) ~{\rm MeV},\,\,\,\,\,\,\,\,\,\,\,\,\,\,\,\,\,\,\,\,\,\,\,\,\,\,\,\,\,\,\,\,\,\,\,\,
f_{K^+\pi^-}=1504(275)~{\rm keV}.
\label{eq:nwa-kpi}
\eeq
Taking into account the correction to the NWA due to the experimental total width $\Gamma_{K\pi}\simeq (270\pm 80)$ MeV\,\cite{PDG}, the mass result becomes:
\beq
M_{K^+\pi^-}^{BW}=1110(135) ~{\rm MeV}.
\label{eq:bw-kpi}
\eeq
\section{The $\eta\pi^0$ molecule}
We shall work with the current:
\beq
{\cal O}_{\eta\pi^0}(x)= \frac{1}{\sqrt{6}}\Big{[}(\bar u\,i\, \gamma_5\,u)  + (\bar d\,i\,\gamma_5\, d)-2(\bar s\, i \,\gamma_5\,s) \Big{]} \otimes\frac{1}{\sqrt{2}}\Big{[} (\bar u\,i\,\gamma_5\,u)  - (\bar d\,\,i\,\gamma_5\, d)\Big{]}
\eeq
The expression of the corresponding spectral function reads:
\bea
\rho^{pert}_{\eta\pi^0}&=&\frac{7}{3^2\times 2^{16}\pi^6}t^4 - \frac{m_s^2}{3\times 2^{11} \pi^6} t^3,\,\,\,\,\,\,\,\,\,\,\,\,\,\,\,\,\,\,\,\,\,\,\,\,\,\,\,\,\,\,\,\,\,\,\,\,\,\,\,\,\,\,\,\,\,\, \rho^{\la G^2\ra }_{\eta\pi^0}=\frac{11}{3}\frac{\la \alpha_s G^2\ra }{  2^{12}\pi^5}\,t^2,\    \nnb\\
\rho^{\la \bar qq\ra}_{\eta\pi^0}&=&\frac{19 [\,m_d\la \bar dd\ra +
\,m_u\la \bar uu\ra ]
+24\, m_s\la \bar ss\ra }{ 3^2\times 2^{10}\pi^4}t^2\nnb\\
\rho^{\la \bar q Gq\ra}_{\eta\pi^0}&=&\frac{121[m_d\la \bar dGd\ra + m_u\la \bar uGu\ra]
+120m_s\la sGs\ra}{3^2\times 2^{10}\pi^4}\,t\nnb\\
\rho^{\la \bar qq\ra^2}_{\eta\pi^0}&=& \rho\frac{  7[ \la \bar uu\ra^2+\la \bar dd\ra^2] +8\la \bar ss\ra^2}{3\times 2^{7}\pi^2}\,t, \,\,\,\,\,\,\,\,\,\,\,\,\,\,\,\,\,\,\,\,\,\,\,\,\,\,\,\,\,\,\,\,\,\,\,\,\,\,\,\,\,\,\,\,\,\, \rho^{\la G^3\ra}_{\eta\pi^0}= {\cal O}(m_q^2\,\la   G^3\ra).
\label{eq:etapi}
\eea
We obtain for the mass and coupling:
\beq
M_{\eta\pi^0}=1040(139) ~{\rm MeV},\,\,\,\,\,\,\,\,\,\,\,\,\,\,\,\,\,\,\,\,\,\,\,\,\,\,\,\,\,\,\,\,\,\,\,\,
f_{\eta\pi^0}=1462(249)~{\rm keV}.
\label{eq:nwa-etapi}
\eeq
\begin{table}[H]
\setlength{\tabcolsep}{0.4pc}
{\footnotesize{
\begin{tabular}{ll ll  ll  ll ll ll ll ll l c}
\hline
\hline
                Currents  
                    &\multicolumn{1}{c}{$\Delta t_c$}
					&\multicolumn{1}{c}{$\Delta \tau$}
					&\multicolumn{1}{c}{$\Delta \Lambda$}
					&\multicolumn{1}{c}{$\Delta PT$}
					&\multicolumn{1}{c}{$\Delta m_q$}
					&\multicolumn{1}{c}{$\Delta \bar{q}q$}
					&\multicolumn{1}{c}{$\Delta \kappa$}					
					&\multicolumn{1}{c}{$\Delta G^2$}
					&\multicolumn{1}{c}{$\Delta \bar q Gq$}
					&\multicolumn{1}{c}{$\Delta G^3$}
					&\multicolumn{1}{c}{$\Delta \bar{q}q^2$}
					
					&\multicolumn{1}{c}{$\Delta OPE$}
					&\multicolumn{1}{c}{Value}
\\
					
\hline
{\bf  Ordinary \boldmath$\bar qq$} &&&&&&&&&&\\
\it Masses [MeV]\\
$\frac{1}{\sqrt{2}}\ga\bar uu\pm\bar dd\dr$&88&4.5&3.0&1.0&0&0&--&1.0&1.0&0&19&24.4&1246(94) \\
$\bar us$&43&4.3&6.0&1.0&3.4&3&13.4&1.1&4.2&0& 23&25.3&1276(58)\\
$\bar ss$&51&2.8&3.5&0.9&7.6&4.2&21.8&1.3&4.9&0&16.5&17.7&1288(62)\\
\it (Couplings$/m_q)\times 10^3$\\
$\frac{1}{\sqrt{2}}\ga\bar uu\pm\bar dd\dr$&38&1.1&3.0&1.5&0&0&--&2.8&0.0&0.0&1.6&0.7&274(38) \\

$\bar us$&23&1.2&2.4&2.1&0.5&0.2&0.8&3&0.15&0&1.2&0.5&264(24)\\

$\bar ss$&17&0.95&2.4&1.9&1.5&0.4&2.5&3.4&0.2&0&1.3&0.5&256(18)\\
\hline
{\bf Molecules} &&&&&&&&&&\\
\it Masses [MeV]\\
$\pi^+\pi^-$&144&5&25&31&--&11&--&0.1&0.4&0&11.5&52&$1017(159)$\\
$K^+K^-$&165&3.9&18.4&115&1.1&11.8&7.4&0.6&3.5&0&13&67&1056(214)\\
$K^+\pi^-$&54&4.5&19.4&105&0.2&10.4&2.5&0.1&1.8&0&13.1&59&1035(134)\\
$\eta\pi$&57&4.5&15.8&109&0.2&10.8&2.8&0.2&1.8&0&10&60&1040(139)\\
\it Couplings [keV]\\
$\pi^+\pi^-$&193&24.5&2.0& 81  &--&87&--&2.5&2.0&0.&70&140&$1657(277) $\\
$K^+K^-$&161&15&0.8&32&4&70&39&2.8&24&0&78&158&1380(255)\\
$K^+\pi^-$&170&16.5&0.2&56&3&73&16&1.9&13&0&78&176&1504(275) \\
$\eta\pi$&165&19&0.1&53&2&71&9.6&2.6&13.7&0&77&142&1462(249) \\
\hline
\hline
\end{tabular}
}}
\vspace*{-0.25cm}
 \caption{\footnotesize Sources of errors and values of the scalar meson masses $M_S$ and couplings $f_S$ for the molecules  and some 4-quark configurations within a NWA. For $\pi^+\pi^-$, the PCAC relation : $(m_u+m_d)\la \bar\psi \psi\ra=-f_\pi^2m_\pi^2$  has been used for the estimate of the $m_\psi\la\bar\psi \psi\ra$ contribution ($f_\pi$=92.3 MeV). The error due to $M_S$ on the coupling has been implicitly included in the one due to $t_c$. We estimate the error due to the truncation of PT series and of the OPE as explained in the text. We take $\vert \Delta\tau\vert\simeq 0.2$ GeV$^{-2}$. Finite width corrections which are tiny are discussed in the text.}
\label{tab:res}
\end{table}
\section{Comments on  molecule states}
\d Within a NWA, the masses of the molecule states are about (230--250) MeV lower than the corresponding $\bar qq$ states. However, by  including the width corrections which act with opposite signs in the two cases, the two predictions 
tend to meet around 1.1 GeV.

\d It is amuzing to observe that the previous value of the $\bar qq$ and molecule masses coincide with the one of the lowest scalar digluonium :
\beq
M^{\rm glue}_{\sigma_B}=1041(111)~{\rm MeV}
\eeq
obtained in Ref.\,\cite{SNGS} indicating that, at this stage, one cannot yet distinguish the $\bar qq$, $\pi\pi$ molecule and gluonium nature of the $\sigma$. 

\d One can notice that the SU(3) breakings to the mass values are small in these two configurations.

\section{The pseudoscalar $\oplus$ scalar four-quark currents}
The four-quark configuration is a ``MESS" as many diquark-anti-diquark currents can describe the four-quark scalar states\,\cite{JAFFE,MARINA,ZHU,STEELE}. In general, the physical state
should be their combination with arbitrary mixing parameters which is (almost) impossible to control. Among these different possibilities, we choose to work with the combination of scalar and pseudoscalar currents which we compare with the combination of vector and axial-vector ones \,\footnote{We shall comment on some choices given in the literature in the next sections.}.
\subsection*{\b The $\bar u\bar d ud$ current}

We consider the current:
\beq
{\cal O}_{\bar u\bar dud}^{S/P}=\epsilon_{abc}\epsilon_{dec}\Big{[} (\bar u_a \gamma_5C \bar d_b^T)\otimes( u _d^TC\gamma_5  d_e)+ r\,( \bar u_a C \bar d_b^T)\otimes( u _d^TC  d_e)\Big{]},
\eeq
where $r$ is an arbitrary mixing parameter.  In Ref.\,\cite{MARINA}, only the scalar current has been considered ($r=0$). In the configuration proposed in Ref.\,\cite{ZHU}, the value of $r$ is $1/\sqrt{2}$. 
We shall work with the normalized current:
\beq
\tilde{\cal O}_{\bar u\bar dud}^{S/P}\equiv {{\cal O}_{\bar u\bar dud}^{S/P}}/{(1+r^2)^{1/2}}\,\,\,\,\,\,\,\,\,\,\,\,\,\,\,\,\,\,\,\,\,\,\,\,{\rm  with:}\,\,\,\,\,\,\,\,\,\,\,\,\,\,\,\,\,\,\,\,\,\,\,\,\la 0\vert \tilde{\cal O}_{\bar u\bar dud}^{S/P}\vert M_{\bar u\bar dud}\ra=\sqrt{2}f_{\bar u\bar dud}M^4_{\bar u\bar dud}. 
\eeq

The QCD expression of the spectral function normalized to $(1+r^2)$ reads at LO\,\footnote{The current: ${\cal O}_{\bar u\bar dud}^{P/S}=( \bar u_a \gamma_5C \bar d_b^T)\otimes( u _a^TC\gamma_5  d_b-u_b^TC\gamma_5 d_a)+k\,( \bar u_a C \bar d_b^T)\otimes( u _a^TC  d_b-u_b^TCd_a)$ gives the same spectral function. For the symmetric current: ${\cal O}_{\bar u\bar dud}^P=( \bar u_a C \bar d_b^T)\otimes (u _a^TC  d_b)$, the contributions of the $G^2$ and the gluon exhange for the mixed condensates are zero.}:
\bea
\rho^{pert}_{\bar u\bar dud}&=&\frac{t^4}{5\times 3\times 2^{12}\pi^6},\,\,\,\,\,\,\,\,\,\,\,\,\,\,\,\,\,\,\,\,\,\,\,\,\,\,\,\,\,
\rho^{\la G^2\ra }_{\bar u\bar dud}=\frac{\la \alpha_s G^2\ra }{ 3\times 2^{9}\pi^5}\,t^2,\nnb\\
\rho^{\la \bar qq\ra}_{\bar u\bar dud}&=&\frac{[ m_d\la \bar dd\ra + 
m_u\la \bar uu\ra](1+r^2)- 2[ m_u\la \bar dd\ra + 
m_d\la \bar uu\ra] (1-r^2)}{ 3\times 2^{6}\pi^4(1+r^2)}\,{t^2}\nnb\\
\rho^{\la \bar q Gq\ra}_{\bar u\bar dud}&=&\frac{[ m_d\la \bar dGd\ra + m_u\la \bar uGu\ra] (1+r^2) +6[ m_u\la \bar dGd\ra + m_d\la \bar uGu\ra](1-r^2) }{3\times 2^{7}\pi^4(1+r^2)}\,t\nnb\\
\rho^{\la \bar qq\ra^2}_{\bar u\bar dud}&=& \frac{\rho\la \bar uu\ra\la \bar dd\ra }{3\times 2^{2}\pi^2}\frac{(1-r^2)}{(1+r^2)}\,t,\,\,\,\,\,\,\,\,\,\,\,\,\,\,\,\,\,\,\,
\rho^{\la G^3\ra}_{\bar u\bar dud}= {\cal O}(m_q^2\,\la g^3  G^3\ra),
\eea
where $\rho$ indicates the deviation from the factorization of the four-quark condensate. The main difference with $\pi^+\pi^-$ molecule  spectral function is the flip of signs of the $D=5,6$ condensate contributions for the pseudoscalar current. 

\begin{figure}[hbt]
\begin{center}
\hspace*{-7cm} {\bf a) \hspace*{8.cm} \bf b)} \\
\includegraphics[width=8cm]{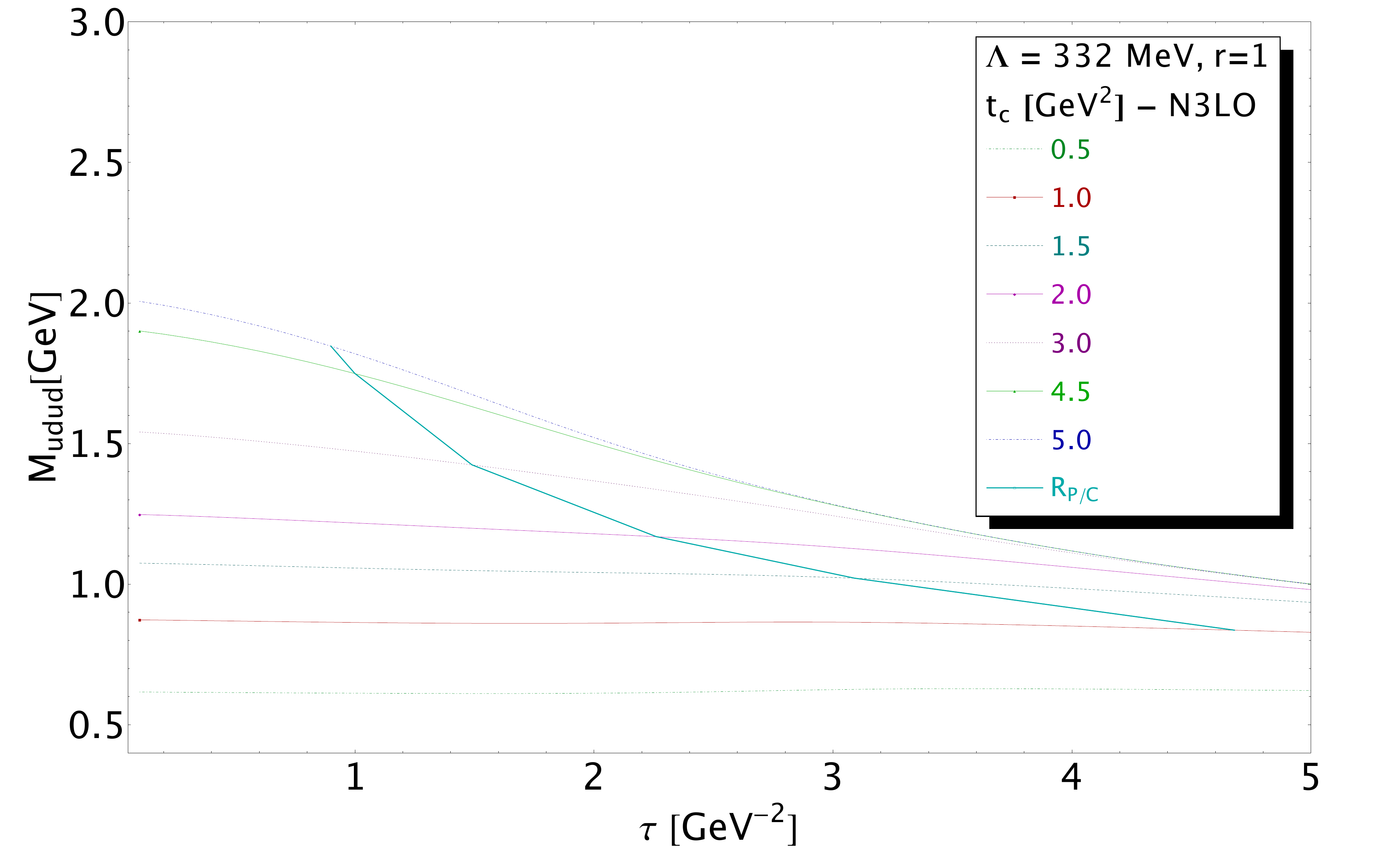}
\includegraphics[width=8.cm]{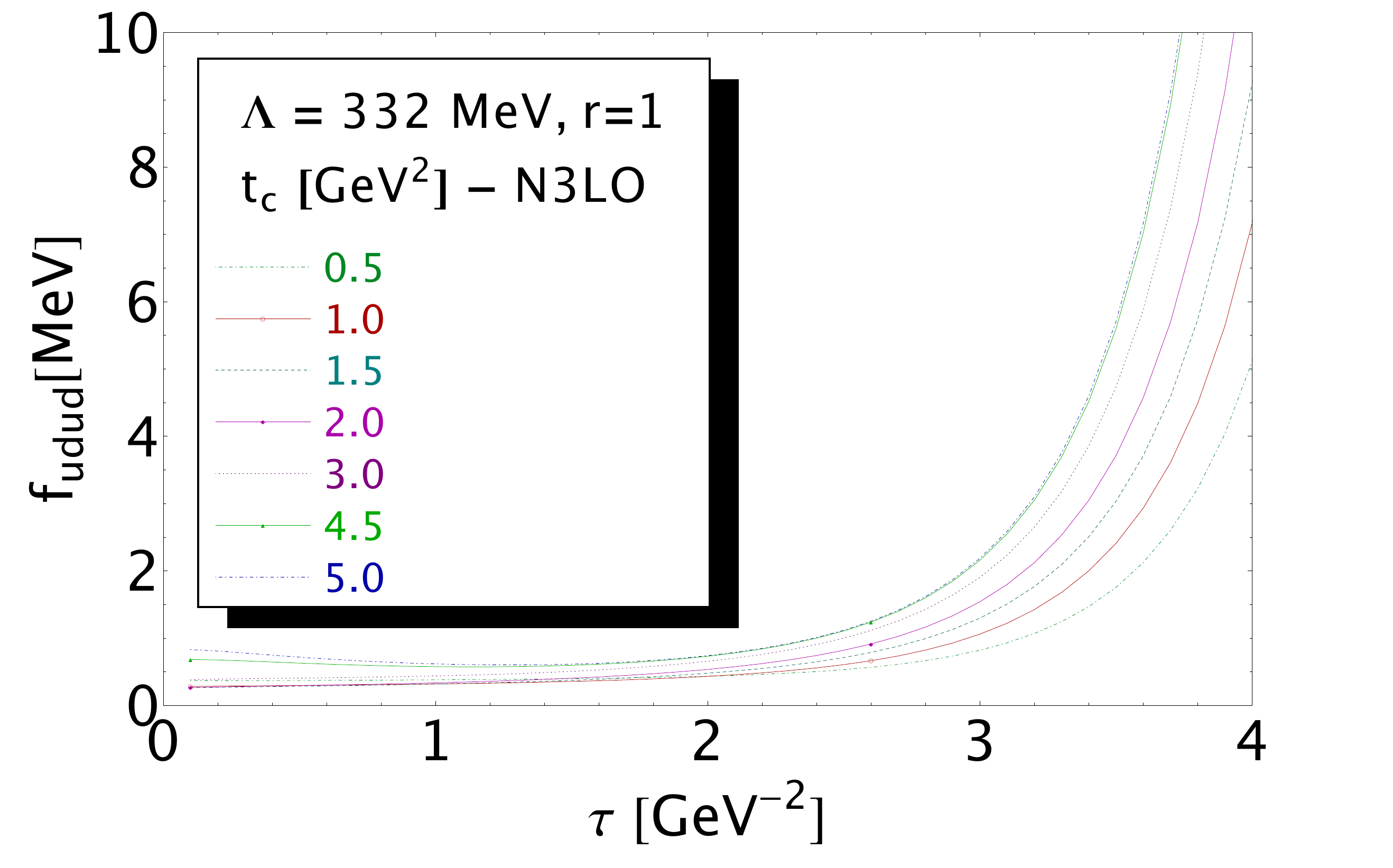} 
\vspace*{-0.5cm}
\caption{\footnotesize   Behaviour of a) mass and  b) coupling of the $\bar u \bar d ud$ four-quark states versus $\tau$ for different values of $t_c$ and for $r=1$. } 
\label{fig:ud-1}
\end{center}
\vspace*{-0.5cm}
\end{figure} 
\begin{figure}[hbt]
\begin{center}
\hspace*{-7cm} {\bf a) \hspace*{8.cm} \bf b)} \\
\includegraphics[width=8cm]{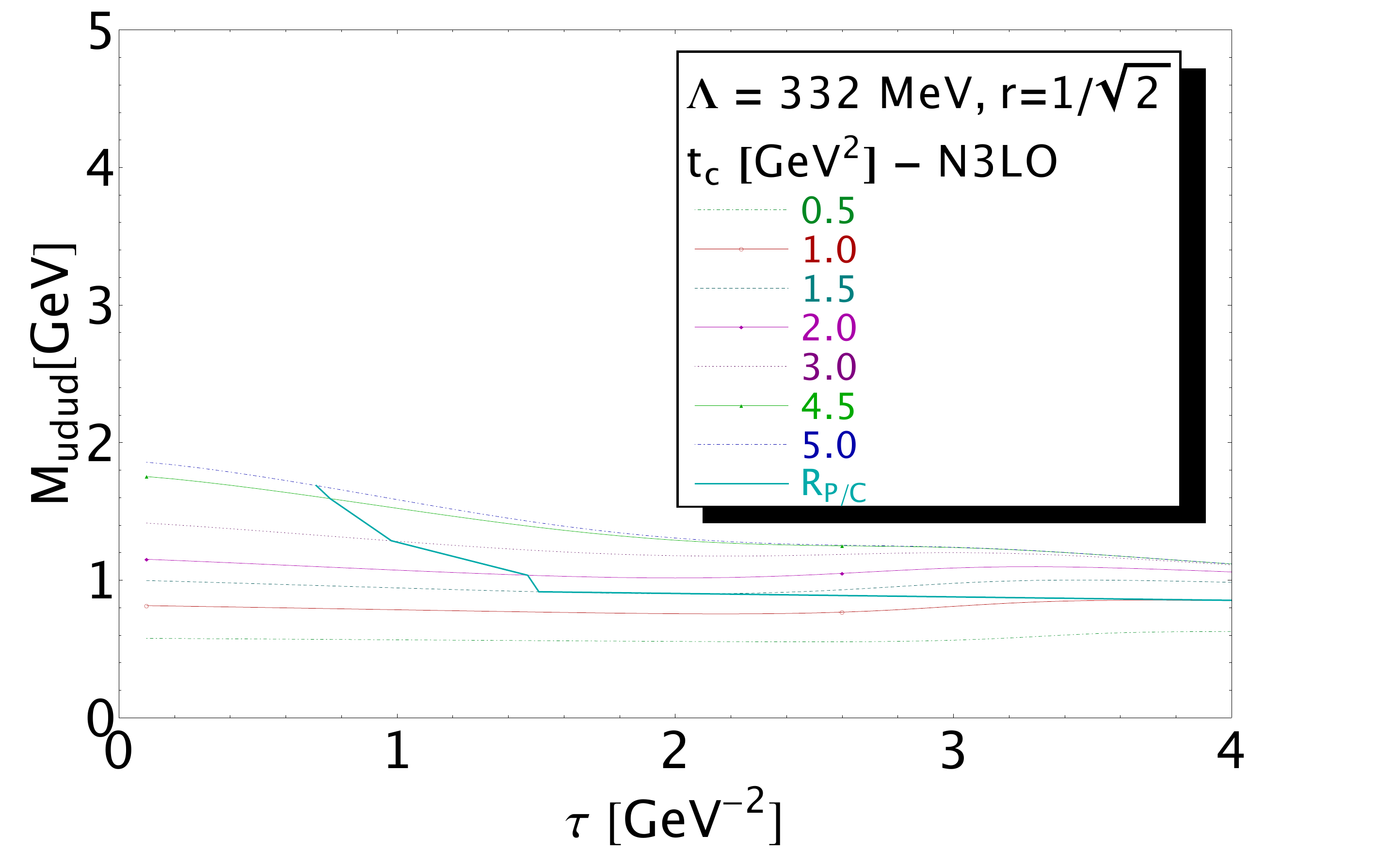}
\includegraphics[width=8.cm]{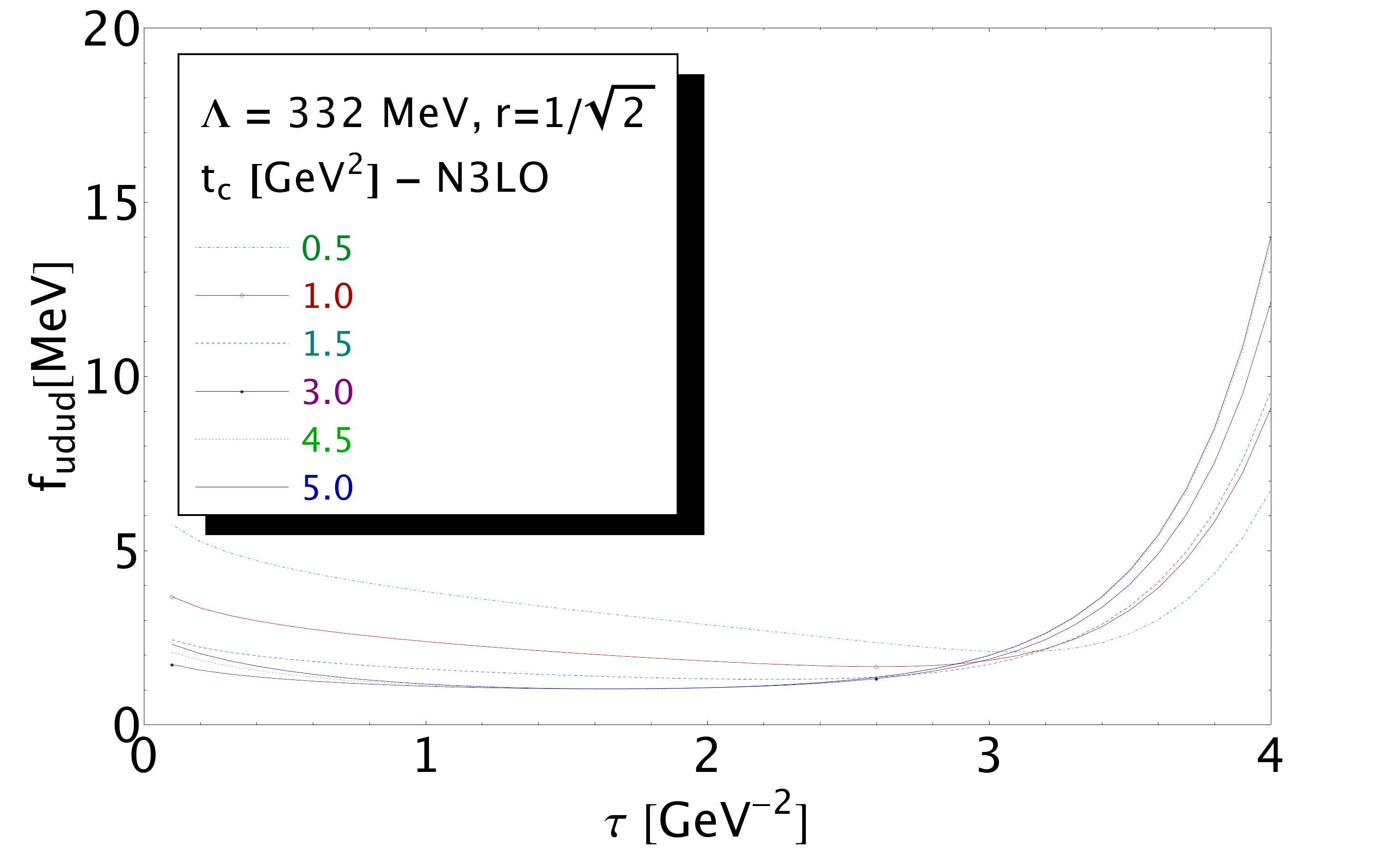} 
\vspace*{-0.5cm}
\caption{\footnotesize   Behaviour of a) mass and  b) coupling of the $\bar u \bar d ud$ four-quark states versus $\tau$ for different values of $t_c$ and for $r=1/\sqrt{2}$. } 
\label{fig:ud-2}
\end{center}
\vspace*{-0.5cm}
\end{figure} 

\begin{figure}[hbt]
\begin{center}
\hspace*{-7cm} {\bf a) \hspace*{8.cm} \bf b)} \\
\includegraphics[width=8cm]{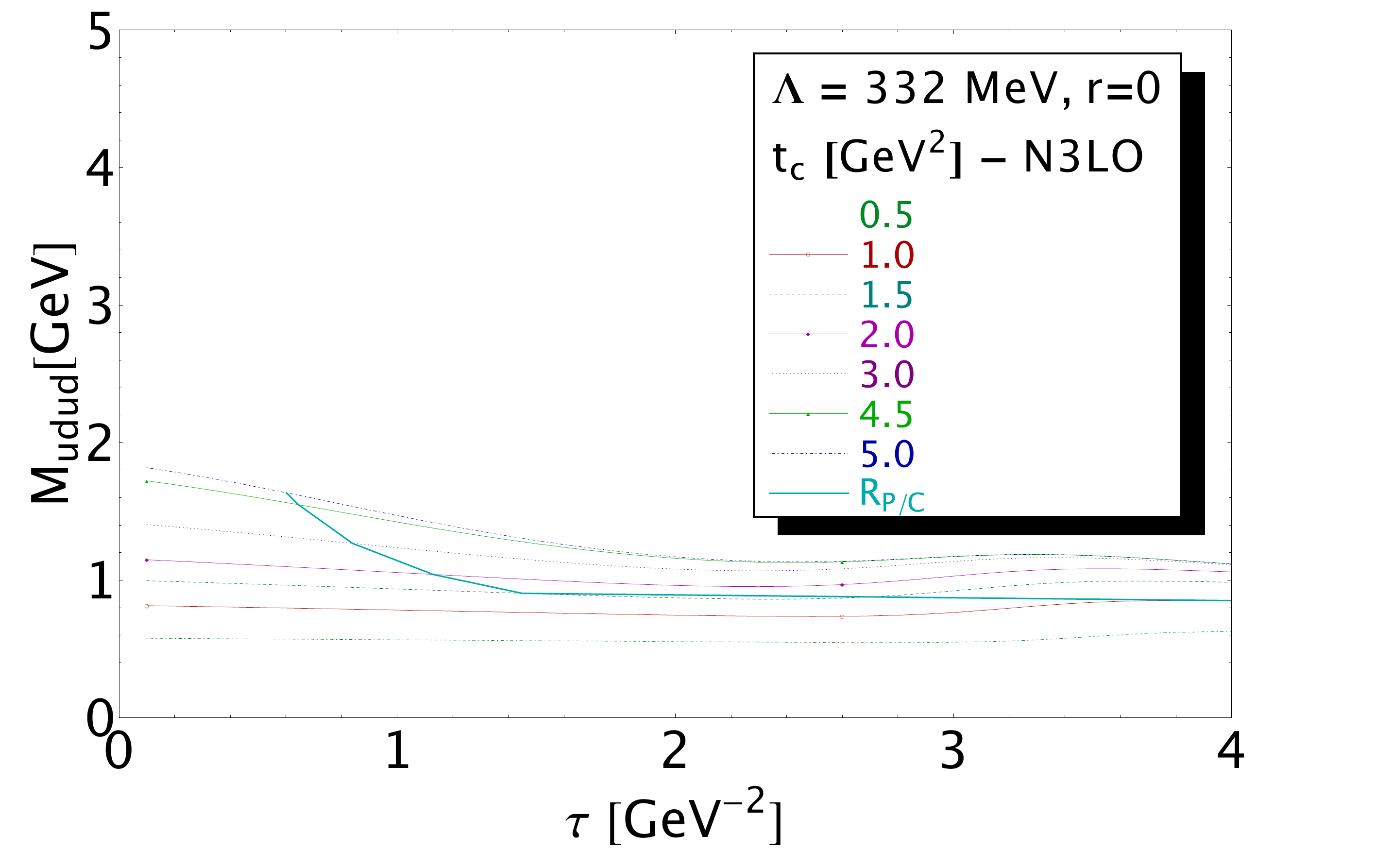}
\includegraphics[width=8.cm]{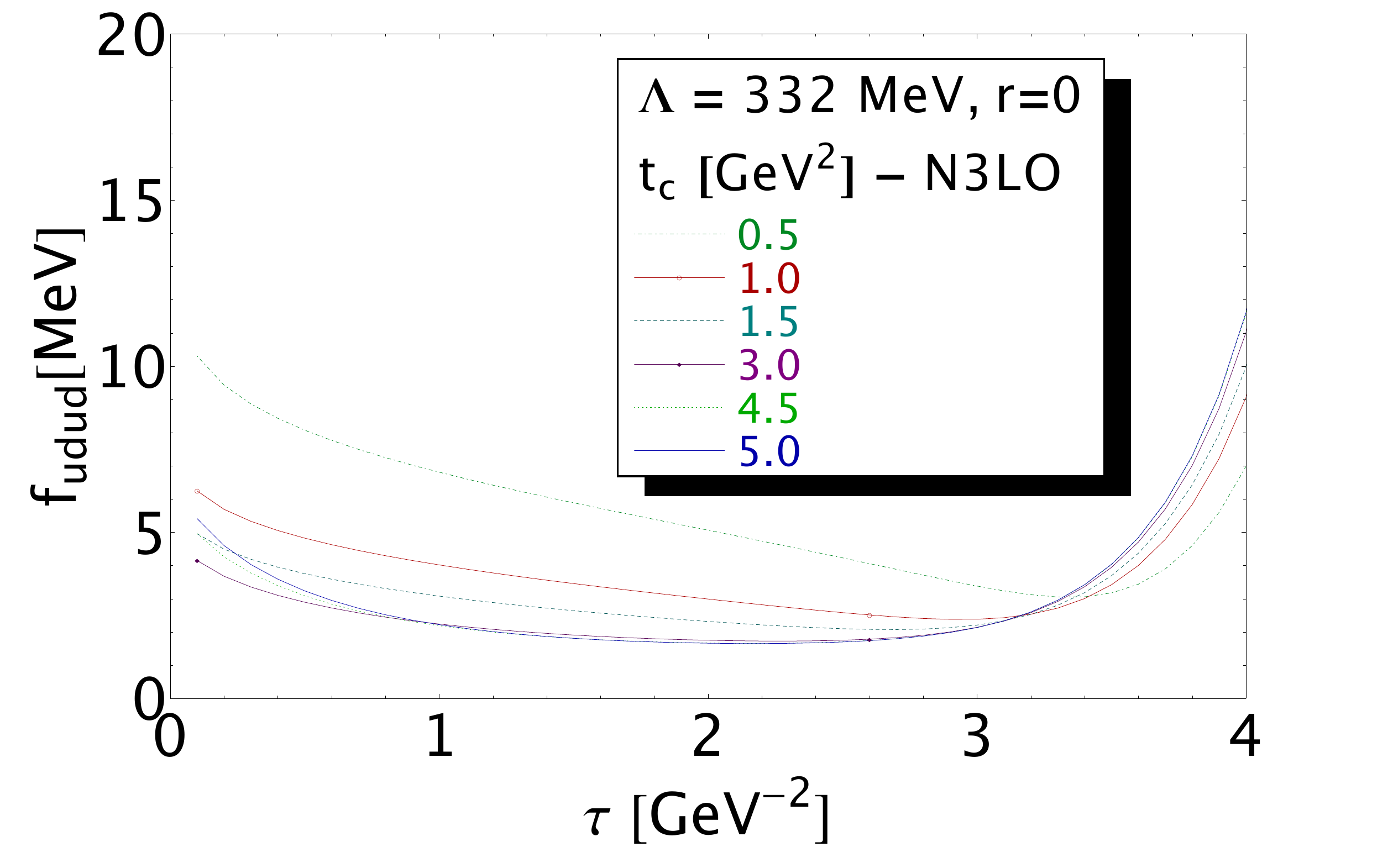} 
\vspace*{-0.5cm}
\caption{\footnotesize   Behaviour of a) mass and  b) coupling of the $\bar u \bar d ud$ four-quark states versus $\tau$ for different values of $t_c$ and for $r=0$. } 
\label{fig:ud-0}
\end{center}
\vspace*{-0.5cm}
\end{figure} 
\d The analysis is shown in Fig.\,\ref{fig:ud-1} to Fig.\,\ref{fig:ud-0} for typical values of the mixing parameter $r$: 1, $1/\sqrt{2}$ and 0, where the OPE is truncated up to dimension $D=6$.  The mass presents stability (plateau) at low values of $t_c$ which becomes an inflexion point for $t_c\geq 3$ GeV$^2$ while the stability of the coupling  depends on\,$r$.  

We also see in these figures that the constraint $R_{P/C}\geq 1$ excludes values of $t_c\leq 1.5$ GeV$^2$ like in the previous sections.

\begin{figure}[hbt]
\begin{center}
\hspace*{-7cm} {\bf a) \hspace*{8.cm} \bf b)} \\
\includegraphics[width=8cm]{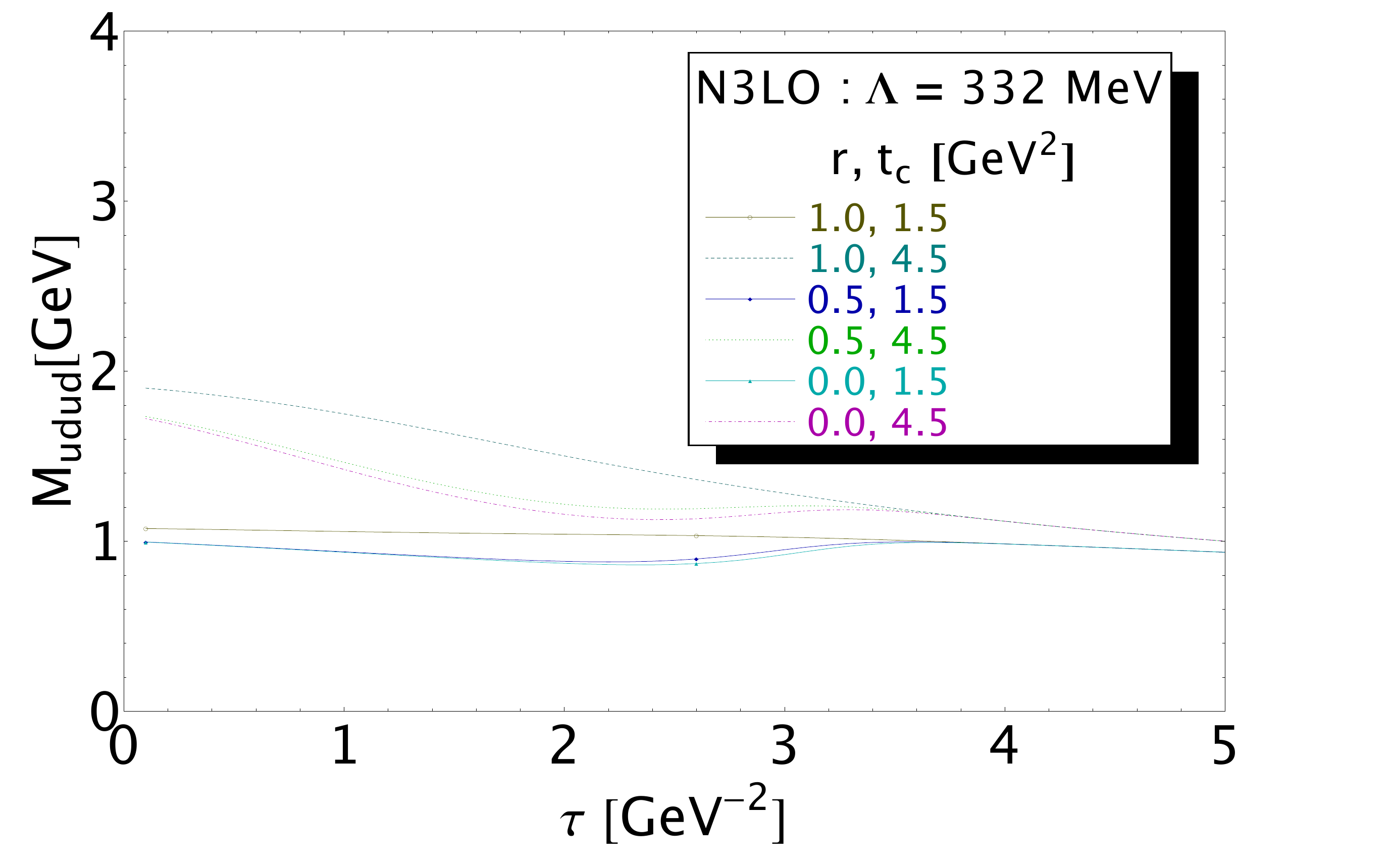}
\includegraphics[width=8.2cm]{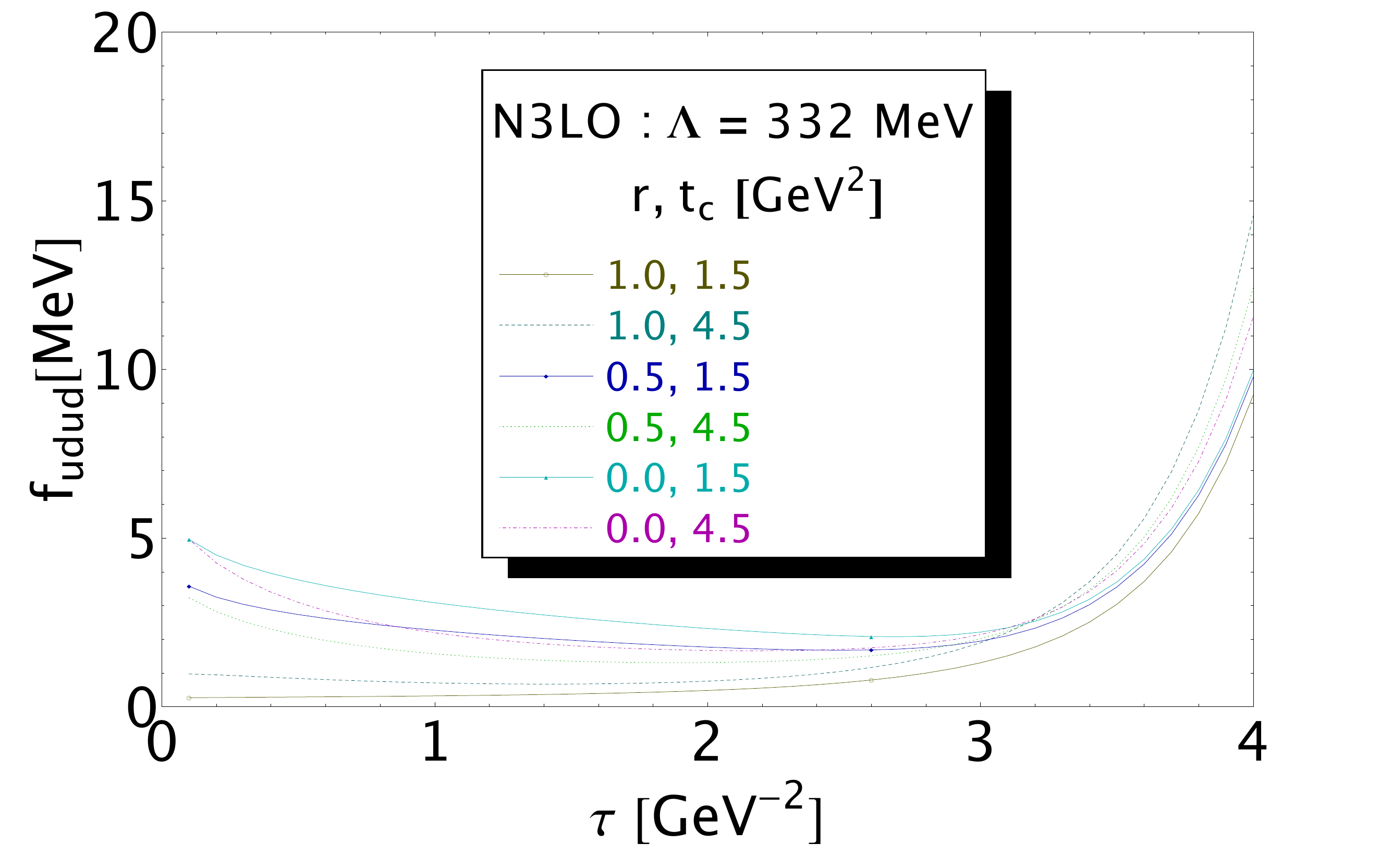} 
\vspace*{-0.5cm}
\caption{\footnotesize   Behaviour of a) mass and  b) coupling of the $\bar u \bar d ud$ four-quark states versus $\tau$ for two extremal values of $t_c$ allowed by the constraint: $R_{P/C}\geq 1$ and for $r=0,~0.5,~1$. } 
\label{fig:ud}
\end{center}
\vspace*{-0.5cm}
\end{figure}
\begin{figure}[H]
\begin{center}
\hspace*{-7cm} {\bf a) \hspace*{8.cm} \bf b)} \\
\includegraphics[width=8cm]{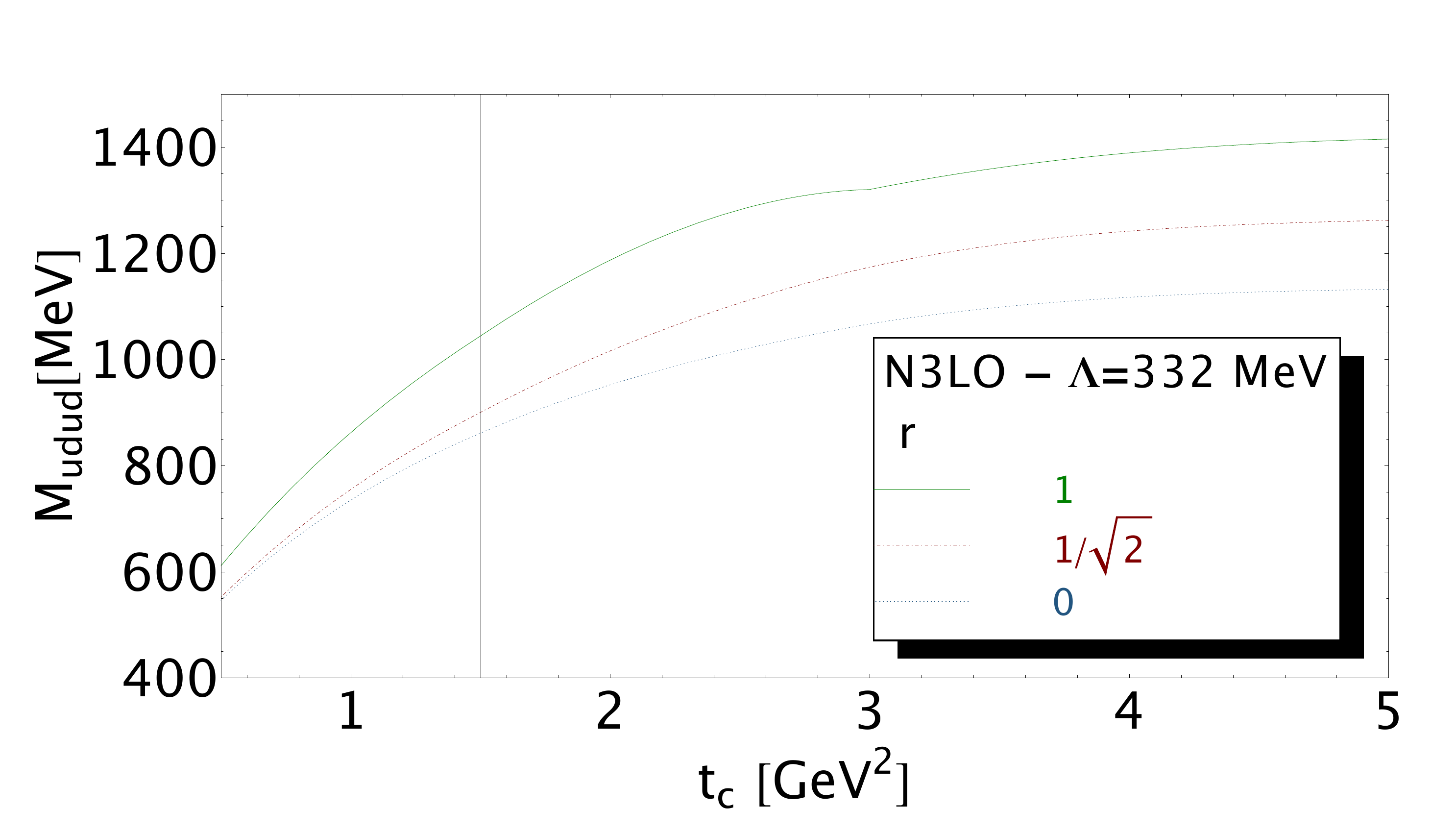}
\includegraphics[width=8.2cm]{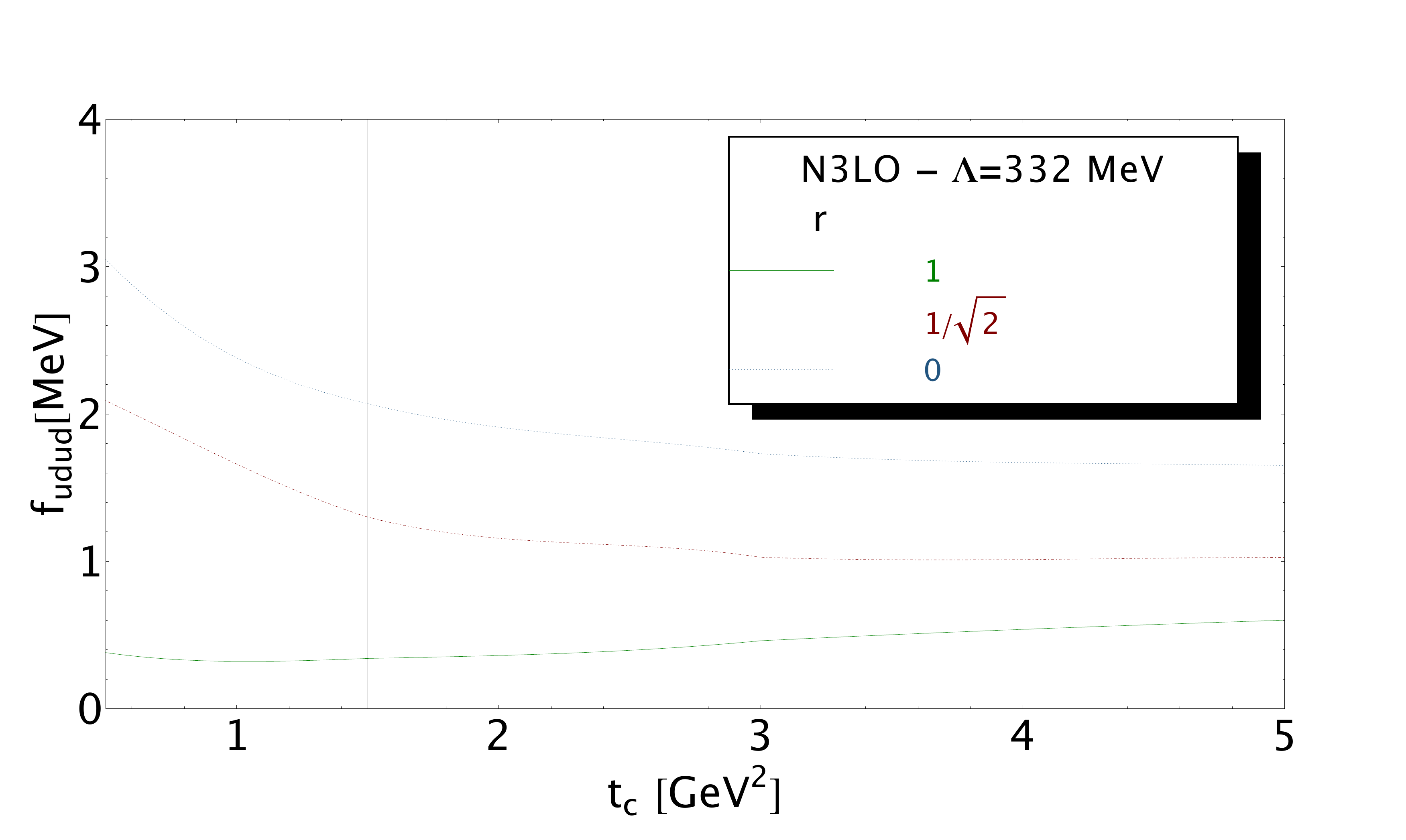} 
\vspace*{-0.5cm}
\caption{\footnotesize   Behaviour of the optimal values in $\tau$ for the  a) mass and  b) coupling of the $\bar u \bar d ud$ four-quark states versus $t_c$ for three values of $r$ (1, $1/\sqrt{2}$, 0) The right side of the vertical line  $t_c=1.5$ GeV$^2$ is the region allowed by the constraint: $R_{P/C}\geq 1$.} 
\label{fig:ud-tc}
\end{center}
\end{figure}

 
\d We show in Fig.\,\ref{fig:ud} the behaviour of the optimal values of the mass versus $\tau$ for two extremal values of $t_c$=1.5 and 4.5 GeV$^2$ and for different values of the mixing parameter $r$.  In Fig.\,\ref{fig:ud-tc}, we show the $t_c$-behaviour of the optimal result in $\tau$ for three typical values of $r$. The vertical value $t_c=1.5$ GeV$^2$ delimits the region $R_{P/C}\geq 1$ where the left region is excluded. This result does not favour the low values of $M_{\sigma}$ found from QCD spectral sum rules in the current literature which are obtained at low values of $t_c\simeq (0.5\sim 1.0)$ GeV$^2$ corresponding to a (misleading) plateau in the sum rule variable $\tau$. 

\begin{figure}[hbt]
\begin{center}
\hspace*{-7cm} {\bf a) \hspace*{8.cm} \bf b)} \\
\includegraphics[width=8.2cm]{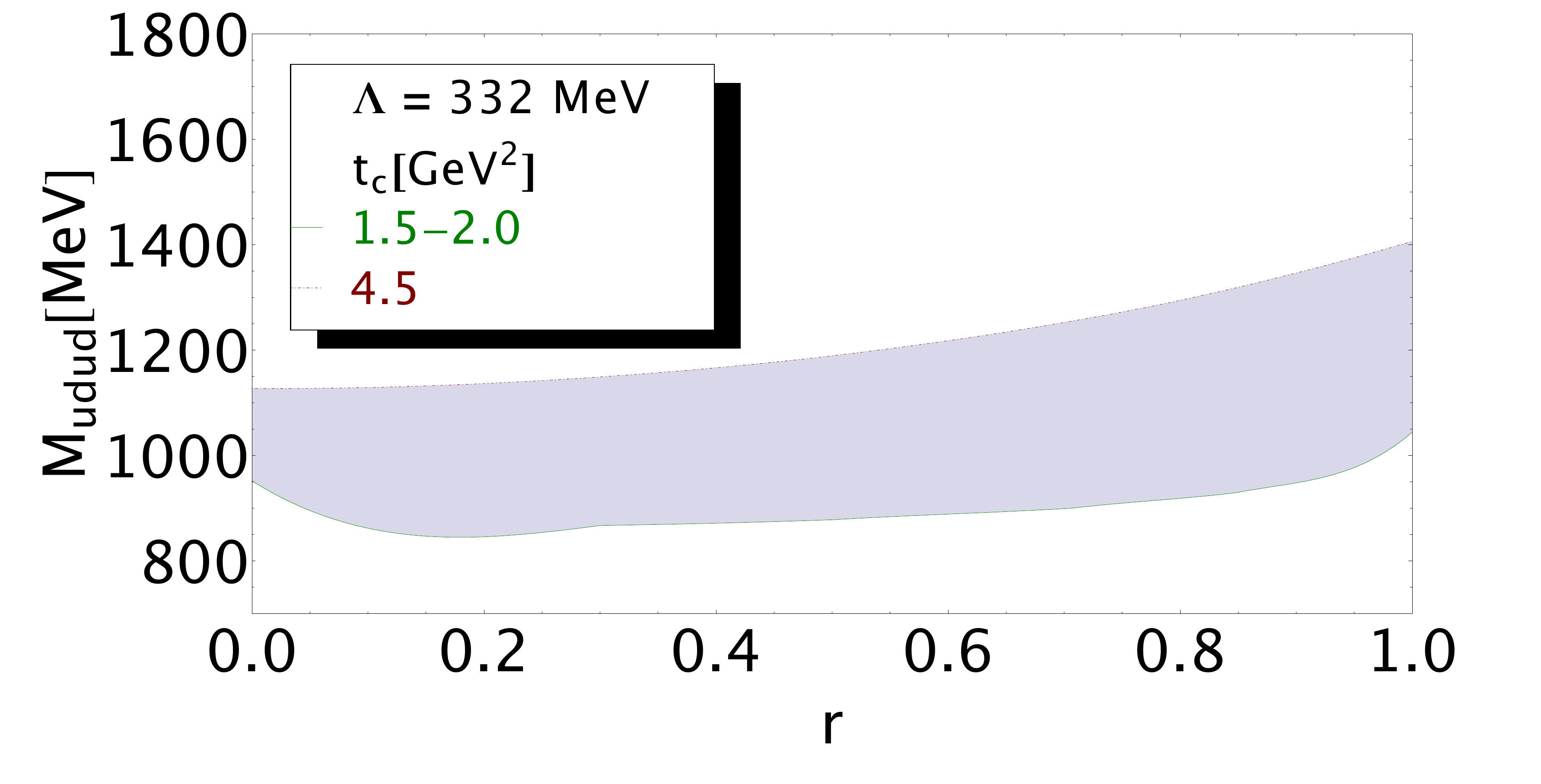}
\includegraphics[width=8.cm]{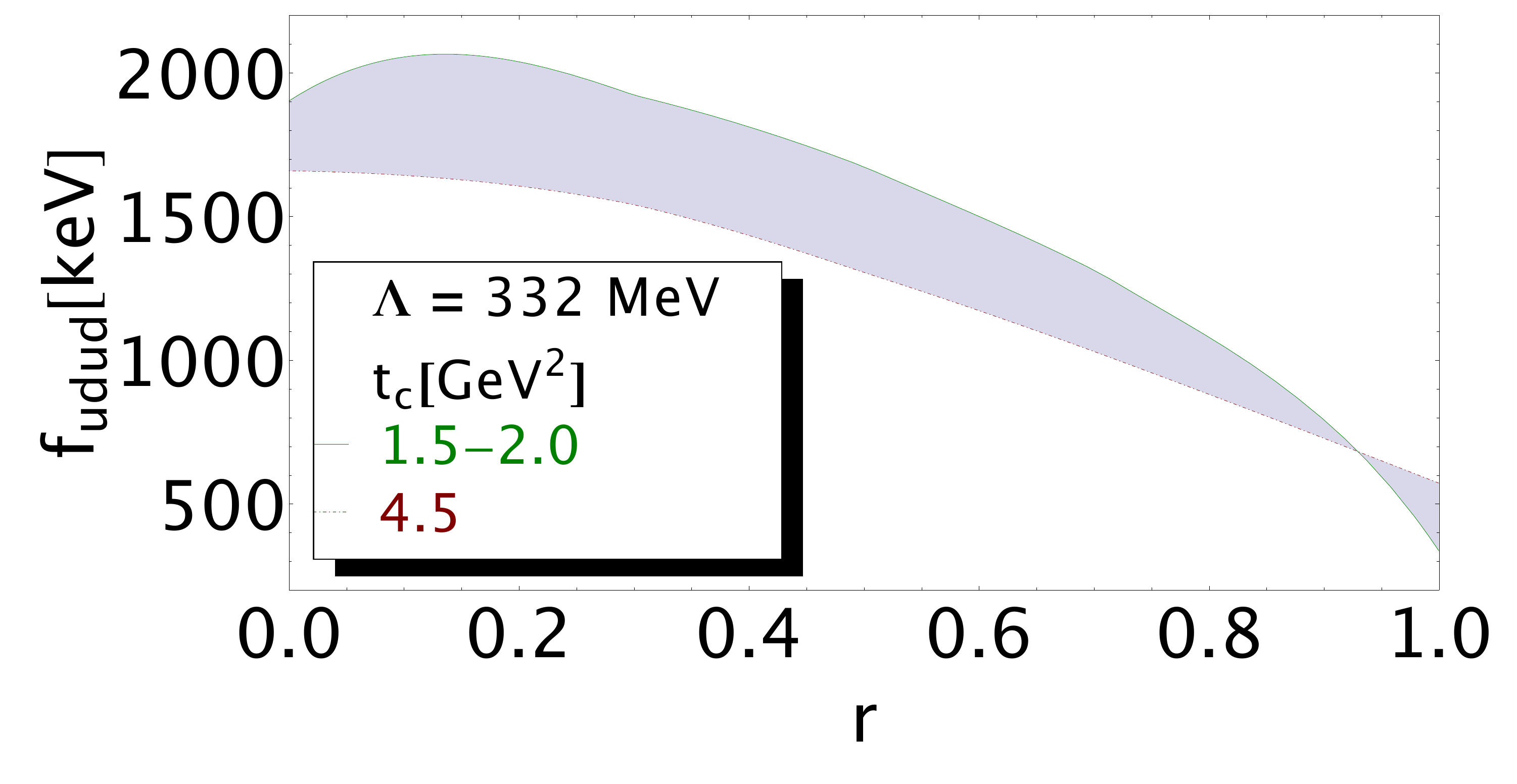} 
\vspace*{-0.5cm}
\caption{\footnotesize   Behaviour of the optimal values in $\tau$ for the  a) mass and  b) coupling of the $\bar u \bar d ud$ four-quark state versus the mixing parameter $r$ for two extremal values of $t_c$ allowed by the constraint: $R_{P/C}\geq 1$.} 
\label{fig:ud-r}
\end{center}
\end{figure}

\d Taking the lower values of  $t_c$=1.5 (resp.2) GeV$^2$ for $ r= 1/\sqrt{2}$, 1 (resp. 0)  allowed by $R_{P/C}\geq 1$ and the higher value of $t_c=$ 4.5 GeV$^2$, we show the $r$-behaviour of the optimal results in $\tau$ in Fig.\,\ref{fig:ud-r}. One can notice that the value of the mass is not strongly affected by the choice of the mixing parameter which is not the case of the coupling. 
  
 \d We consider as a final value the one inside the region $t_c\simeq (1.5\sim 2)- 4.5$ GeV$^2$ allowed by the $R_{P/C}\geq 1$ condition. We obtain:
\bea
M_{\bar u\bar dud}= &&1225(184) ~{\rm MeV}  ,\,\,\,\,\,\,\,\,\,\,\,\,\,\,\,\,\,\,\,\,\,\,\,\,\,\,\,f_{\bar u\bar dud}=\,\,\,\,458(288)~{\rm keV}.\,\,\,\,\,\,\,\,\,\,\,\,\,\,\,\,\,\,\,\,\,\,\,\,\,\,\,\, r=1 \nnb\\
&& 1078(186)~{\rm MeV}  ,\,\,\,\,\,\,\,\,\,\,\,\,\,\,\,\,\,\,\,\,\,\,\,\,\,\,\,\,\,\,\,\,\,\,\,\,\,\,\,\,\,\,\,=\,\,\,\,1160(357)~{\rm keV}.\,\,\,\,\,\,\,\,\,\,\,\,\,\,\,\,\,\,\,\,\,\,\,\,\, \,\,=1/\sqrt{2} \nnb\\
&& 1040(119)~{\rm MeV} ,\,\,\,\,\,\,\,\,\,\,\,\,\,\,\,\,\,\,\,\,\,\,\,\,\,\,\,\,\,\,\,\,\,\,\,\,\,\,\,\,\,\,\,=\,\,\,\,1780(540)~{\rm keV}.\,\,\,\,\,\,\,\,\,\,\,\,\,\,\,\,\,\,\,\,\,\,\,\,\, \,\,=0,
\eea
where the different sources of the errors are given in Table\,\ref{tab:res}. We notice that the errors are dominated by the one due to $t_c$, $\Delta PT$ and (to a lesser extent) by the truncation of the OPE. The one due $\Delta PT$ is mainly due to the  tachyonic gluon mass while the one due to N5LO is relatively small. We have taken their mean. 
\subsection*{\b The $\bar u\bar d us$ current}
\begin{figure}[hbt]
\begin{center}
\hspace*{-7cm} {\bf a) \hspace*{8.cm} \bf b)} \\
\includegraphics[width=8.2cm]{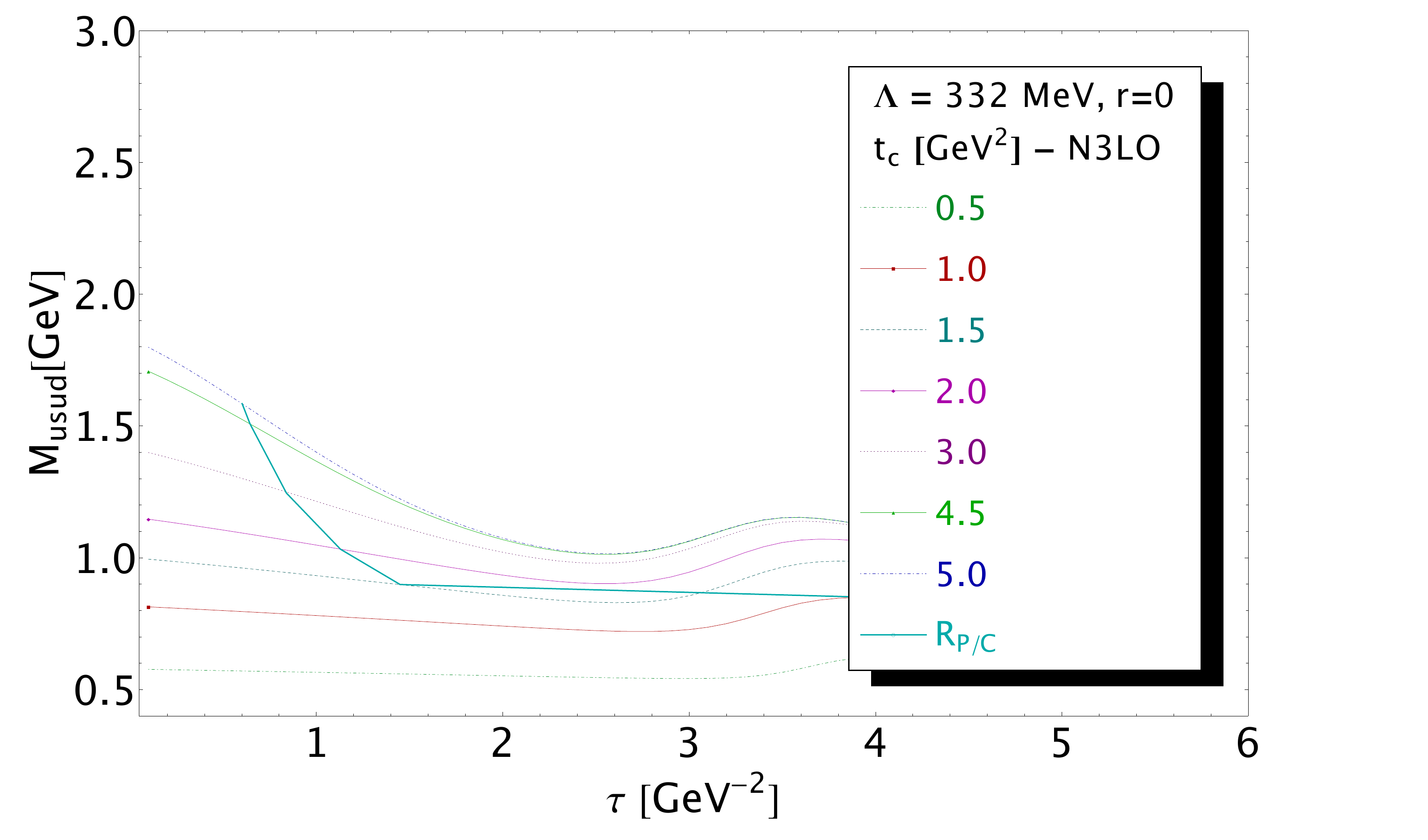}
\includegraphics[width=8.cm]{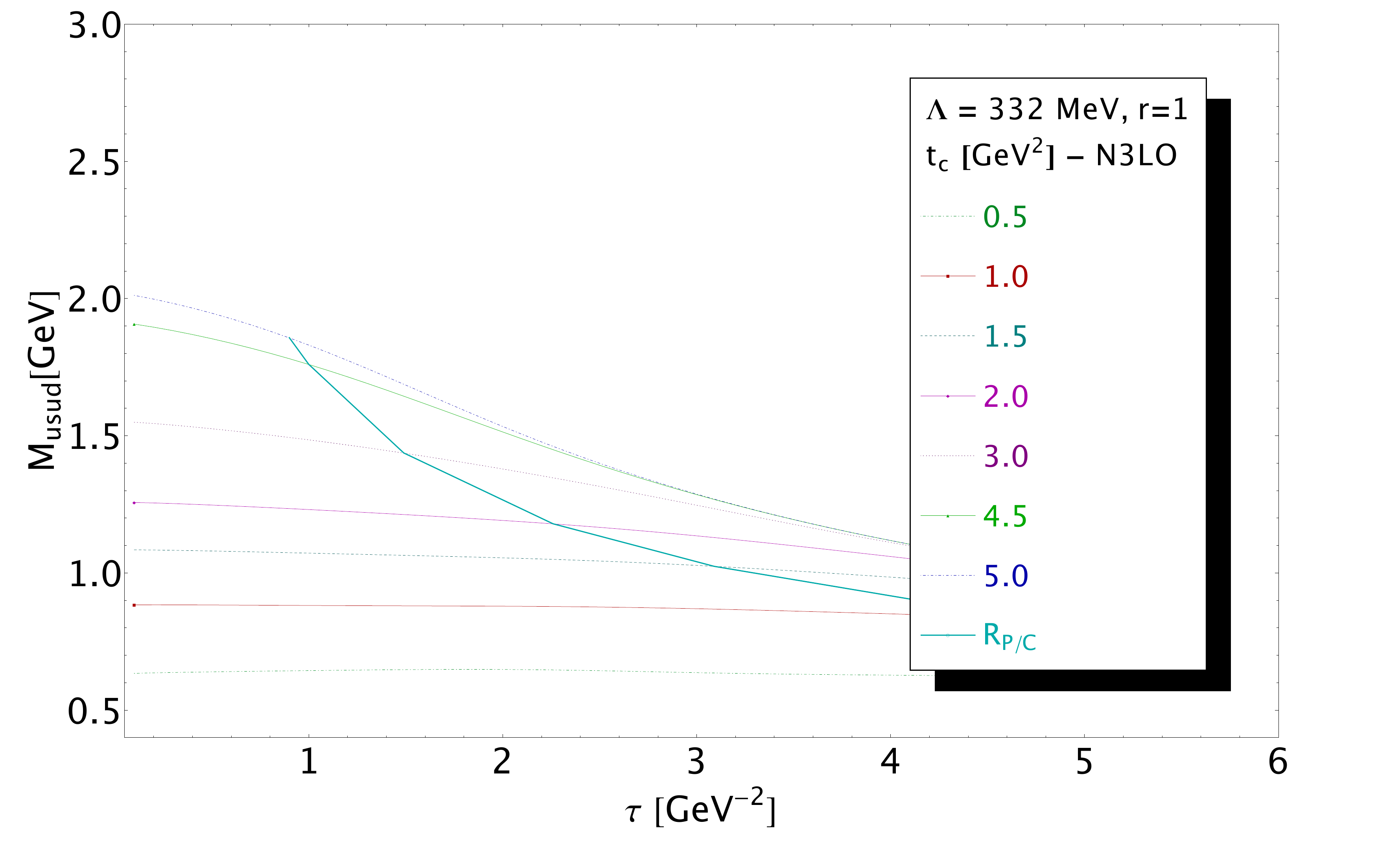} 
\vspace*{-0.5cm}
\caption{\footnotesize   Behaviour of the optimal values in $\tau$ for the  mass  of the $\bar u \bar s ud$ four-quark state : a) $r=0$,  b) $r=1$.} 
\label{fig:us-mass}
\end{center}
\end{figure}

We consider the current :
\beq
{\cal O}_{\bar u\bar dus}^{S/P}=\epsilon_{abc}\epsilon_{dec}\Big{[}( \bar u_a \gamma_5 C\bar d_b^T)\otimes( u _d^TC\gamma_5 s_e)+ r\,( \bar u_a C \bar d_b^T)\otimes( u _d^TC  s_e)\Big{]}.
\eeq
which is expected to have the quantum numbers of the $K^*_0(1430)$. The coresponding spectral function normalized to $(1+r^2)$ reads:
\bea
\rho^{pert}_{\bar u\bar dus}&=&\frac{t^4}{5\times 3\times 2^{12}\pi^6}-\frac{m_s^2}{3\times 2^{10}\pi^6}t^3,\,\,\,\,\,\,\,\,\,\,\,\,\,\,\,\,\,\,\,\,\,\,\,\,\,\,\,\,\,
\rho^{\la G^2\ra }_{\bar u\bar dus}=\frac{\la \alpha_s G^2\ra }{ 3\times 2^{9}\pi^5}\,t^2,\nnb\\
\rho^{\la \bar qq\ra}_{\bar u\bar dus}&=&\frac{[2m_u\la \bar uu\ra+ m_d\la \bar dd\ra + 
m_s\la\bar ss\ra]- 2[ (m_d+m_s)\la \bar uu\ra+ 
m_u(\la \bar ss\ra+\la \bar dd\ra)] (1-r^2)/(1+r^2)}{ 3\times 2^{7}\pi^4}\,{t^2}\nnb\\
\rho^{\la \bar q Gq\ra}_{\bar u\bar dus}&=&\frac{[ 2m_u\la \bar uGu\ra+m_d\la \bar dGd\ra + m_s\la\bar sGs\ra] +6[ m_u(\la \bar dGd\ra +\la \bar sGs\ra)+(m_d+m_s)\la\bar uGu\ra](1-r^2) /(1+r^2) }{3\times 2^{8}\pi^4}\,t\nnb\\
\rho^{\la \bar qq\ra^2}_{\bar u\bar dus}&=& \frac{\rho\la \bar uu\ra(\la \bar dd\ra +\la\bar ss\ra)}{3\times 2^{3}\pi^2}\frac{(1-r^2)}{(1+r^2)}\,t,\,\,\,\,\,\,\,\,\,\,\,\,\,\,\,\,\,\,\,\,\,\,\,\,\,\,
\rho^{\la G^3\ra}_{\bar u\bar dus}= {\cal O}(m_q^2\,\la g^3  G^3\ra),
\eea
The analysis and the shape of different curves are very similar to the case of the $f_0/a_0(980)$ and will not be shown. The $R_{P/C}\geq 1$ condition also 
restricts the value of $t_c$ to be in the range (2-4.5) GeV$^2$ for $r=0$ as explictily shown in Fig.\,\ref{fig:us-mass} a)  and (1.5-4.5) GeV$^2$ for $0<r\leq 1$ as illustrated in Fig.\,\ref{fig:us-mass} b) for $r=1$.
\\

\begin{figure}[hbt]
\begin{center}
\hspace*{-7cm} {\bf a) \hspace*{8.cm} \bf b)} \\
\includegraphics[width=8.2cm]{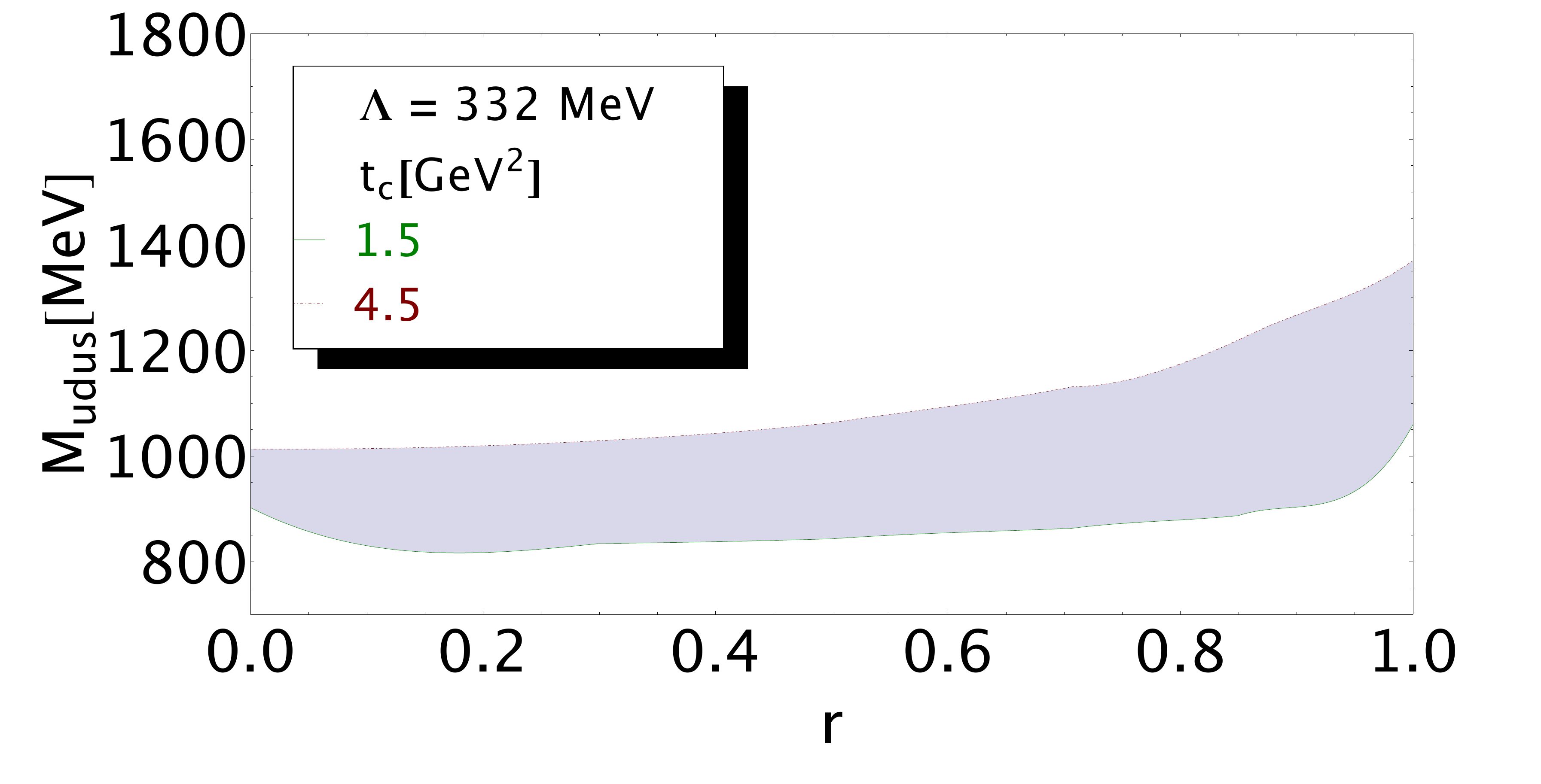}
\includegraphics[width=8.cm]{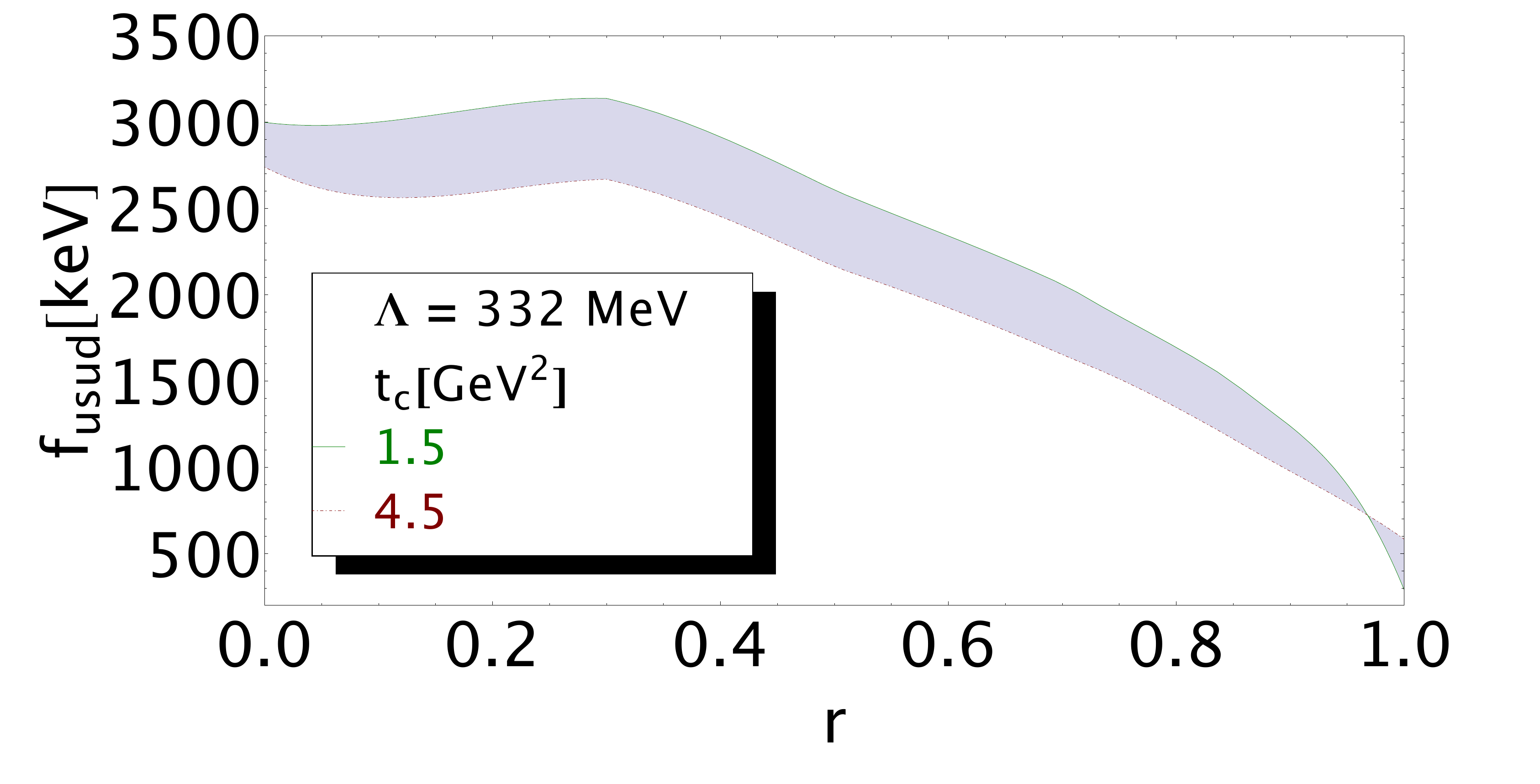} 
\vspace*{-0.5cm}
\caption{\footnotesize   Behaviour of the optimal values in $\tau$ for the  a) mass and  b) coupling of the $\bar u \bar s ud$ four-quark state versus the mixing parameter $r$ for two extremal values of $t_c$ allowed by the constraint: $R_{P/C}\geq 1$.} 
\label{fig:us-r}
\end{center}
\end{figure}

We show in Fig.\,\ref{fig:us-r} the $r$-behaviour of the optimal results in $\tau$ for the previous range of $t_c$. We obtain for $\tau\simeq 2.6,\, 2.4,\,2.3$ GeV$^{-2}$ for $r=0,\,1/\sqrt{2},1$:
\bea
M_{\bar u\bar sud}= &&1215(157) ~{\rm MeV}  ,\,\,\,\,\,\,\,\,\,\,\,\,\,\,\,\,\,\,\,\,\,\,\,\,\,\,\,\,\,f_{\bar u\bar sud}=\,\,\,\,422(131)~{\rm keV}.     \,\,\,\,\,\,\,\,\,\,\,\,\,\,\,\,\,\,\,\,\,\,\,\,\,\,\, r=1 \nnb\\
&& 997(147)~{\rm MeV}  ,\,\,\,\,\,\,\,\,\,\,\,\,\,\,\,\,\,\,\,\,\,\,\,\,\,\,\,\,\,\,\,\,\,\,\,\,\,\,\,\,\,\,\,\,\,\,\,=\,\,\,\,\,1835(382)~{\rm keV}.    \,\,\,\,\,\,\,\,\,\,\,\,\,\,\,\,\,\,\,\,\,\,\,\,\,\,=1/\sqrt{2} \nnb\\
&&957(123)~{\rm MeV} ,\,\,\,\,\,\,\,\,\,\,\,\,\,\,\,\,\,\,\,\,\,\,\,\,\,\,\,\,\,\,\,\,\,\,\,\,\,\,\,\,\,\,\,\,\,\,\,=\,\,\,\,2902(638)~{\rm keV}.   \,\,\,\,\,\,\,\,\,\,\,\,\,\,\,\,\,\,\,\,\,\,\,\,\, \,\,=0.
\eea
Taking into account the correction to the NWA due to the experimental total width $\Gamma_{K\pi}\simeq (270\pm 80)$ MeV\,\cite{PDG}, the mass result becomes:
\bea
M_{\bar u\bar sud}^{BW}= &&1236(157) ~{\rm MeV}  ,\,\,\,\,\,\,\,\,\,\,\,\,\,\,\,\,\,\,\,\,\,\,\,\,\,\,\,\,\,\,\,\,\,\,\,\,\,\,\,\,\,\,\,\,\,\,\,\,\,\,\,\,\,\,\,\,\, \,\,\,\,r=1 \nnb\\
&& 1086(147)~{\rm MeV}  ,\,\,\,\,\,\,\,\,\,\,\,\,\,\,\,\,\,\,\,\,\,\,\,\,\,\,\,\,\,\,\,\,\,\,\,\,\,\,\,\,\,\,\,\,\,\,\,\,\,\,\,\,\,\,\,\,\,\,\,\,\,\,\,\,\,\,=1/\sqrt{2} \nnb\\
&&1071(123)~{\rm MeV} ,\,\,\,\,\,\,\,\,\,\,\,\,\,\,\,\,\,\,\,\,\,\,\,\,\,\,\,\,\,\,\,\,\,\,\,\,\,\,\,\,\,\,\,\,\,\,\,\,\,\,\,\,\,\,\,\,\,\,\,\,\,\,\,\,\,\,=0.
\eea

\subsection*{\b The $\bar u\bar s ds$ current}
We consider the current:
\beq
{\cal O}_{\bar u\bar sds}^{S/P}=\epsilon_{abc}\epsilon_{dec}\Big{[}( \bar u_a\gamma_5  C\bar s_b^T)\otimes( d _d^TC\gamma_5 s_e)+ r\,( \bar u_a C \bar s_b^T)\otimes( d _d^TC  s_e)\Big{]}.
\eeq

\begin{figure}[hbt]
\begin{center}
\hspace*{-7cm} {\bf a) \hspace*{8.cm} \bf b)} \\
\includegraphics[width=8.2cm]{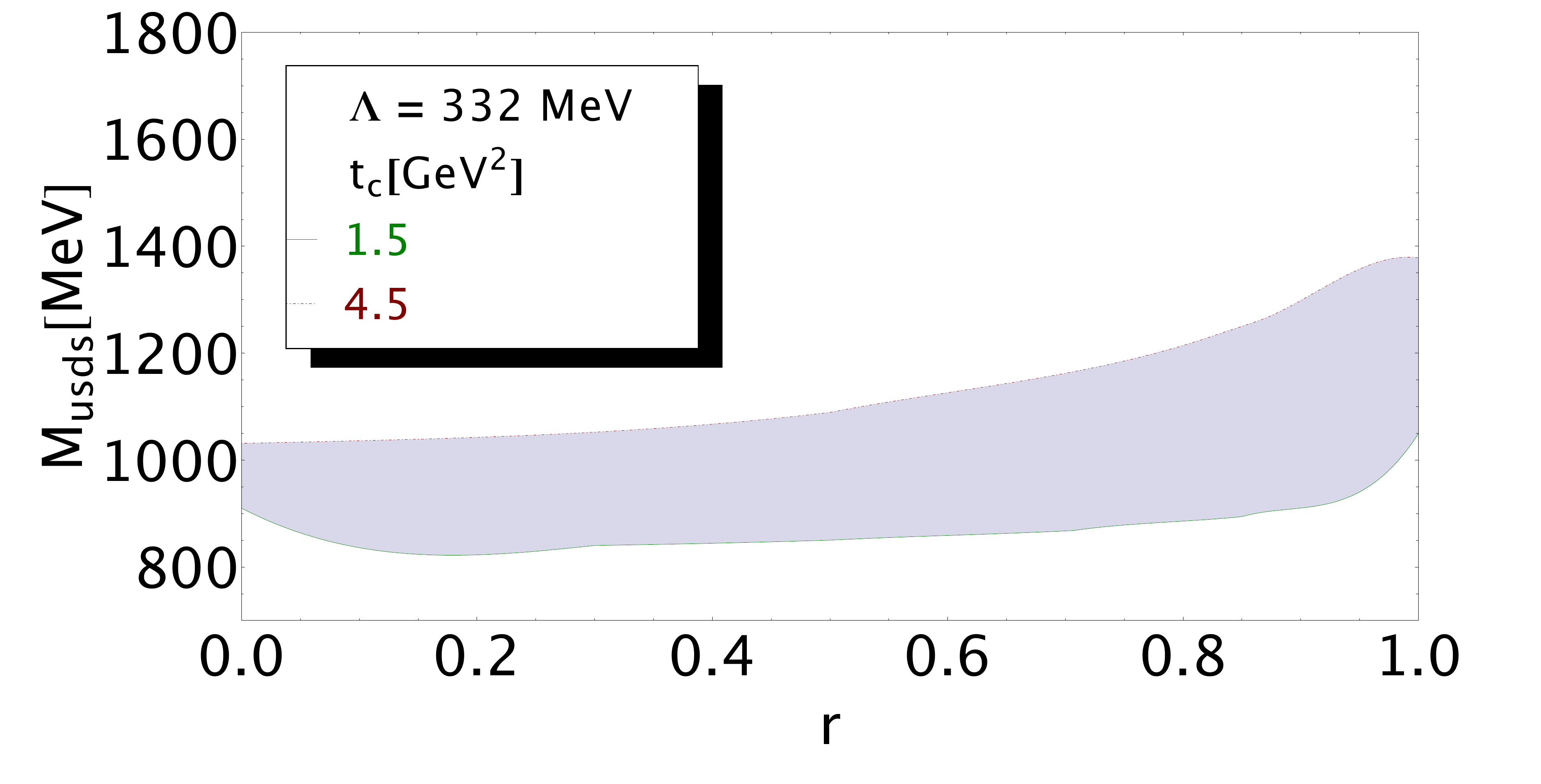}
\includegraphics[width=8.cm]{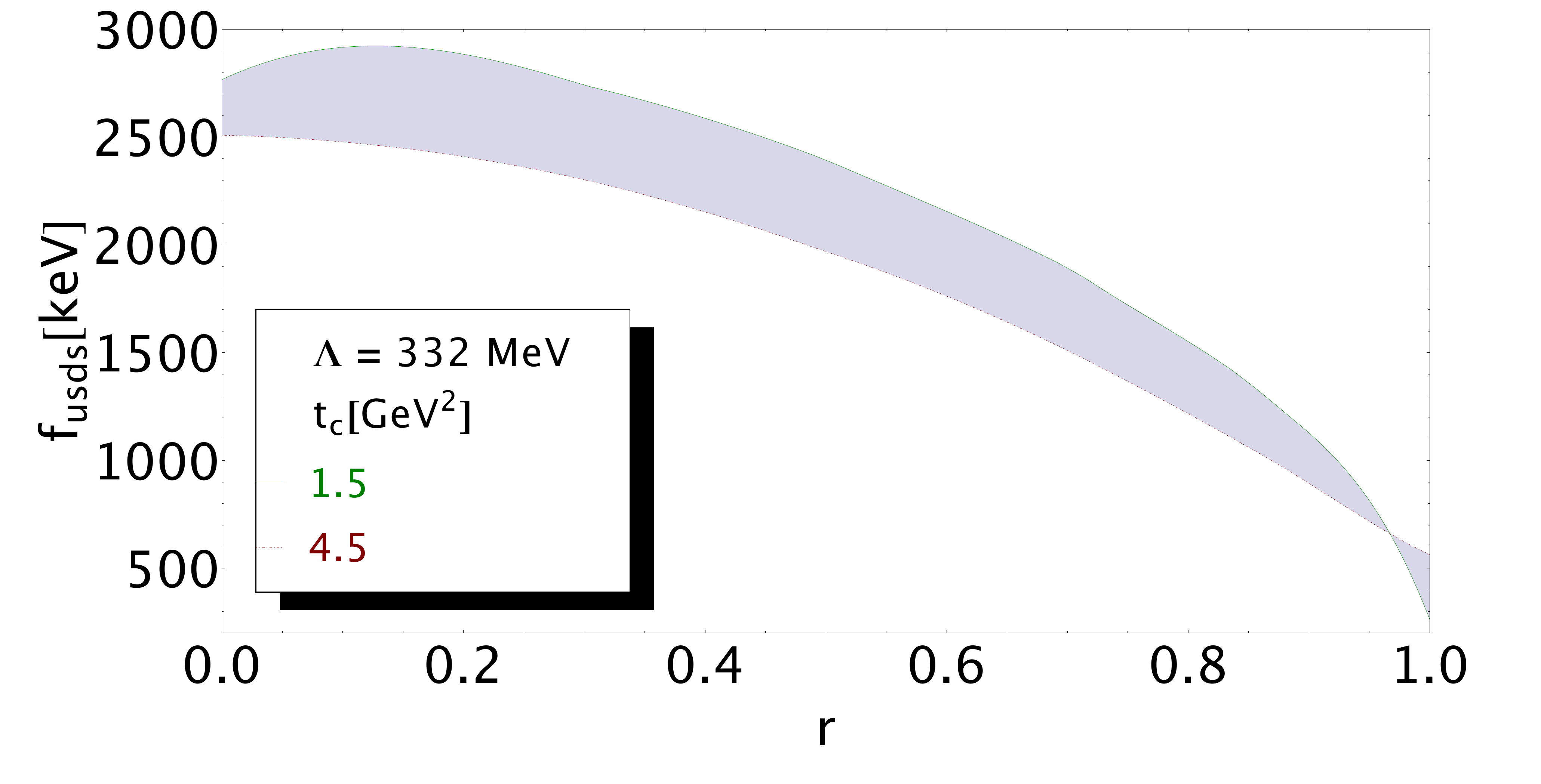} 
\vspace*{-0.5cm}
\caption{\footnotesize   Behaviour of the optimal values in $\tau$ for the  a) mass and  b) coupling of the $\bar u \bar s ds$ four-quark state versus the mixing parameter $r$ for two extremal values of $t_c$ allowed by the constraint: $R_{P/C}\geq 1$.} 
\label{fig:ss-r}
\end{center}
\end{figure}

Its QCD spectral function normalized to $(1+r^2)$ reads:
\bea
\rho^{pert}_{\bar u\bar sds}&=&\frac{t^4}{5\times 3\times 2^{12}\pi^6}-\frac{m_s^2}{3\times 2^{9}\pi^6}t^3,\,\,\,\,\,\,\,\,\,\,\,\,\,\,\,\,\,\,\,\,\,\,\,\,\,\,\,\,\,
\rho^{\la G^2\ra }_{\bar u\bar sds}=\frac{\la \alpha_s G^2\ra }{ 3\times 2^{9}\pi^5}\,t^2,\nnb\\
\rho^{\la \bar qq\ra}_{\bar u\bar sds}&=&\frac{[ m_d\la \bar dd\ra + 
m_u\la \bar uu\ra+2m_s\la\bar ss\ra]- 2[ m_s(\la \bar dd\ra +\la \bar uu\ra)+   
(m_d+m_u)\la \bar ss\ra] (1-r^2)/(1+r^2)}{ 3\times 2^{7}\pi^4}\,{t^2}\nnb\\
\rho^{\la \bar q Gq\ra}_{\bar u\bar sds}&=&\frac{[ m_d\la \bar dGd\ra + m_u\la \bar uGu\ra+2m_s\la\bar sGs\ra]  +6[ m_s(\la \bar dGd\ra +\la \bar uGu\ra)+(m_d+m_u)\la\bar sGs\ra](1-r^2)/(1+r^2) }{3\times 2^{8}\pi^4}\,t\nnb\\
\rho^{\la \bar qq\ra^2}_{\bar u\bar sds}&=& \frac{\rho\la \bar ss\ra(\la \bar dd\ra +\la\bar uu\ra)}{3\times 2^{3}\pi^2}\frac{(1-r^2)}{(1+r^2)}\,t,\,\,\,\,\,\,\,\,\,\,\,\,\,\,\,\,\,\,\,\,\,\,\,\,\,\,
\rho^{\la G^3\ra}_{\bar u\bar sds}= {\cal O}(m_q^2\,\la g^3  G^3\ra),
\eea
The analysis and the shape of different curves are very similar to the case of the $\sigma$ and will not be reported here. The $R_{P/C}\geq 1$ condition 
restricts the value of $t_c$ to be in the range (2-4.5) GeV$^2$ for $r=0$ and (1.5-4.5) GeV$^2$ for $0<r\leq 1$. We show in Fig.\,\ref{fig:ss-r} the $r$-behaviour of the optimal results in $\tau$ for the previous range of $t_c$. We obtain for $\tau\simeq 2.5,\, 2.4,\,2.8$ GeV$^{-2}$ for $r=0,\,1/\sqrt{2},\,1$:  
\bea
M_{\bar u\bar sds}= &&1214(166) ~{\rm MeV}  ,\,\,\,\,\,\,\,\,\,\,\,\,\,\,\,\,\,\,\,\,\,\,\,\,\,\,\,\,\,f_{\bar u\bar sds}=\,\,\,\,413(153)~{\rm keV},\,\,\,\,\,\,\,\,\,\,\,\,\,\,\,\,\,\,\,\,\,\,\,\,\,r=1, \nnb\\
&& 1012(170)~{\rm MeV}  ,\,\,\,\,\,\,\,\,\,\,\,\,\,\,\,\,\,\,\,\,\,\,\,\,\,\,\,\,\,\,\,\,\,\,\,\,\,\,\,\,\,\,\,=\,\,\,\,1683(342)~{\rm keV},\,\,\,\,\,\,\,\,\,\,\,\,\,\,\,\,\,\,\,\,\,\,\,\,\,  \,\,=1/\sqrt{2}, \nnb\\
&& \,\,971(112)~{\rm MeV} ,\,\,\,\,\,\,\,\,\,\,\,\,\,\,\,\,\,\,\,\,\,\,\,\,\,\,\,\,\,\,\,\,\,\,\,\,\,\,\,\,\,\,\,\,=\,\,\,\,2637(514)~{\rm keV},\,\,\,\,\,\,\,\,\,\,\,\,\,\,\,\,\,\,\,\,\,\,\,\,\, \,\,=0.
\eea

\begin{table}[H]
\setlength{\tabcolsep}{0.48pc}
{\footnotesize{
\begin{tabular}{ll ll  ll  ll ll ll ll ll l c}
\hline
\hline
                Currents  
                    &\multicolumn{1}{c}{$\Delta t_c$}
					&\multicolumn{1}{c}{$\Delta \tau$}
					&\multicolumn{1}{c}{$\Delta \Lambda$}
					&\multicolumn{1}{c}{$\Delta PT$}
					&\multicolumn{1}{c}{$\Delta m_q$}
					&\multicolumn{1}{c}{$\Delta \bar{q}q$}
					&\multicolumn{1}{c}{$\Delta \kappa$}					
					&\multicolumn{1}{c}{$\Delta G^2$}
					&\multicolumn{1}{c}{$\Delta \bar q Gq$}
					&\multicolumn{1}{c}{$\Delta G^3$}
					&\multicolumn{1}{c}{$\Delta \bar{q}q^2$}
					
					&\multicolumn{1}{c}{$\Delta OPE$}
					&\multicolumn{1}{c}{Value}
\\
					
\hline
\boldmath$ 
{\cal O}_{\bar u\bar dud}^{S/P}$& \\
\it Masses [MeV]\\
1 &181&13&3&28&0&0&--&1&0.2&0&0&--&1225(184)\\
$1/\sqrt{2}$&177&6.1&22&31&0&10.2&--&0.2&0.4&0&10.3&32&1078(184)\\
0 &87&5.3&25&31&0&10.5&--&0&0&0&11.5&60&1040(114)\\
\it Couplings [keV]\\
1 &124&14&12&5&0&0&--&2.6&0&0&0&--&449(126)\\
$1/\sqrt{2}$&140&12&6.5&26&0&52&--&2.5&1.5&0&53&54&1160(170)\\
0&122&29&7.5&119&0&94&--&1.5&1.5&0&95&152&1780(266) \\
\boldmath$ 
{\cal O}_{\bar u\bar sud}^{S/P}$& \\
\it Masses [MeV]\\
1&155&26&2&3.5&0.7&1.2&1&0.8&1.7&0&0&1.9&1215(157)\\
$1/\sqrt{2}$&134&7&25&31&0.1&11&2.6&0.2&0.4&0&26&33&997(147)\\
0&56&8&21&62&0.5&7&2.7&0.2&0.3&0&23&84&957(123)\\
\it Couplings [keV]\\
1&129&13&8.5&5&4&1.7&1.7&4&2.3&0&0.4&8.5&422(131)\\
$1/\sqrt{2}$&203&28&5.5&118&1&90&20.5&1.5&5.5&0&189&214&1835(382)\\
0&169&60&7.5&355&1.5&140&33.5&1&9.5&0&293&377&2902(638)\\
\boldmath $ 
{\cal O}_{\bar u\bar sds}^{S/P}$& \\
\it Masses [MeV]\\
1&164&25&1&2.5&0.5&2.0&2.0&0.15&2.2&0&0&2&1214(166)\\
$1/\sqrt{2}$&153&5&25&30&0.2&11.5&6.5&0.4&2.1&0&24.3&56&1012(170)\\
0&60&8&27&37&0&9.5&5.6&0.15&1.1&0&22.4&79&971(112)\\
\it Couplings [keV]\\
1&150&16&12&11&3.5&3.5&5&4&4.5&0&0&16&413(153)\\
$1/\sqrt{2}$&192&25&4&98&2.5&81&45.5&1.5&10&0&178&171&1683(342)\\
0&129&56&11&297&2.5&113&79&0.7&38&0&278&241&2637(514)\\

\hline
\hline
\end{tabular}
}}
 \caption{\footnotesize The same caption as for Table\,\ref{tab:res} but for the four-quark states in the Scalar $\oplus$ Pseudoscalar configurations and for  three typical values of the mixing parameter $r=1,~1/\sqrt{2},~0$. Note that the lower value of $t_c$ allowed by $R_{P/C}\geq 1$ is 2 GeV$^2$ for $r=0$ and 1.5 GeV$^2$ for $r=1/\sqrt{2}, ~1.$}
\label{tab:res-4q}
\end{table}

\section{The vector\,$\oplus$\,axial-vector currents}
\subsection*{\b The $\bar u\bar d us$ current in QCD}
\d We consider the current:
\beq
{\cal O}_{\bar u\bar dud}^{V/A}=\frac{1}{\sqrt{2}}\Big{[}( \bar u_a \gamma_\mu\gamma_5 C\bar d_b^T)\otimes( u _a^TC\gamma^\mu\gamma_5  q_b-u_b^TC\gamma^\mu\gamma_5q_a)+r\,( \bar u_a \gamma_\mu C \bar d_b^T)\otimes( u _a^TC \gamma^\mu q_b+u_b^TC\gamma^\mu q_a)\Big{]}.
\eeq
In Refs.\,\cite{ZHU,STEELE}, the value $r=1/\sqrt{2}$ has been taken which corresponds to the choice $3_c$ and $\bar 6_c$ in the colour representation (denoted $A_6$ and $V_3$ currents). In our analysis, we leave $r$ as a free parameter inside the range 0 to 1. 
\subsection*{\d QCD lowest order expression up to $D=6$}
The expression of the spectral function normalized to $(1+2r^2)$ reads:
\bea
\rho^{pert}_{\bar u\bar dus}&=&\frac{t^3}{5\times 3\times 2^{11}\pi^6} 
\Bigg{[}\big{[}t-20 (2 m_u^2 + m_d^2 + m_s^2)\big{]}-\frac{(1 - 2 r^2)}{(1+2r^2)} 10m_u (m_d + m_s)\Bigg{]},\nnb\\
\rho^{\la \bar qq\ra}_{\bar u\bar dus}&=&\frac{\big{[}(3+2r^2)m_u+2m_d+(1-2r^2)m_s\big{]}\la \bar qq\ra +[(1-2r^2)m_u+(1+2r^2)m_s]\la\bar ss\ra}{ 3\times 2^{6}\pi^4(1+2r^2)}\,{t^2},\nnb\\
\rho^{\la G^2\ra }_{\bar u\bar dus}&=&\frac{\la \alpha_s G^2\ra }{ 3\times 2^{9}\pi^5}\frac{(1+6r+5r^2)}{(1+2r^2)}\,t^2,\nnb\\
\rho^{\la \bar q Gq\ra}_{\bar u\bar dus}&=
&-\Big{[}{\la\bar qGq\ra \big{[}(5 - 36 r - 11 r^2) m_u +  2(2-9r -2 r^2) m_d + 
   3 (1 + r^2) m_s\big{]}}\,\nnb\\
   &&+
  {\la\bar sGs\ra \big{[}(1 - 18 r - 7 r^2) m_s + 3 (1 + r^2) m_u)\big{]}}\Big{]}
\frac{t}{3\times 2^{8}\pi^4(1+2r^2) }\,\nnb\\
\rho^{\la \bar qq\ra^2}_{\bar u\bar dus}&=&- \frac{\rho[\la \bar qq\ra^2+\la \bar qq\ra\la\bar ss\ra]}{24\,\pi^2}\frac{(1-2r^2)}{(1+2r^2)}\,t,\,\,\,\,\,\,\,\,\,\,\,\,\,\,\,\,\,\,\,\,\,\,\,\,\,\,
\rho^{\la G^3\ra}_{\bar u\bar dus}= {\cal O}(m_q^2\,\la g^3  G^3\ra),
\label{eq:spec-VA}
\eea
where $\la\bar qq\ra \equiv \la\bar uu\ra\simeq \la\bar dd\ra$.  
\subsection*{\d Truncation of the OPE}
 We truncate the series at the dimension-six condensate contributions due to the less controlled values of the higher dimension condensates and to the mixing of different operators which appear after the uses of the equation of motion\,\cite{SNT}. We estimate the unknown higher dimension contributions (some classes of dimension-eight contributions are given in Ref.\,\cite{ZHU}) by using the scaling factor in Eq.\,\ref{eq:ope}.

\subsection*{\d Approximate higher order PT QCD corrections}
We introduce the higher order PT corrections from the convolution integral\,\cite{PICH,SNPIVO}:
\beq
\hspace*{-0.1cm}\frac{1}{ \pi}{\rm Im}\, \psi_{\bar u\bar d ud}(t)= k_{V/A}\,\int_{0}^tdt_1\,
\int_{0}^{(\sqrt{t}-\sqrt{t_1})^2}\hspace*{-0.8cm}dt_2~\lambda^{1/2}
\Bigg{[}\ga \frac{t_1}{ t}+ \frac{t_2}{ t}-1\dr^2+8\frac{t_1t_2}{t^2}\Bigg{]}\times \frac{1}{\pi}{\rm Im} \,\Pi_{V/A}(t_1) \frac{1}{\pi} {\rm Im}\,\Pi_{V/A}(t_2),
\eeq
where :
\beq
k_{V/A}=\frac{1}{96\pi^2}
\eeq
is an appropriate normalisation factor. $\Pi_{V/A}(t)$ is the two-point function associated to the vector $V\equiv \bar u\gamma_\mu d$ and axial-vector  $A\equiv \bar u\gamma_\mu\gamma_5 d$ currents. Its spectral function is known to order $\alpha_s^4$ in the massless quark limit. It reads for three flavours and in the $\overline{MS}$-scheme\,\footnote{For reviews, see e.g.\,\cite{CHETV,SNB1}.}\,:
\beq
 \frac{1}{\pi}{\rm Im} \,\Pi_{V/A}(t)=\frac{1}{4\pi^2}\Big{[}1+a_s+ 1.623\,a_s^2 - 6.370\,a_s^3 - 106.8798\,a_s^4 +1092\,a_s^5\Big{]}.
\eeq
We have estimated the last $a_s^5$ term assuming a geometric growth of the $a_s$ coefficient while the alternate sign is fixed assuming that the series reaches its asymptotics at this order. 

\subsection*{\b The $\bar u\bar d ud$ state }
 The QCD expression can be deduced from Eq.\,\ref{eq:spec-VA} by  replacing the quark $s$ by $d$.
\subsection*{\d LSR analysis  within a NWA}
\begin{figure}[hbt]
\begin{center}
\hspace*{-7cm} {\bf a) \hspace*{8.4cm} \bf b)} \\
\includegraphics[width=8.4cm]{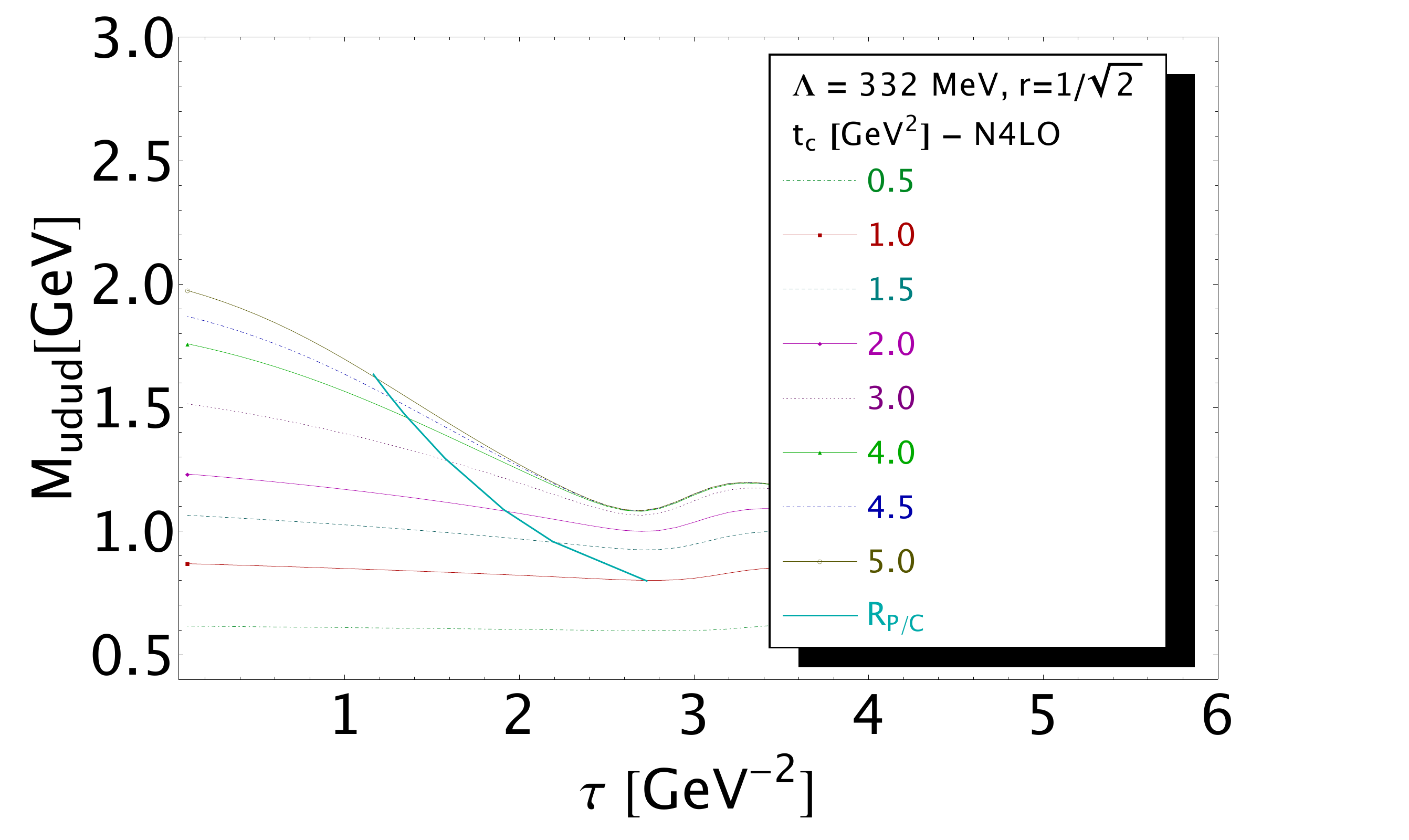}
\includegraphics[width=7.9cm]{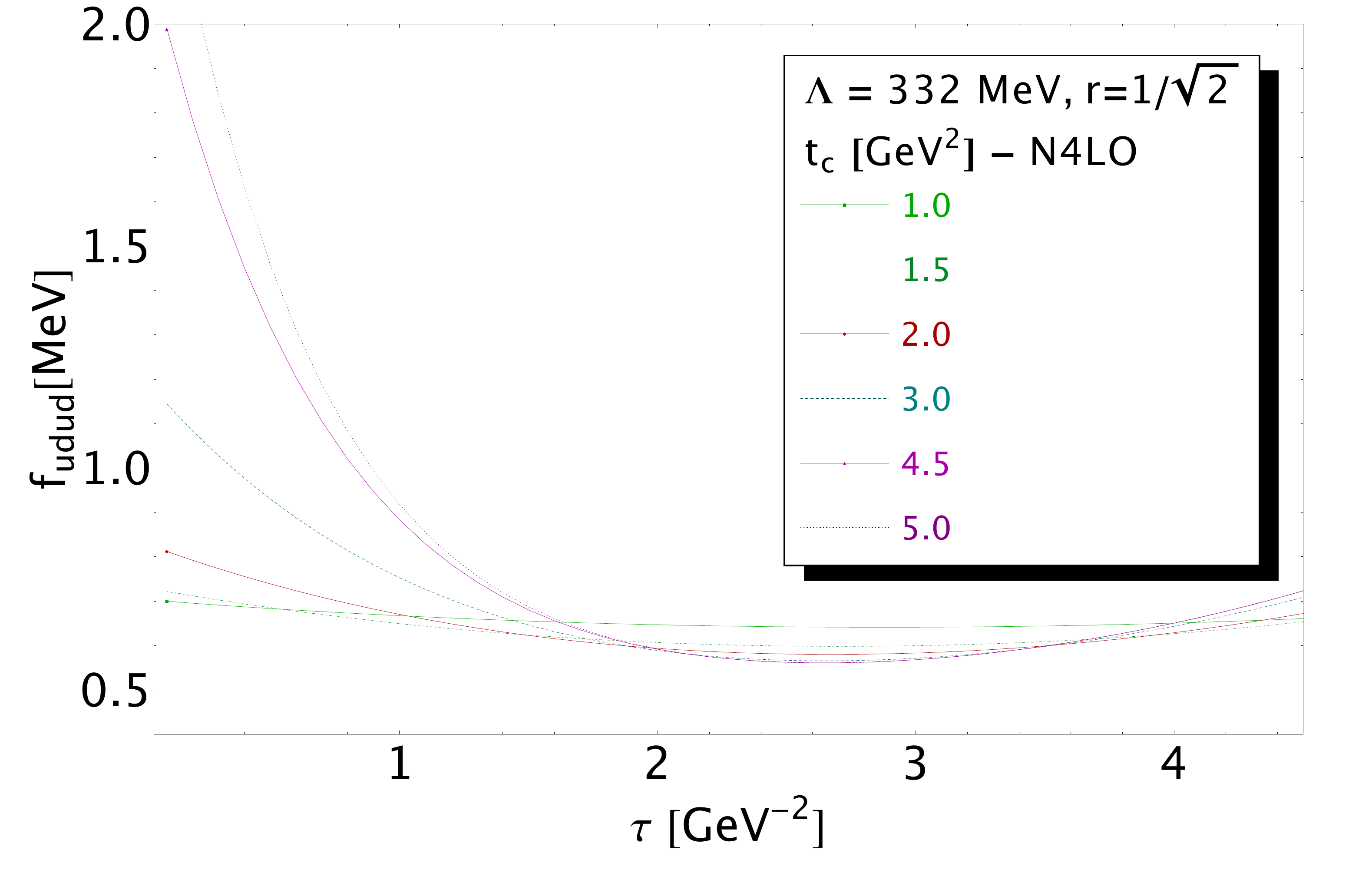} 
\vspace*{-0.5cm}
\caption{\footnotesize   Behaviour of a) mass and  b) coupling of the $\bar u \bar d ud$ four-quark states versus $\tau$ for different values of $t_c$ and for $r=1/\sqrt{2}$ in the case of vector $\oplus$ axial-vector current. } 
\label{fig:ud-2-VA}
\end{center}
\vspace*{-0.5cm}
\end{figure}

 --  We show the analysis in Figs.\,\ref{fig:ud-1-VA} and \ref{fig:ud-2-VA} for $r=1$ and $1/\sqrt{2}$. The analysis for $0\leq r< 1/\sqrt{2}$ is not conclusive within our truncation of the OPE as we do not have $\tau$-stability. This feature is expected from the negative contribution of the dimension-six condensate (see Eq.\,\ref{eq:spec-VA}). 
  
 -- One can notice from the figures that the stability is reached for large values of $\tau\simeq (2.3-3.7)$ GeV$^{-2}$ for the coupling and (2.8-2.9) GeV$^{-2}$ for the mass due to the relative small contributions of the $\la\bar qq\ra^2$ condensate which vanish for $r=1/\sqrt{2}$. This fact is reflected on the large errors due to the truncation of the OPE and of the N5LO PT corrections as given in Table\,\ref{tab:res-4q-VA}. We obtain within a NWA:
\bea
M^{V/A}_{\bar u\bar dud}= &&831(141) ~{\rm MeV}  ,\,\,\,\,\,\,\,\,\,\,\,\,\,\,\,\,\,\,\,\,\,\,\,\,\,\,\,\,\,f_{\bar u\bar dud}=\,\,\,\,938(278)~{\rm keV}.\,\,\,\,\,\,\,\,\,\,\,\,\,\,\,\,\,\,\,\,\,\, r=1, \nnb\\
&& 941(193)~{\rm MeV}  ,\,\,\,\,\,\,\,\,\,\,\,\,\,\,\,\,\,\,\,\,\,\,\,\,\,\,\,\,\,\,\,\,\,\,\,\,\,\,\,\,\,\,\,\,=\,\,\,\,601(66)~{\rm keV},\,\,\,\,\,\,\,\,\,\,\,\,\,\,\,\,\,\,\,\,\,\,\,\,\,  \,\,\,=1/\sqrt{2},
\label{eq:4ud}
\eea
for $t_c=(1-4.5)$ GeV$^2$ and $\tau \simeq (2.8\sim 3.7)$ GeV$^{-2}$: $t_c=1$ GeV$^2$ corresponds to the beginning of $\tau$-stability allowed by $R_{P/C}$ and $t_c=4.5$ GeV$^2$ is the beginning of $t_c$-stability. 
\subsection*{\d Effects of the higher order  PT corrections in the LSR analysis}
\begin{figure}[hbt]
\begin{center}
\hspace*{-7cm} {\bf a) \hspace*{8.4cm} \bf b)} \\
\includegraphics[width=8.cm]{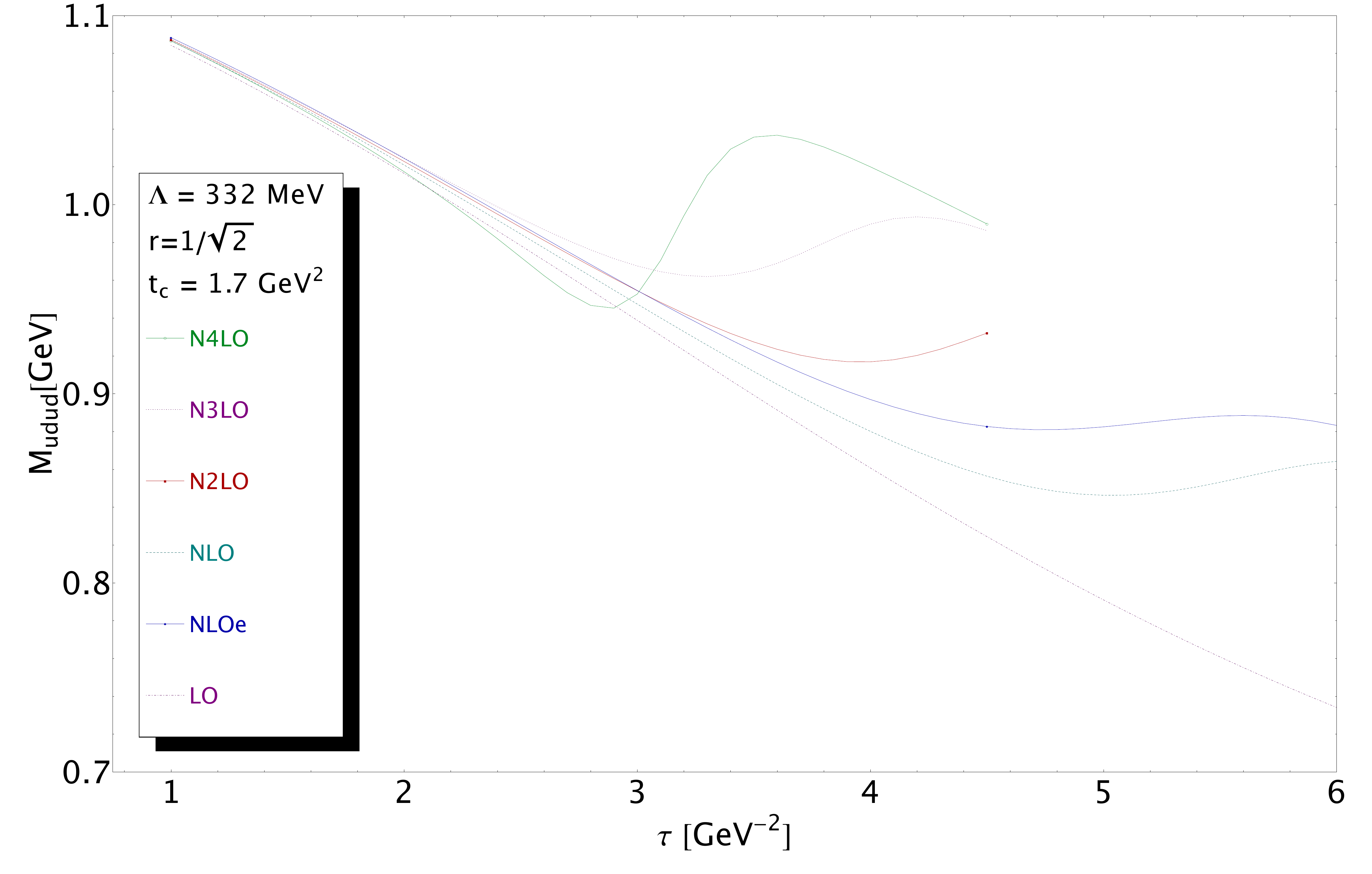}
\includegraphics[width=8.2cm]{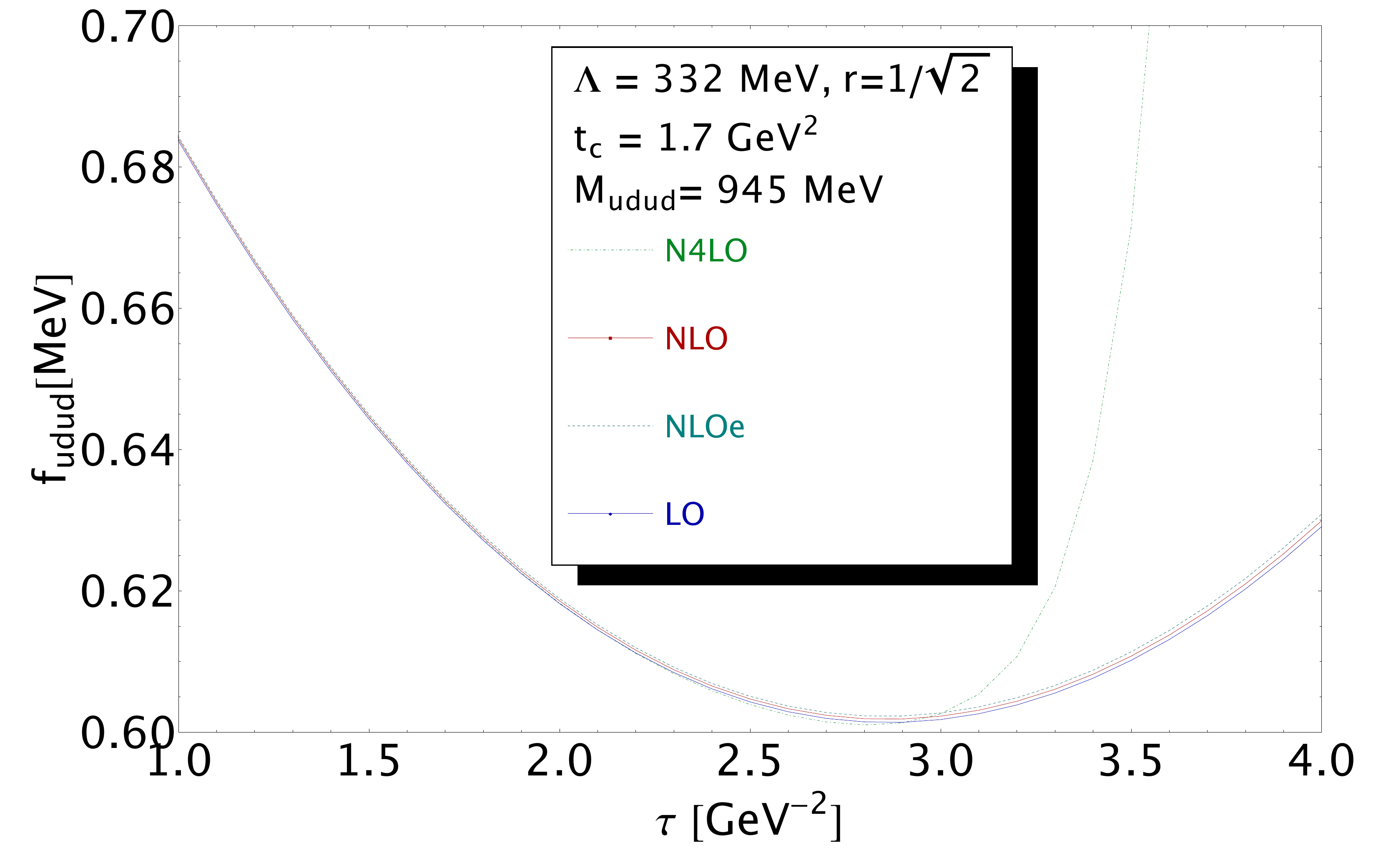} 
\vspace*{-0.5cm}
\caption{\footnotesize   Behaviour of a) mass and  b) coupling of the $\bar u \bar d ud$ four-quark states for different truncation of the PT series versus $\tau$ for $t_c$ =1 GeV$^2$, $r=1/\sqrt{2}$ in the case of vector $\oplus$ axial-vector current. NLOe is the exact result from \cite{STEELE}.} 
\label{fig:pert-2-VA}
\end{center}
\vspace*{-0.5cm}
\end{figure}

-- Our approximate NLO coefficient is 1.72 smaller than the exact NLO result of Ref.\,\cite{STEELE}. The difference is due to the fact that the exact result takes into account the contribution of the off-diagonal diagram and the non-leading term in $1/N_c$.   We compare the effect of different truncation of the PT series in Fig.\,\ref{fig:pert-2-VA} for the case $r=1/\sqrt{2}$ where we have fixed $t_c\simeq 1.7$ GeV$^2$ for approximately reproducing the central values of the mass and coupling obtained previously:
\beq
M^{V/A}_{\bar u\bar dud}\vert_{1.7}\simeq 945~{\rm GeV}  ,\,\,\,\,\,\,\,\,\,\,\,\,\,\,\,\,\,\,\,\,\,\,\,\,\,\,\,\,\,f^{V/A}_{\bar u\bar dud}\vert_{1.7}\simeq \,\,\,\,601~{\rm keV}.
 \eeq
We find that the use of the factorized $\alpha_s$ correction underestimates the value of the mass by about 35 MeV compared to the exact result of 881 MeV. However, this shift  is smaller than the error of the mass determination. 
 
 -- We notice, like in the previous cases, that the inclusion of higher order PT corrections improve the predictions of the LSR as the  stability is obtained at lower values of $\tau$. From NLO to N4LO, the value of the mass shifts by about 64 MeV.  The effects of the $\alpha_s^n$ corrections to the coupling are marginal. 
\subsection*{\d Finite width corrections}
  We proceed as in the previous sections and find that the masses become:
\bea
M_{\bar u\bar dud}^{V/A}\vert_{BW}= &&1014(141) ~{\rm MeV}  ,\,\,\,\,\,\,\,\,\,\,\,\,\,\,\,\,\,\,\,\,\,\,\,\,\,\,\,\,\,\,\,\,\,\,\,\,\,\,\,\,\,\,\,\,\,\,\,\,\,\,\,\,\,\,\,\,\, \,\,\,\,r=1, \nnb\\
&& 1050(193)~{\rm MeV}  ,\,\,\,\,\,\,\,\,\,\,\,\,\,\,\,\,\,\,\,\,\,\,\,\,\,\,\,\,\,\,\,\,\,\,\,\,\,\,\,\,\,\,\,\,\,\,\,\,\,\,\,\,\,\,\,\,\,\,\,\,\,\,\,\,\,=1/\sqrt{2}.
\eea
  where we have used as input the experimental width $\Gamma_\sigma^{pole}$ 520 MeV and $\Gamma_\sigma^{BW}$ = 700 MeV. 
\begin{figure}[hbt]
\begin{center}
\hspace*{-7cm} {\bf a) \hspace*{8.cm} \bf b)} \\
\includegraphics[width=8.2cm]{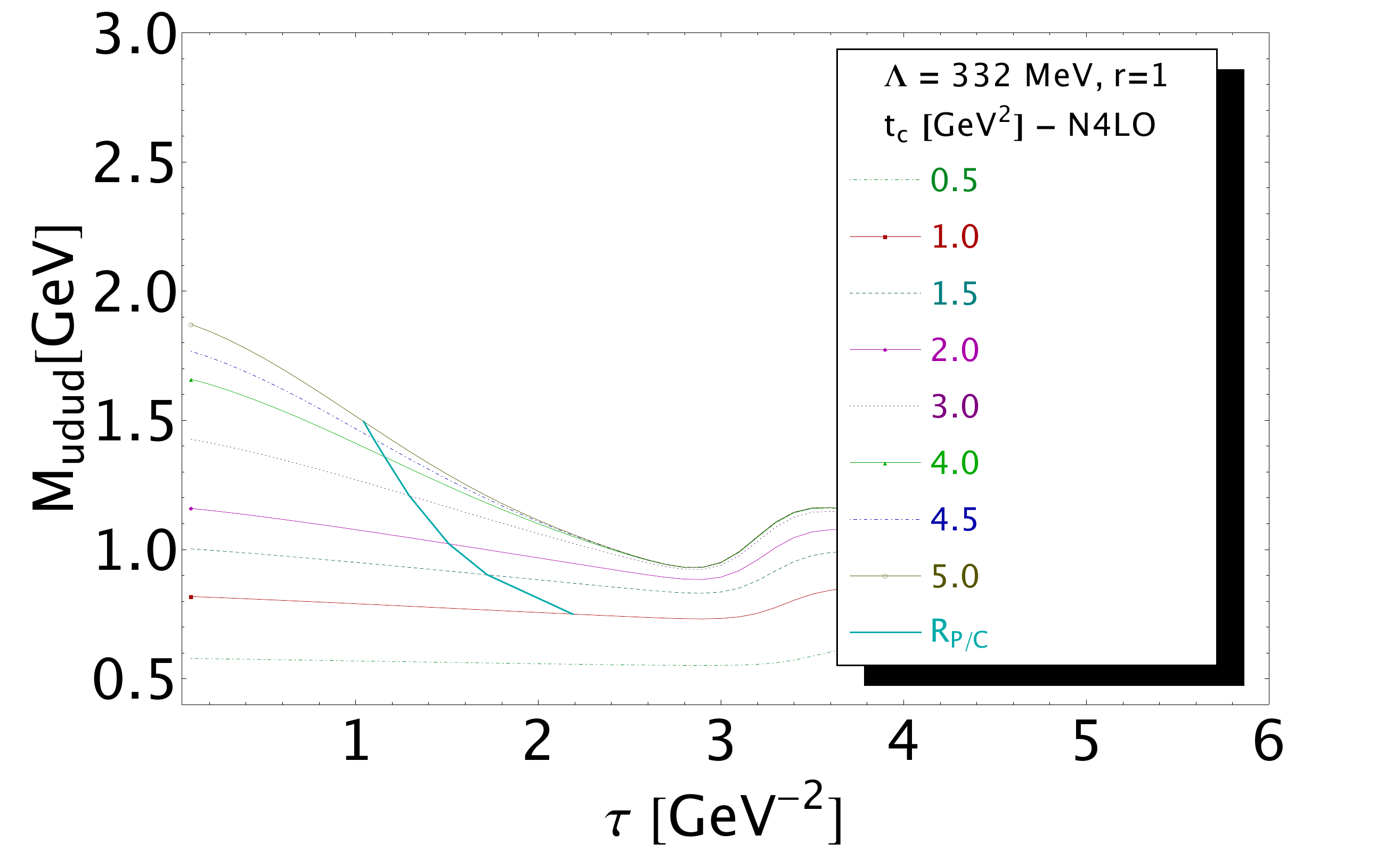}
\includegraphics[width=8.cm]{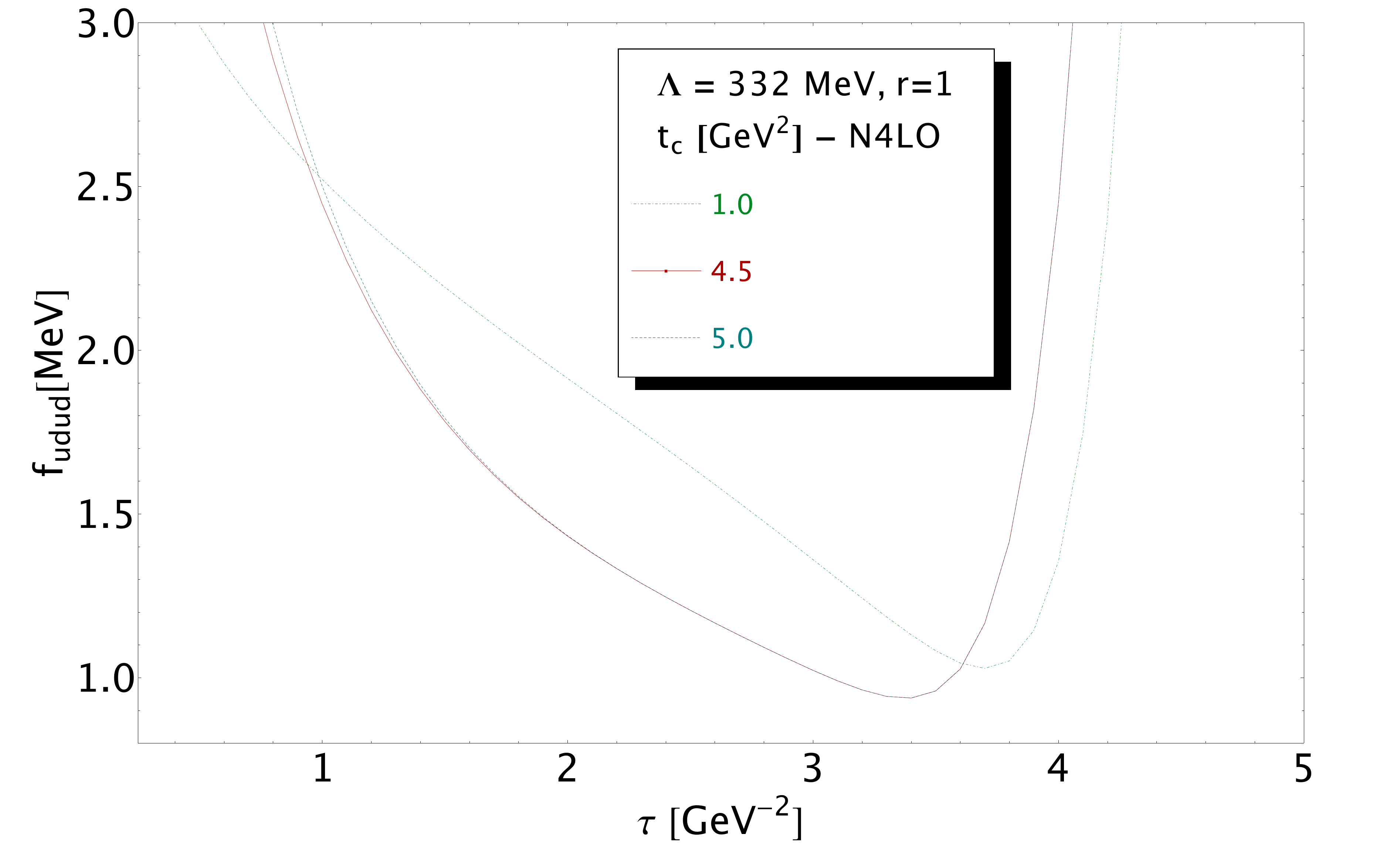} 
\vspace*{-0.5cm}
\caption{\footnotesize   Behaviour of a) mass and  b) coupling of the $\bar u \bar d ud$ four-quark states versus $\tau$ for different values of $t_c$ and for $r=1$ in the case of vector $\oplus$ axial-vector current. } 
\label{fig:ud-1-VA}
\end{center}
\vspace*{-0.5cm}
\end{figure} 
\subsection*{The $\bar u\bar d us$ state}
\begin{figure}[hbt]
\begin{center}
\includegraphics[width=8.cm]{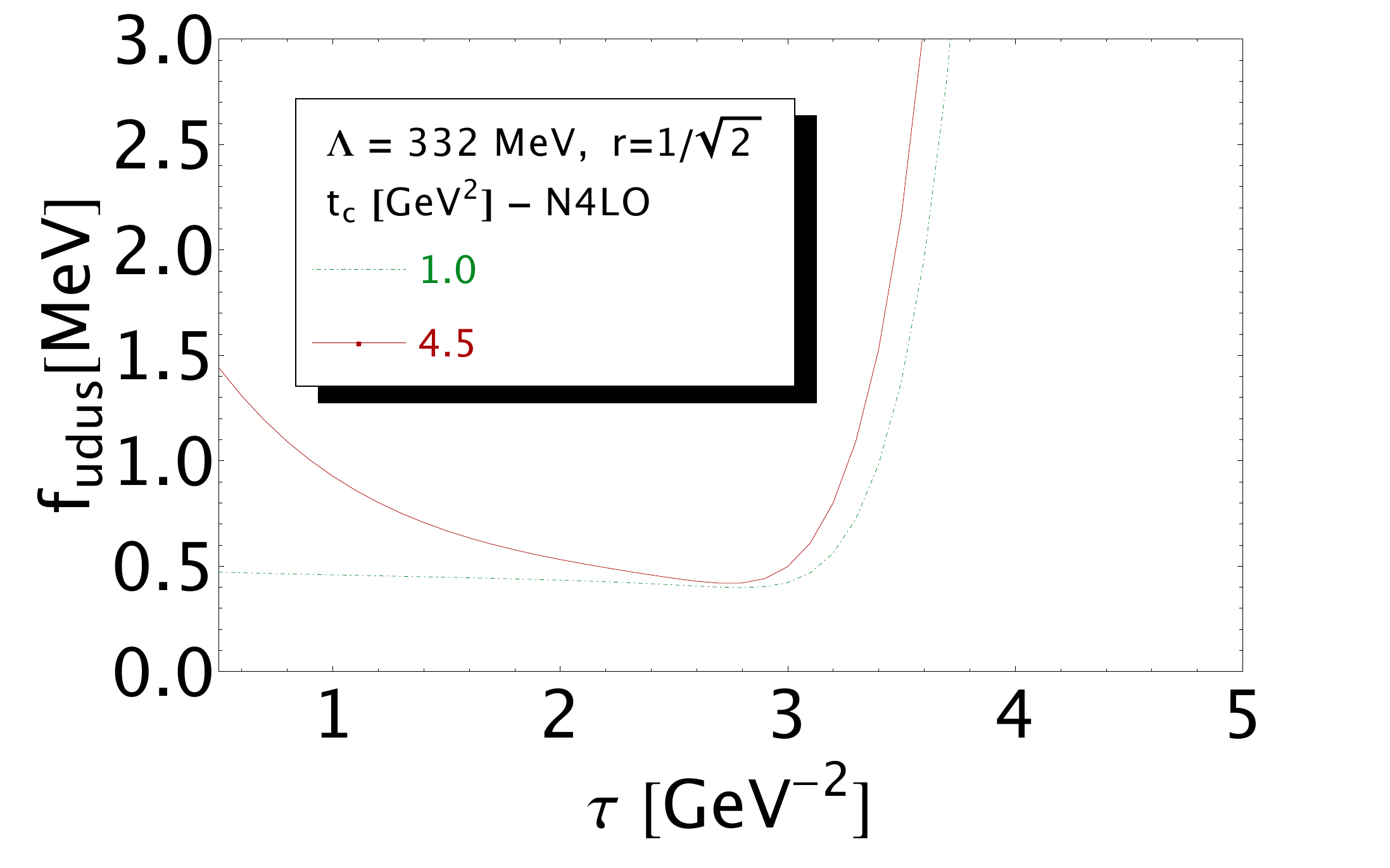}
\vspace*{-0.5cm}
\caption{\footnotesize   Behaviour of the coupling of the $\bar u \bar d us$ four-quark states versus $\tau$ for two extremal values of $t_c$ and  $r=1/\sqrt{2}$ in the case of vector $\oplus$ axial-vector current.} 
\label{fig:4us-coupling-2-VA}
\end{center}
\vspace*{-0.5cm}
\end{figure}

The QCD expression of the spectral function is given in Eq.\,\ref{eq:spec-VA}.  The analysis and the shape of the curves are very similar to the case of the $\bar u\bar d ud$ state and will not be shown as well as the range of $t_c=(1-4.5)$ GeV$^2$ allowed by the $R_{P/C}$ condition. We obtain:
\bea
M^{V/A}_{\bar u\bar dus}= &&834(141) ~{\rm MeV}  ,\,\,\,\,\,\,\,\,\,\,\,\,\,\,\,\,\,\,\,\,\,\,\,\,\,\,\,\,\,f^{V/A}_{\bar u\bar dus}=\,\,\,\,1042(295)~{\rm keV}.\,\,\,\,\,\,\,\,\,\,\,\,\,\,\,\,\,\,\, r=1, \nnb\\
&& 978(185)~{\rm MeV}  ,\,\,\,\,\,\,\,\,\,\,\,\,\,\,\,\,\,\,\,\,\,\,\,\,\,\,\,\,\,\,\,\,\,\,\,\,\,\,\,\,\,\,\,\,=\,\,\,\,408(57)~{\rm keV},\,\,\,\,\,\,\,\,\,\,\,\,\,\,\,\,\,\,\,\,\,\,\,\,\,  \,\,\,=1/\sqrt{2}.
\eea
The sources of errors are given in Table\,\ref{tab:res-4q-VA} where we have added to the QCD errors of $\bar u\bar dud$ the ones due to $SU(3)$ breakings. One can notice the  relative small error due to $t_c$ and $\tau$ for the coupling compared to the  one of $\bar u\bar dud$ which is due to the vicinity of the two minimum for $t_c=1$ and 4.5 GeV$^2$ as shown in Fig.\ref{fig:4us-coupling-2-VA}. The small $SU(3)$ breakings  increase the central value of the mass by 3 (resp. 37) MeV for $r$=1 (resp. 1/$\sqrt{2}$).  

\subsection*{\b The $\bar u\bar s ds$ state}
\d The QCD expression of the spectral function normalized to $(1+2r^2)$ reads:
\bea
\rho^{pert}_{\bar u\bar sds}&=&\frac{t^3}{5\times 3\times 2^{11}\pi^6} 
\Bigg{[}\big{[}t-20 (m_u^2 + m_d^2 + 2m_s^2)\big{]}-\frac{(1 - 2 r^2)}{(1+2r^2)} 10\,m_s (m_u + m_d)\Bigg{]},\nnb\\
\rho^{\la \bar qq\ra}_{\bar u\bar sds}&=&\frac{(m_u + m_d)\la \bar qq \ra+ 2 m_s \la \bar ss\ra +(1-2r^2)\big{[}(m_u+m_d)\la\bar ss\ra+2m_s\la \bar qq \ra\big{]}}{ 3\times 2^{6}\pi^4(1+2r^2)}\,{t^2},\nnb\\
\rho^{\la G^2\ra }_{\bar u\bar sds}&=&\frac{\la \alpha_s G^2\ra }{ 3\times 2^{9}\pi^5}\frac{(1+6r+5r^2)}{(1+2r^2)}\,t^2,\nnb\\
\rho^{\la \bar q Gq\ra}_{\bar u\bar sds}&=
&-\Big{[}{\la\bar qGq\ra \big{[}    (m_u+m_d) (1 - 18 r - 7 r^2)   +6m_s(1+r^2)\big{]}}\,\nnb\\
   &&+
  {\la\bar sGs\ra \big{[} 3(m_u+m_d)(1+r^2)+2m_s (1 - 18 r - 7 r^2)  \big{]}}\Big{]}
\frac{t}{3\times 2^{8}\pi^4(1+2r^2) }\,\nnb\\
\rho^{\la \bar qq\ra^2}_{\bar u\bar sds}&=&- \frac{\rho[\la \bar qq\ra\la\bar ss\ra]}{12\,\pi^2}\frac{(1-2r^2)}{(1+2r^2)}\,t,\,\,\,\,\,\,\,\,\,\,\,\,\,\,\,\,\,\,\,\,\,\,\,\,\,\,
\rho^{\la G^3\ra}_{\bar u\bar sds}= {\cal O}(m_q^2\,\la g^3  G^3\ra),
\label{eq:4ss-VA}
\eea
where $\la\bar qq\ra \equiv \la\bar uu\ra\simeq \la\bar dd\ra$.  

\begin{figure}[hbt]
\begin{center}
\hspace*{-7cm} {\bf a) \hspace*{8.4cm} \bf b)} \\
\includegraphics[width=8.cm]{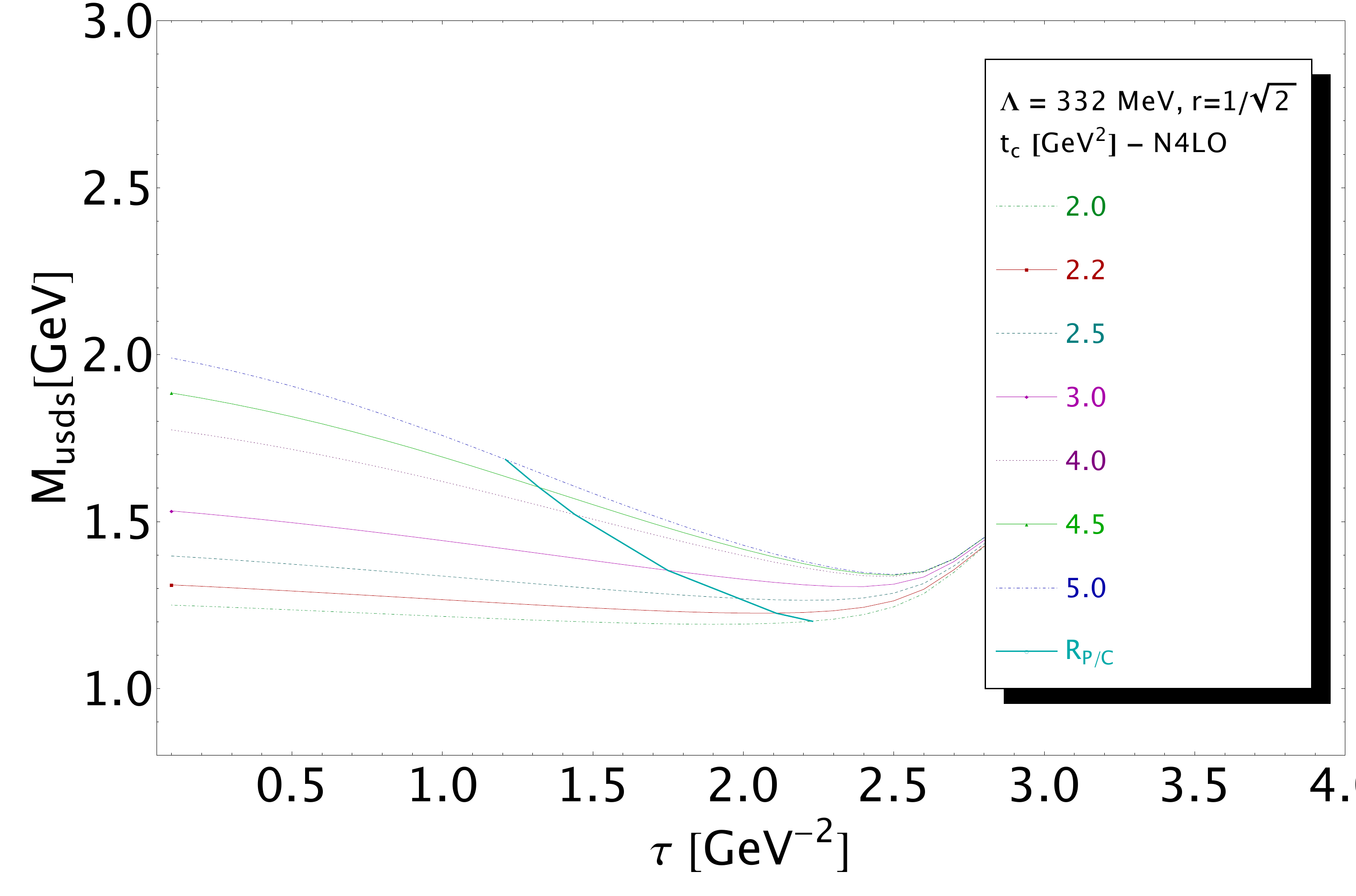}
\includegraphics[width=8.2cm]{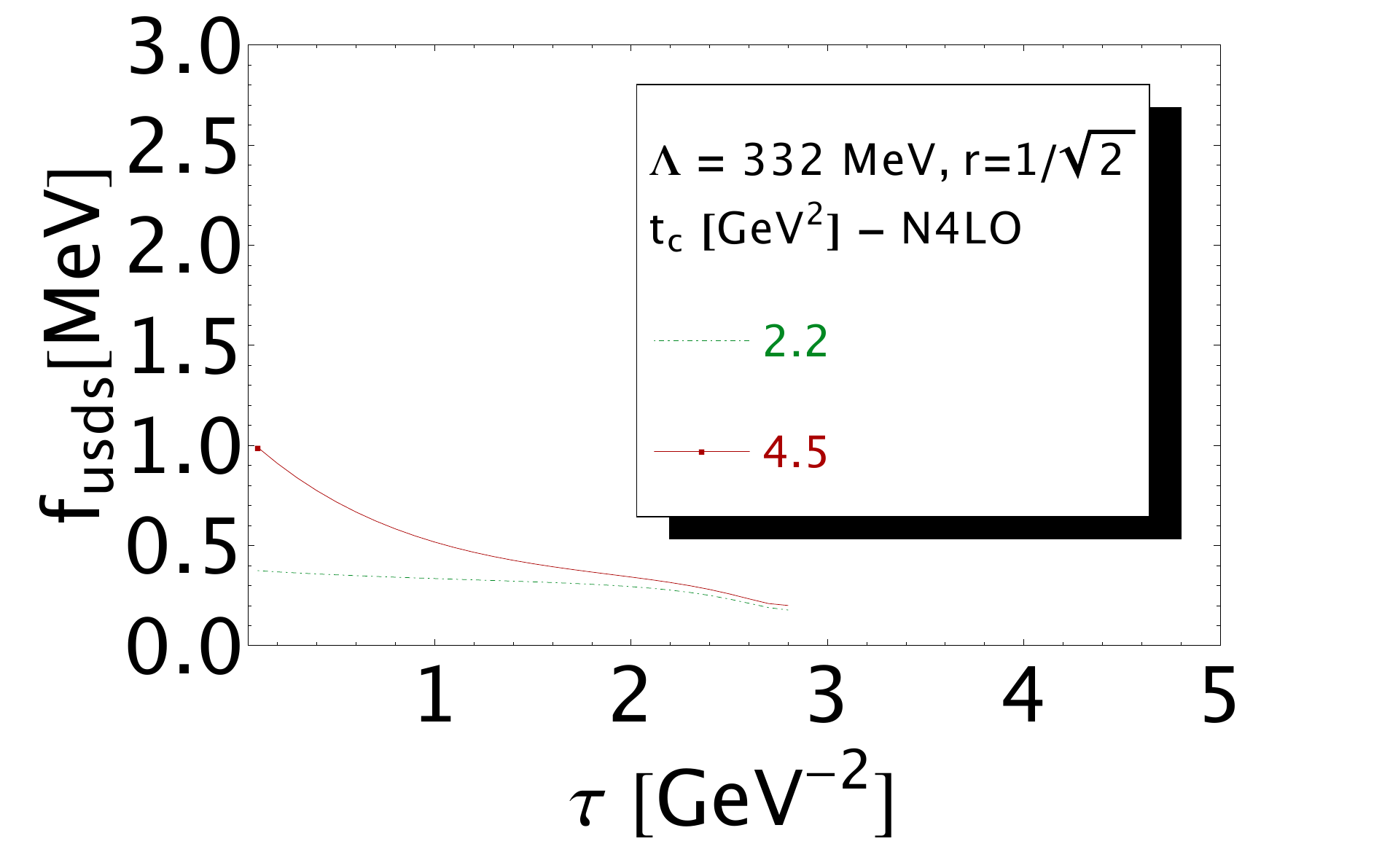} 
\vspace*{-0.5cm}
\caption{\footnotesize   Behaviour of a) mass and  b) coupling of the $\bar u \bar sds$ four-quark states versus $\tau$ for different values of $t_c$  and for $r=1/\sqrt{2}$ in the case of vector $\oplus$ axial-vector current. } 
\label{fig:4ss-2-VA}
\end{center}
\vspace*{-0.5cm}
\end{figure}

\d The analysis and the shape of the curves are very similar to the case of the $\bar u\bar d ud$ state except that the $R_{P/C}\geq 1$ condition requires the range $t_c=(2-4.5)$ GeV$^2$ as shown in Fig.\ref{fig:4ss-2-VA} for $r=1/\sqrt{2}$. 
\bea
M^{V/A}_{\bar u\bar sds}= &&840(140) ~{\rm MeV}  ,\,\,\,\,\,\,\,\,\,\,\,\,\,\,\,\,\,\,\,\,\,\,\,\,\,\,\,\,\,\,\,\,f_{\bar u\bar sds}=\,\,\,\,829(282)~{\rm keV}.\,\,\,\,\,\,\,\,\,\,\,\,\,\,\,\,\,\,\,\,\,\, r=1, \nnb\\
&& 1282(145)~{\rm MeV}  ,\,\,\,\,\,\,\,\,\,\,\,\,\,\,\,\,\,\,\,\,\,\,\,\,\,\,\,\,\,\,\,\,\,\,\,\,\,\,\,\,\,\,\,\,=\,\,\,\,354(69)~{\rm keV},\,\,\,\,\,\,\,\,\,\,\,\,\,\,\,\,\,\,\,\,\,\,\,\,\,  \,\,\,=1/\sqrt{2}.
\eea
The sources of errors are given in Table\,\ref{tab:res-4q-VA} where we have added to the QCD errors of $\bar u\bar dud$ the ones due to $SU(3)$ breakings which are negligible. 

\d  One can notice the  relative small error due to $t_c$ and $\tau$ for the coupling compared to the  one of $\bar u\bar dud$ which is due to the vicinity of the two minimum for $t_c=2$ and 4.5 GeV$^2$ as shown in Fig.\,\ref{fig:4ss-2-VA} for $r=1/\sqrt{2}$. The shape of the coupling is similar to the case of the $\bar u\bar d ud$ state shown in Fig.\ref{fig:ud-1-VA}b). 

\d The {\it anomalously} large value of the mass for $r=1/\sqrt{2}$ is essentially due to the range of $t_c$-value required by the $R_{P/C}$ condition. 

\begin{table}[hbt]
\setlength{\tabcolsep}{0.48pc}
{\footnotesize{
\begin{tabular}{ll ll  ll  ll ll ll ll ll l c}
\hline
\hline
                Currents  
                    &\multicolumn{1}{c}{$\Delta t_c$}
					&\multicolumn{1}{c}{$\Delta \tau$}
					&\multicolumn{1}{c}{$\Delta \Lambda$}
					&\multicolumn{1}{c}{$\Delta PT$}
					&\multicolumn{1}{c}{$\Delta m_q$}
					&\multicolumn{1}{c}{$\Delta \bar{q}q$}
					&\multicolumn{1}{c}{$\Delta \kappa$}					
					&\multicolumn{1}{c}{$\Delta G^2$}
					&\multicolumn{1}{c}{$\Delta \bar q Gq$}
					&\multicolumn{1}{c}{$\Delta G^3$}
					&\multicolumn{1}{c}{$\Delta \bar{q}q^2$}
					
					&\multicolumn{1}{c}{$\Delta OPE$}
					&\multicolumn{1}{c}{Value}
\\
					
\hline
\boldmath$ 
{\cal O}_{\bar u\bar dud}^{V/A}$& \\
\it Masses [MeV]\\
1 &100&3.5&16&97&0&4.5&--&1.8&0&0&0&9.3&831(141)\\
$1/\sqrt{2}$&141&14&15.8&130&0&0.8&--&0.9&2.5&0&0&3.6&941(193)\\
\it Couplings [keV]\\
1 &45&16.5&4&269&0&19.5&--&16.5&2.9&0&2.1&45.5&983(278)\\
$1/\sqrt{2}$&40&1.5&0&20&0&1.6&--&48&5.6&0&0&7&601(66)\\
\boldmath$ 
{\cal O}_{\bar u\bar sud}^{V/A}$& \\
\it Masses [MeV]\\
1 &99&18&16&97&0.1&4.5&1&1.8&0&0&0&9.3&834(141)\\
$1/\sqrt{2}$&129&14.5&15.8&130&0.1&0.8&10&0.9&2.5&0&0&3.6&978(185)\\
\it Couplings [keV]\\
1 &108&16.5&4&269&0.1&19.5&1.7&16.5&2.9&0&2.1&45.5&1042(295)\\
$1/\sqrt{2}$&11&2.2&0&20&0.1&1.6&19.5&48&5.6&0&0&7&408(57)\\
\boldmath$ 
{\cal O}_{\bar u\bar sds}^{V/A}$& \\
\it Masses [MeV]\\
1 &97&23&16&97&0.1&4.5&3&1.8&0&0&0&9.3&840(140)\\
$1/\sqrt{2}$&57&21&15.8&130&0.1&0.8&10&0.9&2.5&0&0&3.6&1282(145)\\
\it Couplings [keV]\\
1 &63&21.5&4&269&0.3&19.5&4&16.5&2.9&0&2.1&45.5&829(282)\\
$1/\sqrt{2}$&40&13&0&20&0.15&1.6&16&48&5.6&0&0&7&354(69)\\


\hline
\hline
\end{tabular}
}}
 \caption{\footnotesize The same caption as for Table\,\ref{tab:res} but for the four-quark states in the Vector $\oplus$ Axial-Vector configurations and for two typical values of the mixing parameter $r=1$ and $1/\sqrt{2}$. Note that the lower value of $t_c$ allowed by $R_{P/C}\geq 1$ is 1 GeV$^2$.}
\label{tab:res-4q-VA}
\end{table}
\section{First radial excitations}
In this section, we attempt to extract the masses and couplings  of the 1st radial excitations for different assignements. We limit ourselves to the case of $SU(2)_F$ by observing that $SU(3)$ breakings shift only slightly the values of these observables and cannot be seen within the errors of the determination. In so doing, we shall substract the contribution of the lowest ground states obtained in previous section from the sum rules and work with the 1st radial excitation $\oplus$ QCD continuum in a higher range of $t_c$-values. This procedure is more helpful than a direct extraction of two resonance parameters where extracting simultaneously  the masses and decay constants of the ground state and 1st radial excitation from ${\cal R}_{10}$ is hopeless. 
One could also work with higher ratio of moments more sensitive to the higher states  like in Ref.\,\cite{SNGS} for the case of multiple resonances. Unfortunately, within our truncation of the OPE up to $D=6$ condensates,  the next moment  ${\cal R}_{21}$ cannot help as it does not present a $\tau$-minimum but a slight inflexion point.  Therefore, the obtained result is less accurate than the one from ${\cal R}_{10}$.  We also emphasize that high moments which are more sensitive to higher radial excitations cannot help to improve the mass and coupling of the ground state which is one of the main goal of this work. 
\subsection*{\b The $\bar ud$ current}
\begin{figure}[hbt]
\begin{center}
\hspace*{-7cm} {\bf a) \hspace*{8.cm} \bf b)} \\
\includegraphics[width=8.cm]{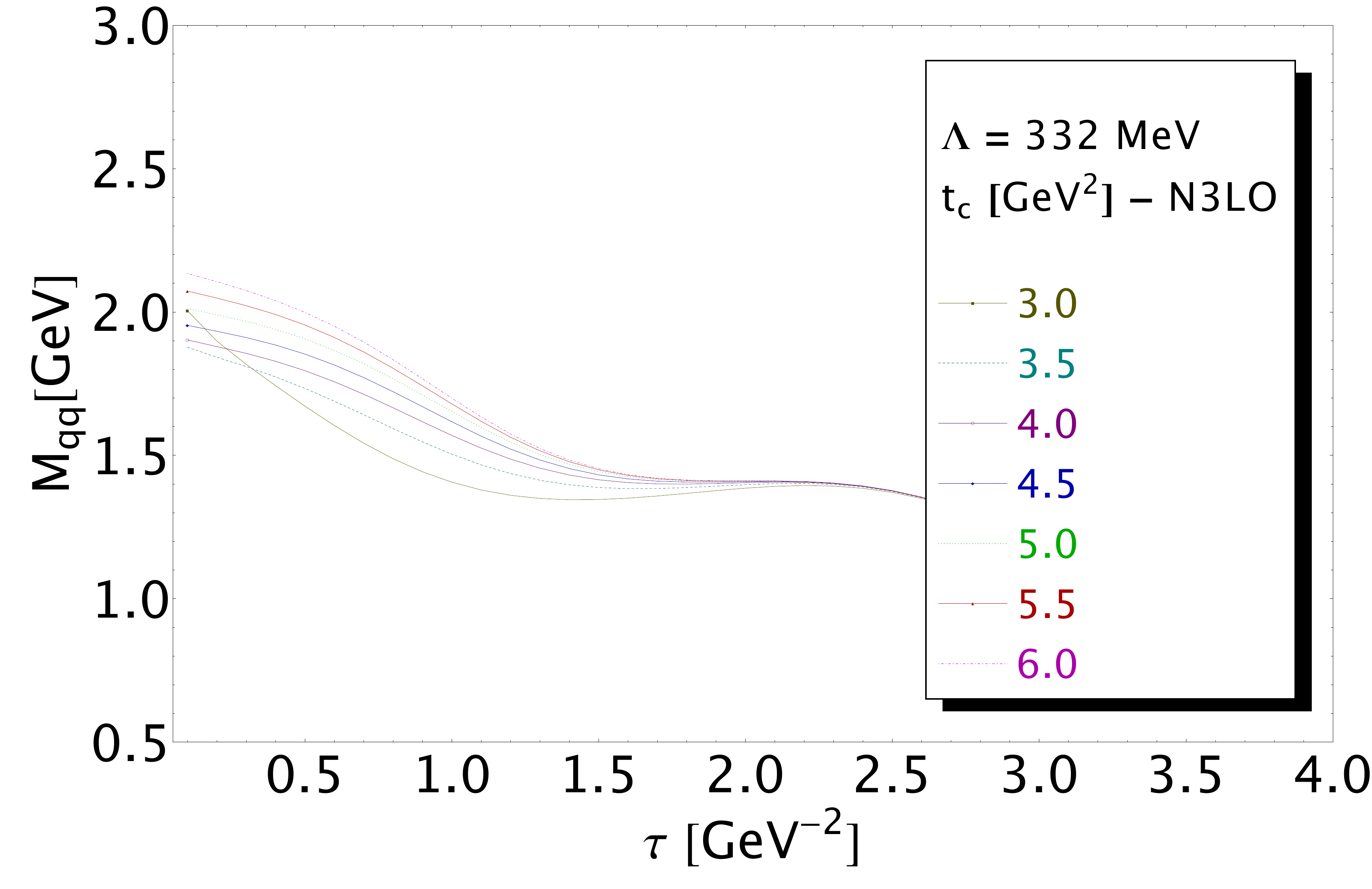}
\includegraphics[width=8.2cm]{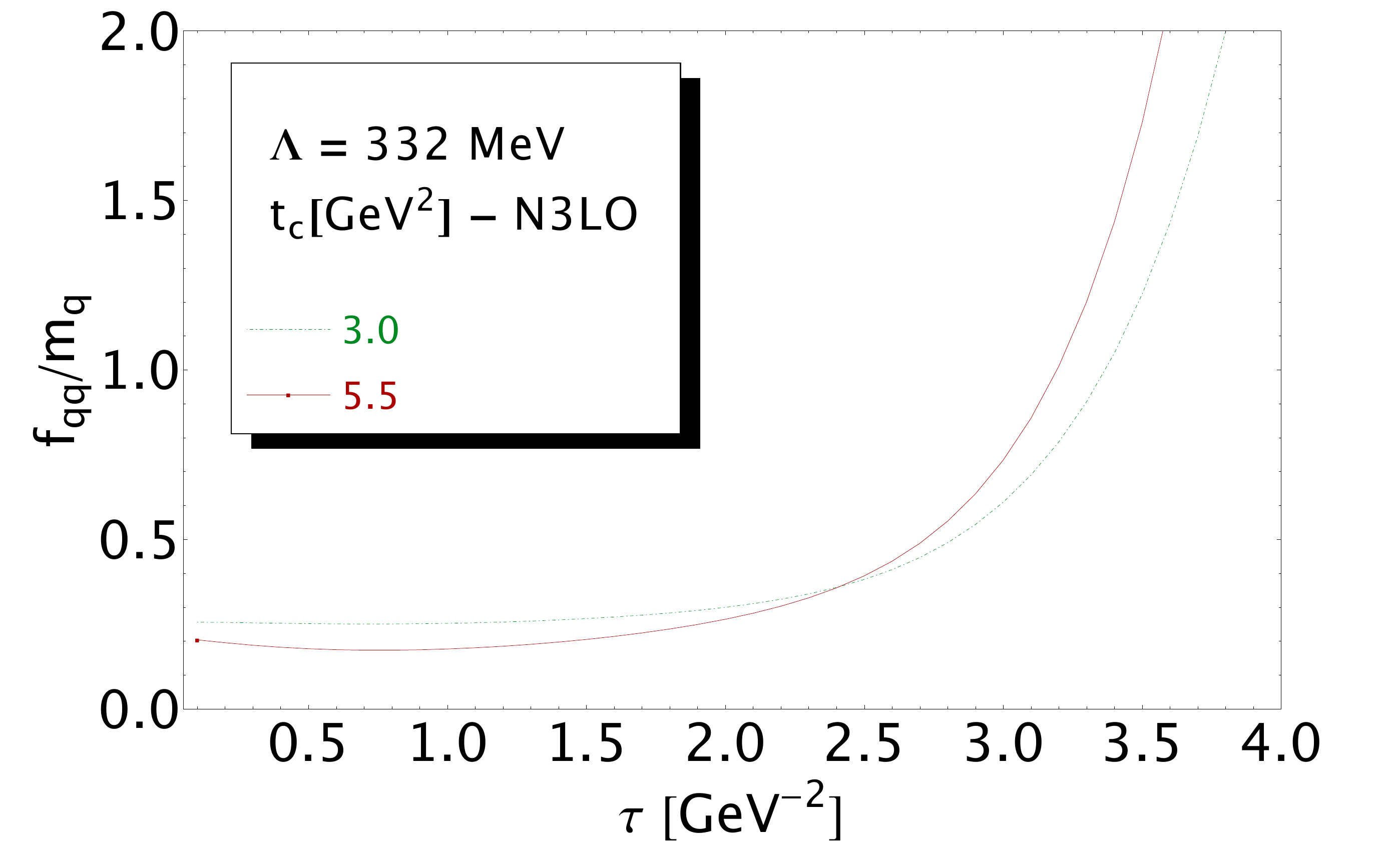} 
\vspace*{-0.5cm}
\caption{\footnotesize   Behaviour of the optimal values in $\tau$ for the  a) mass and  b) coupling of the $\bar u \bar d$  state versus $\tau$ for different values of $t_c$.} 
\label{fig:ud-rad}
\end{center}
\end{figure}

We show the result of the analysis in Fig.\,\ref{fig:ud-rad} where the contribution of the ground state has been subtracted from the sum rule.  One obtains:
\beq
M^{(1)}_{\bar ud}=1378(186)~{\rm MeV}  ,\,\,\,\,\,\,\,\,\,\,\,\,\,\,\,\,\,\,\,\,\,\,\,\,\,\,\,\,\,f^{(1)}_{\bar ud}=\,\,\,\,212(38)~{\rm keV}.
\label{eq:ud-rad}
\eeq
Compared to the case of the ground state, the optimal value of the mass and coupling  is obtained at lower values of $\tau$ as intuitively expected.The mass shift from the ground state is about 350 MeV which is relatively small compared to the case of the $\rho$-meson of 680 MeV.
\begin{figure}[hbt]
\begin{center}
\hspace*{-7cm} {\bf a) \hspace*{8.cm} \bf b)} \\
\includegraphics[width=8.2cm]{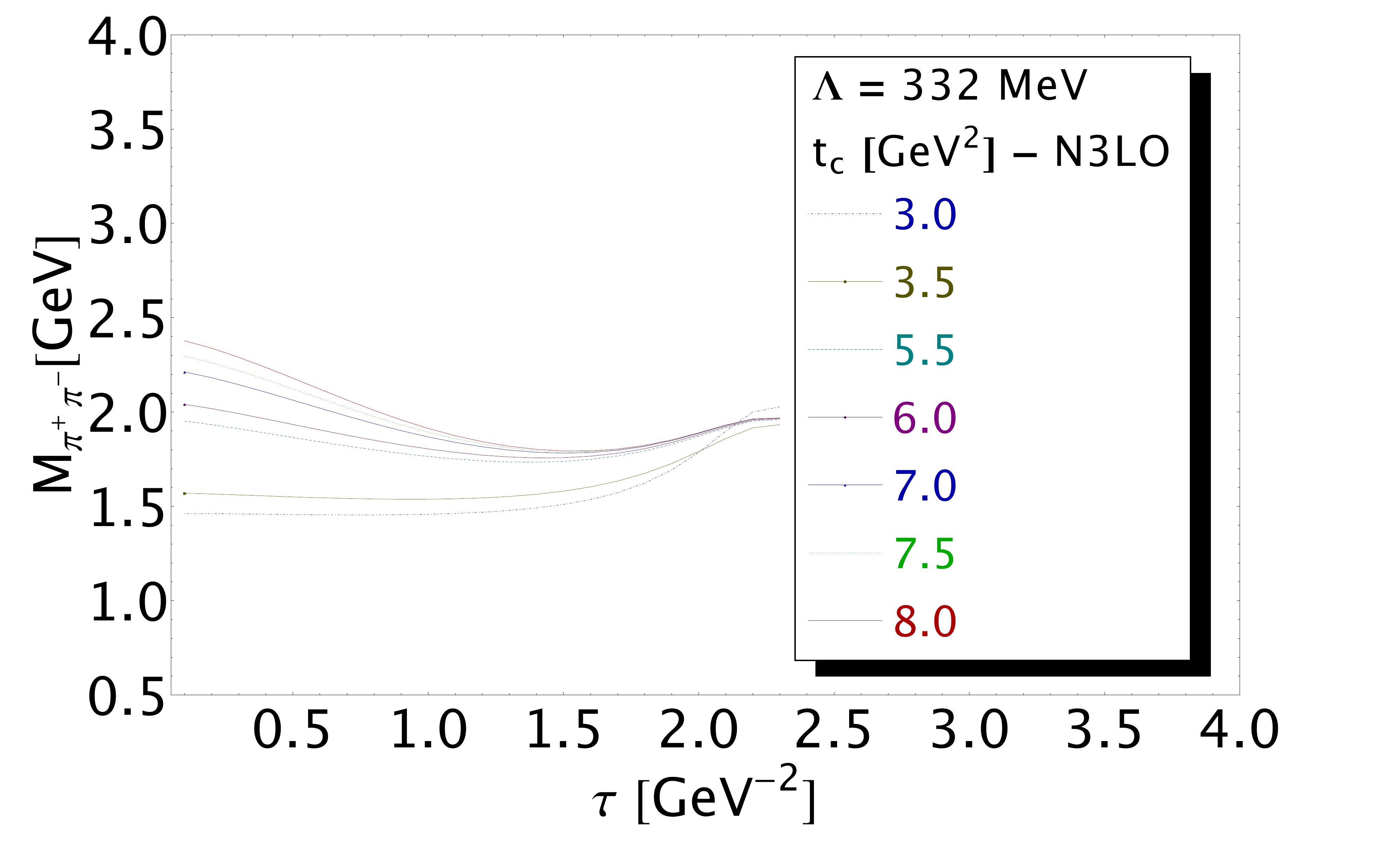}
\includegraphics[width=7.8cm]{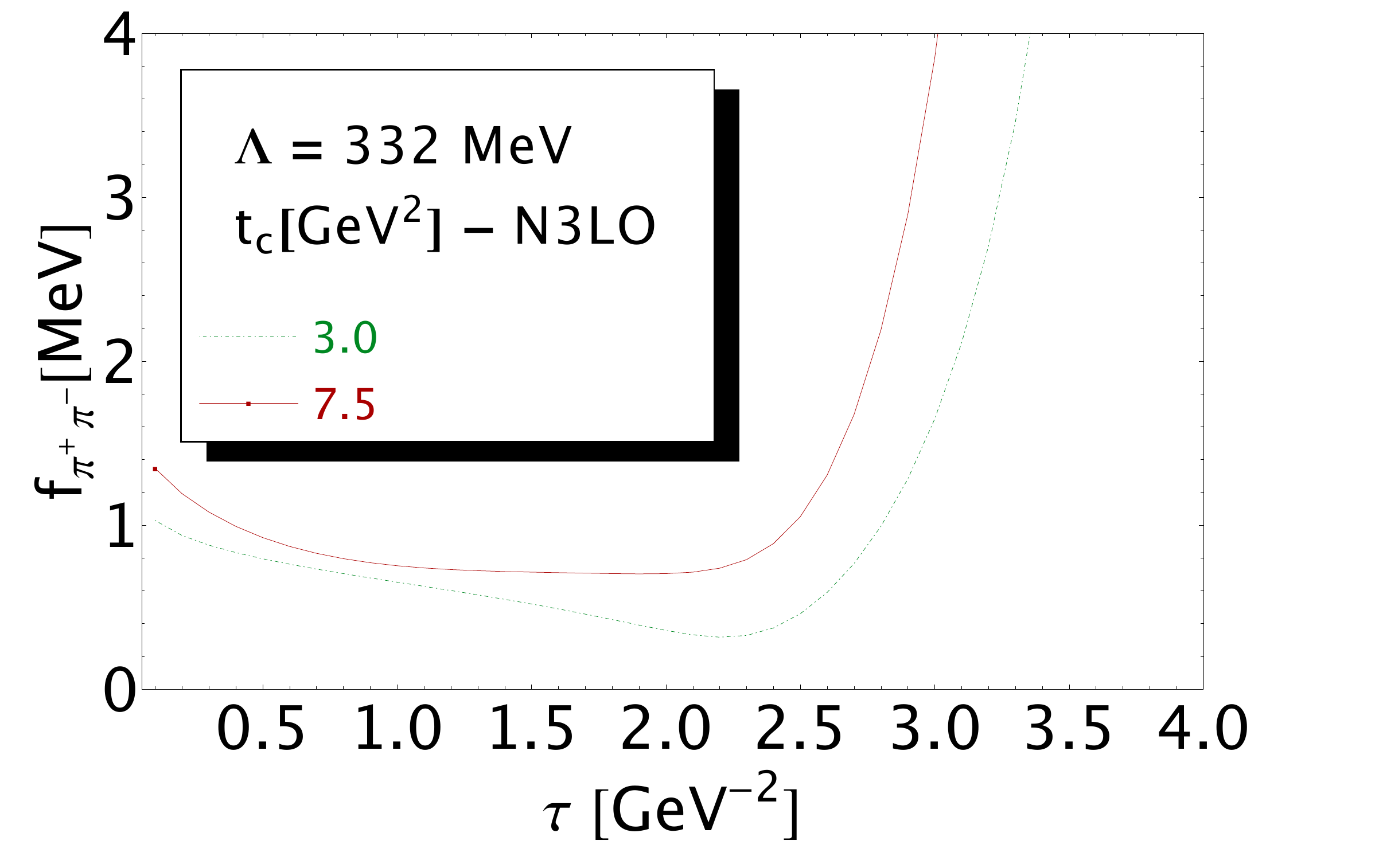} 
\vspace*{-0.5cm}
\caption{\footnotesize   Behaviour of the optimal values in $\tau$ for the  a) mass and  b) coupling of the $\pi^+\pi^-$  molecule state versus $\tau$ for different values of $t_c$.} 
\label{fig:pipi-rad}
\end{center}
\vspace*{-0.5cm}
\end{figure} 
\subsection*{\b The $\pi^+\pi^-$ molecule current}
The analysis is shown in Fig.\,\ref{fig:pipi-rad}. We deduce:
\beq
M^{(1)}_{\pi^+\pi^-}=1621(514)~{\rm MeV}  ,\,\,\,\,\,\,\,\,\,\,\,\,\,\,\,\,\,\,\,\,\,\,\,\,\,\,\,\,\,f^{(1)}_{\pi^+\pi^-}=\,\,\,\,665(338)~{\rm keV},
\eeq
where one should notice that the optimal value for $t_c=3$ GeV$^2$ is taken at the inflexion point $\tau=1.1$ GeV$^{-2}$. 
\subsection*{\b The $\bar u\bar dud$ pseudo (scalar) four-quark current}
\begin{figure}[hbt]
\vspace*{-0.25cm}
\begin{center}
\hspace*{-7cm} {\bf a) \hspace*{8.cm} \bf b)} \\
\includegraphics[width=7.6cm]{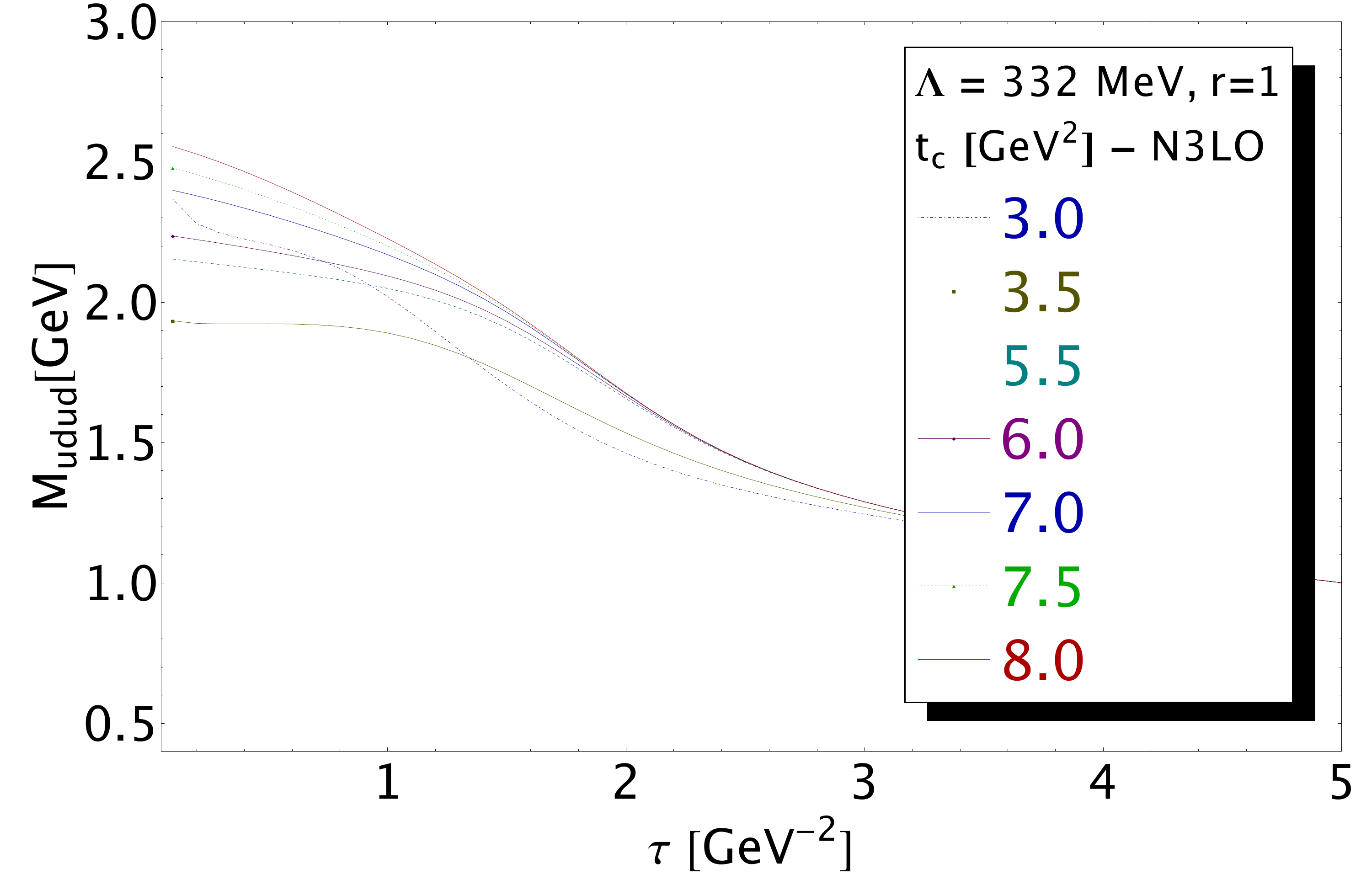}
\includegraphics[width=8.cm]{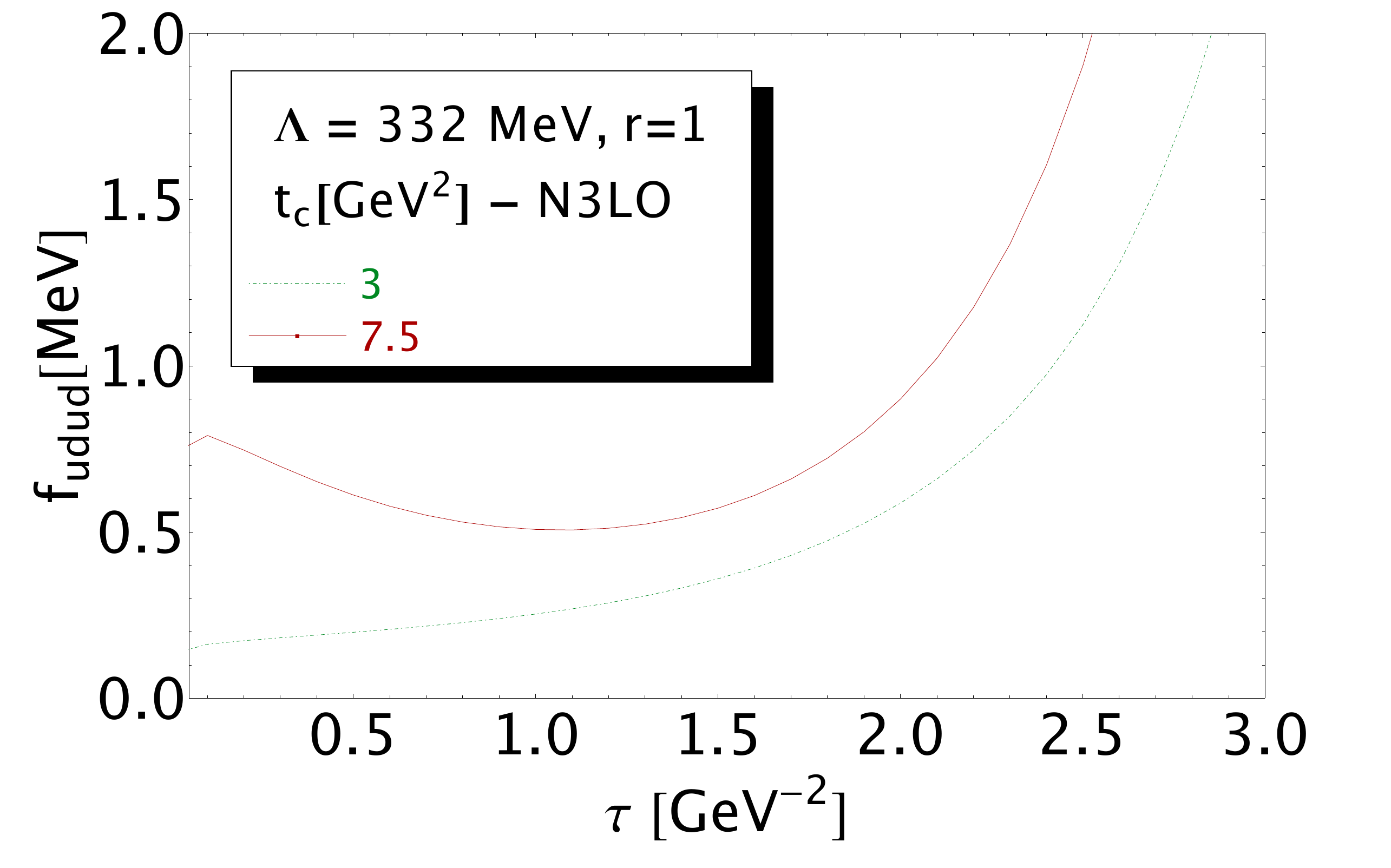} 
\vspace*{-0.5cm}
\caption{\footnotesize   Behaviour of the optimal values in $\tau$ for the  a) mass and  b) coupling of the $\bar u\bar d ud$ 1st radial state versus $\tau$ for different values of $t_c$ for  $r=1$ for the (pseudo)scalar currents.} 
\label{fig:4q1-rad}
\end{center}
\vspace*{-0.5cm}
\end{figure} 
\begin{figure}[hbt]
\vspace*{-0.25cm}
\begin{center}
\hspace*{-7cm} {\bf a) \hspace*{8.cm} \bf b)} \\
\includegraphics[width=8.cm]{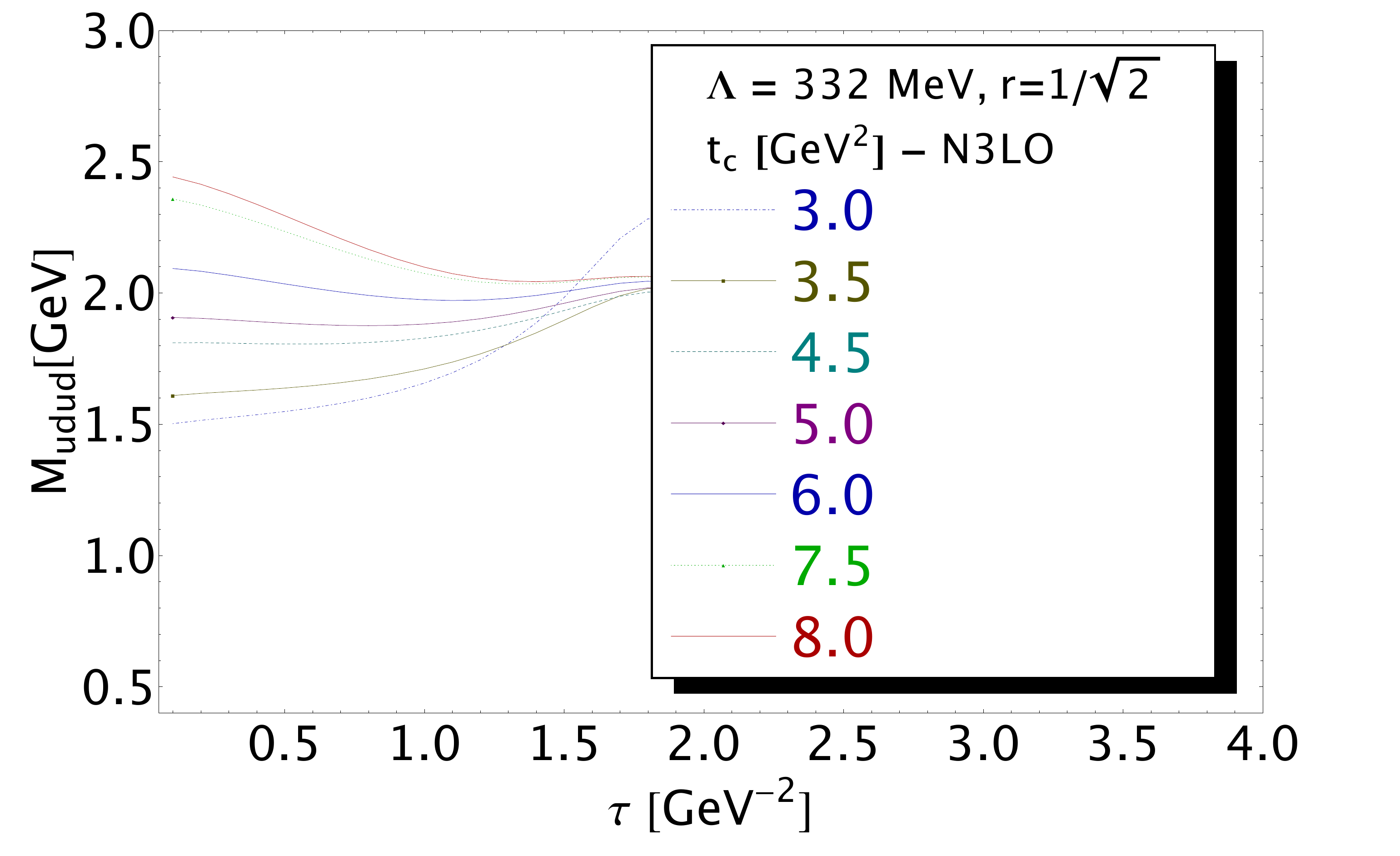}
\includegraphics[width=7.8cm]{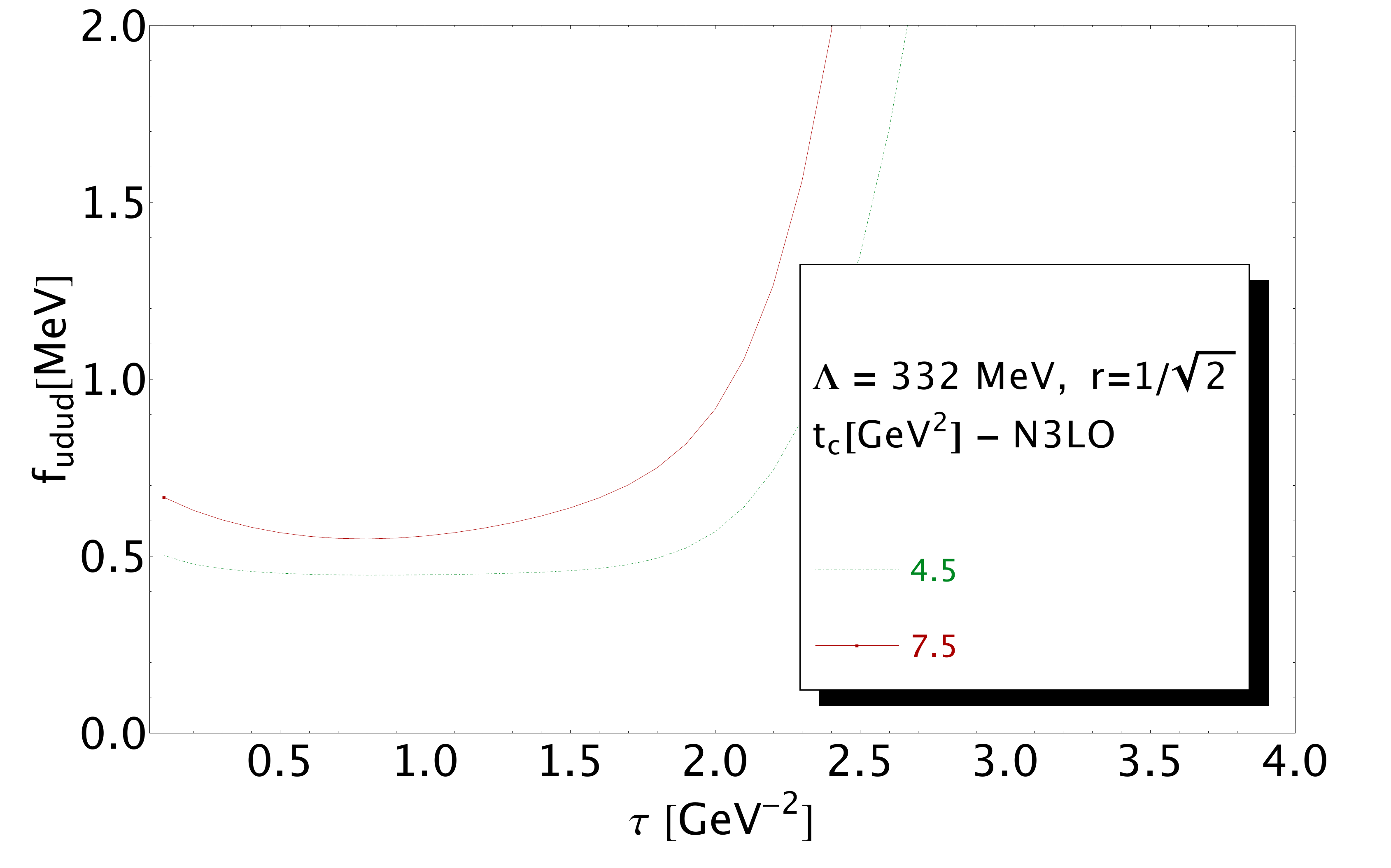} 
\vspace*{-0.5cm}
\caption{\footnotesize   Behaviour of the optimal values in $\tau$ for the  a) mass and  b) coupling of the $\bar u\bar d ud$ 1st radial state versus $\tau$ for different values of $t_c$ for $r=1/\sqrt{2}$ for the (pseudo)scalar currents.} 
\label{fig:4q2-rad}
\end{center}
\end{figure}
\begin{figure}[hbt]
\begin{center}
\hspace*{-7cm} {\bf a) \hspace*{8.cm} \bf b)} \\
\includegraphics[width=8.2cm]{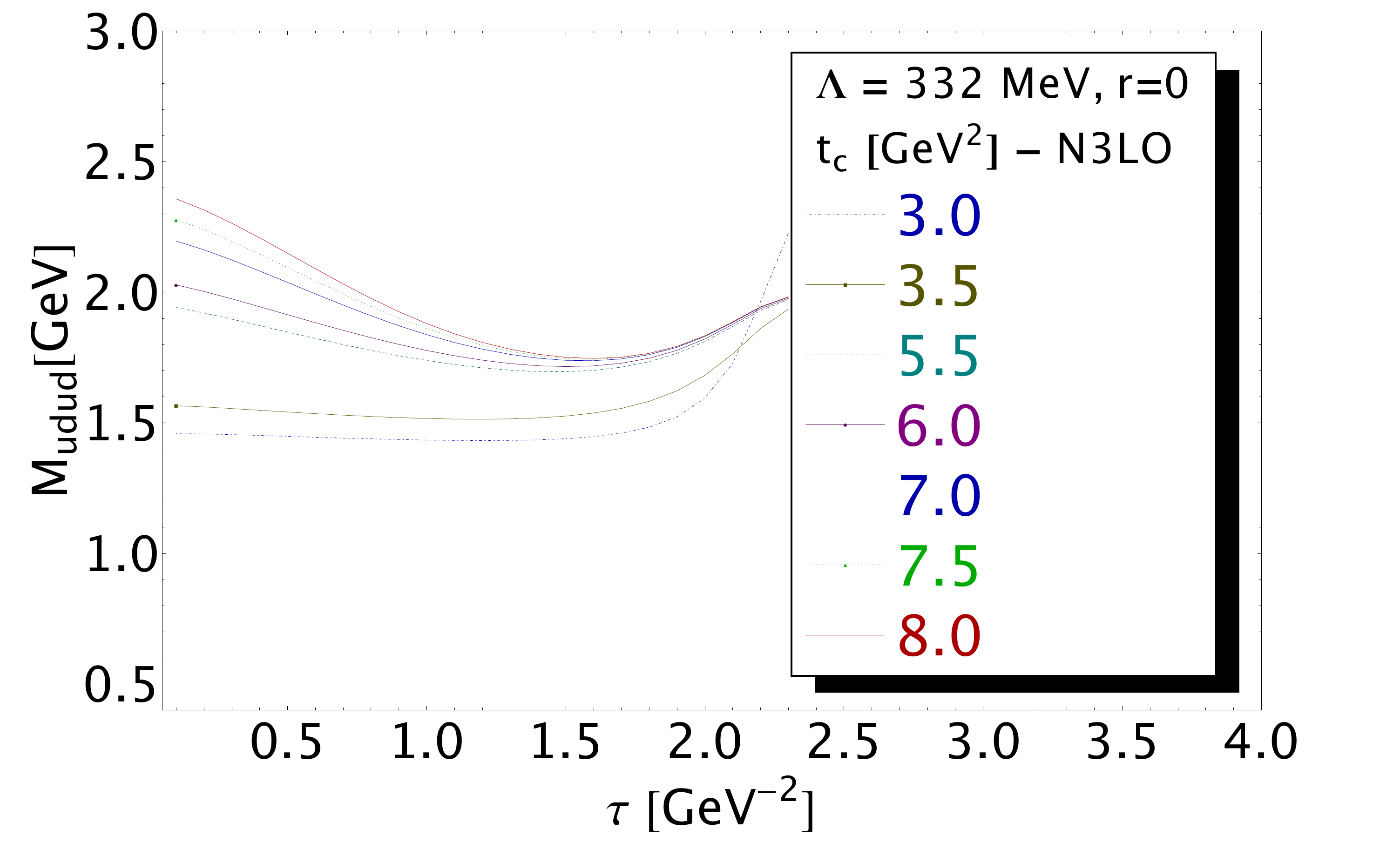}
\includegraphics[width=8.cm]{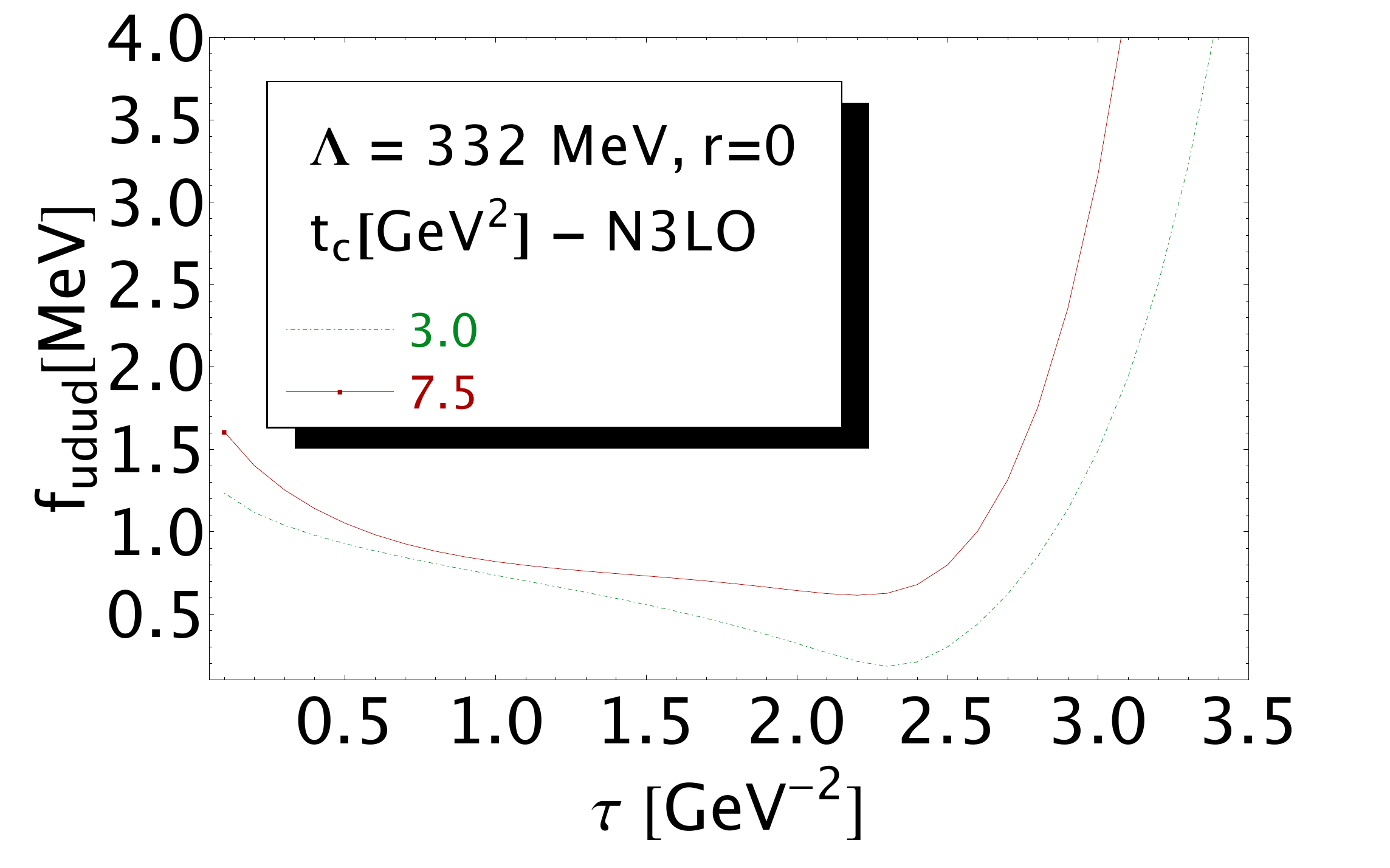} 
\vspace*{-0.5cm}
\caption{\footnotesize   Behaviour of the optimal values in $\tau$ for the  a) mass and  b) coupling of the $\bar u\bar d ud$ state versus $\tau$ for different values of $t_c$ for $r=0$ for the (pseudo)scalar currents.} 
\label{fig:4q0-rad}
\end{center}
\end{figure}

 
We show the analysis for three typical values of the mixing parameter in Fig.\ref{fig:4q1-rad} to \ref{fig:4q0-rad}.  
We obtain:
\bea
M^{S/P(1)}_{\bar u\bar dud}= &&1670(168) ~{\rm MeV}  ,\,\,\,\,\,\,\,\,\,\,\,\,\,\,\,\,\,\,\,\,\,\,\,\,\,\,\,f^{S/P(1)}_{\bar u\bar dud}=\,\,\,\,381(128)~{\rm keV}.\,\,\,\,\,\,\,\,\,\,\,\,\,\,\,\,\,\,\,\,\,\,\,\,\,r=1,\nnb\\
&& 1920(317)~{\rm MeV}  ,\,\,\,\,\,\,\,\,\,\,\,\,\,\,\,\,\,\,\,\,\,\,\,\,\,\,\,\,\,\,\,\,\,\,\,\,\,\,\,\,\,\,\,\,\,\,\,=\,\,\,\,498(97)~{\rm keV}.\,\,\,\,\,\,\,\,\,\,\,\,\,\,\,\,\,\,\,\,\,\,\,\,\, \,\,\,\,\,\,=1/\sqrt{2}, \nnb\\
&&1588(511)~{\rm MeV} ,\,\,\,\,\,\,\,\,\,\,\,\,\,\,\,\,\,\,\,\,\,\,\,\,\,\,\,\,\,\,\,\,\,\,\,\,\,\,\,\,\,\,\,\,\,\,\,=\,\,\,\,777(189)~{\rm keV}.\,\,\,\,\,\,\,\,\,\,\,\,\,\,\,\,\,\,\,\,\,\,\,\,\, \,\,\,=0.
\eea
One should note, like in the case of $\pi^+\pi^-$, that the optimal value of the coupling for $r=0$ and $t_c=3$ GeV$^2$  is taken at the inflexion point $\tau=1.3$ GeV$^{-2}$. One should also note that the relatively large value of the mass for $r=\sqrt{2}$ is due to the fact that we have taken the minimum value of $t_c=4.5$ GeV$^2$ where the coupling starts to have stability instead of 3 GeV$^2$ in the other channels $r=0,1$. 

\subsection*{\b The $\bar u\bar dud$ axial(vector) four-quark currents}
\begin{figure}[hbt]
\begin{center}
\hspace*{-7cm} {\bf a) \hspace*{8.cm} \bf b)} \\
\includegraphics[width=8.2cm]{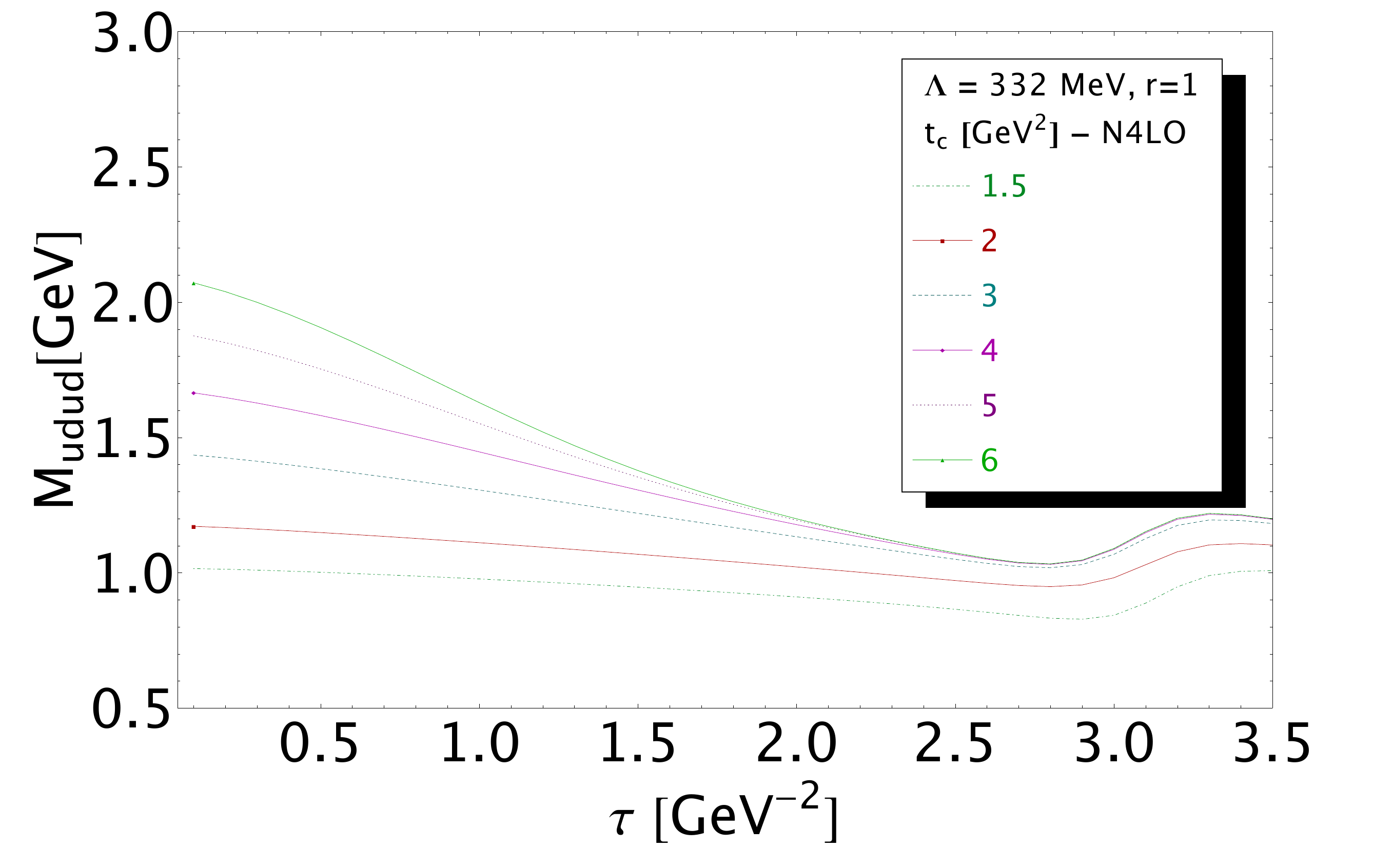}
\includegraphics[width=8.cm]{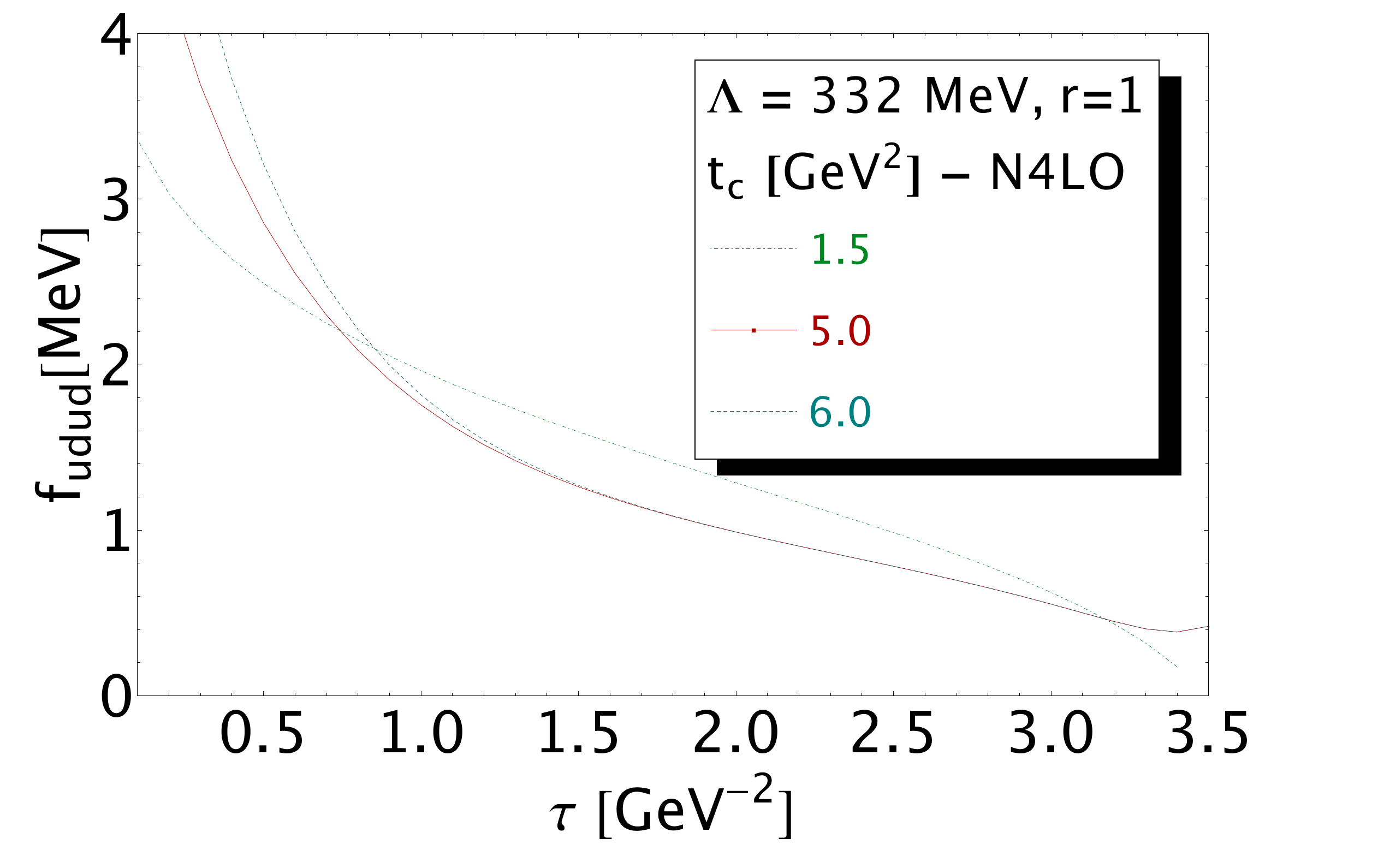} 
\vspace*{-0.5cm}
\caption{\footnotesize   Behaviour of the optimal values in $\tau$ for the  a) mass and  b) coupling of the 1st radial excitation of the $\bar u\bar d ud$ state versus $\tau$ for different values of $t_c$ for $r=1$ for the axial (vector) currents.} 
\label{fig:4q-rad-VA-1}
\end{center}
\end{figure}

\begin{figure}[H]
\begin{center}
\hspace*{-7cm} {\bf a) \hspace*{8.cm} \bf b)} \\
\includegraphics[width=8.2cm]{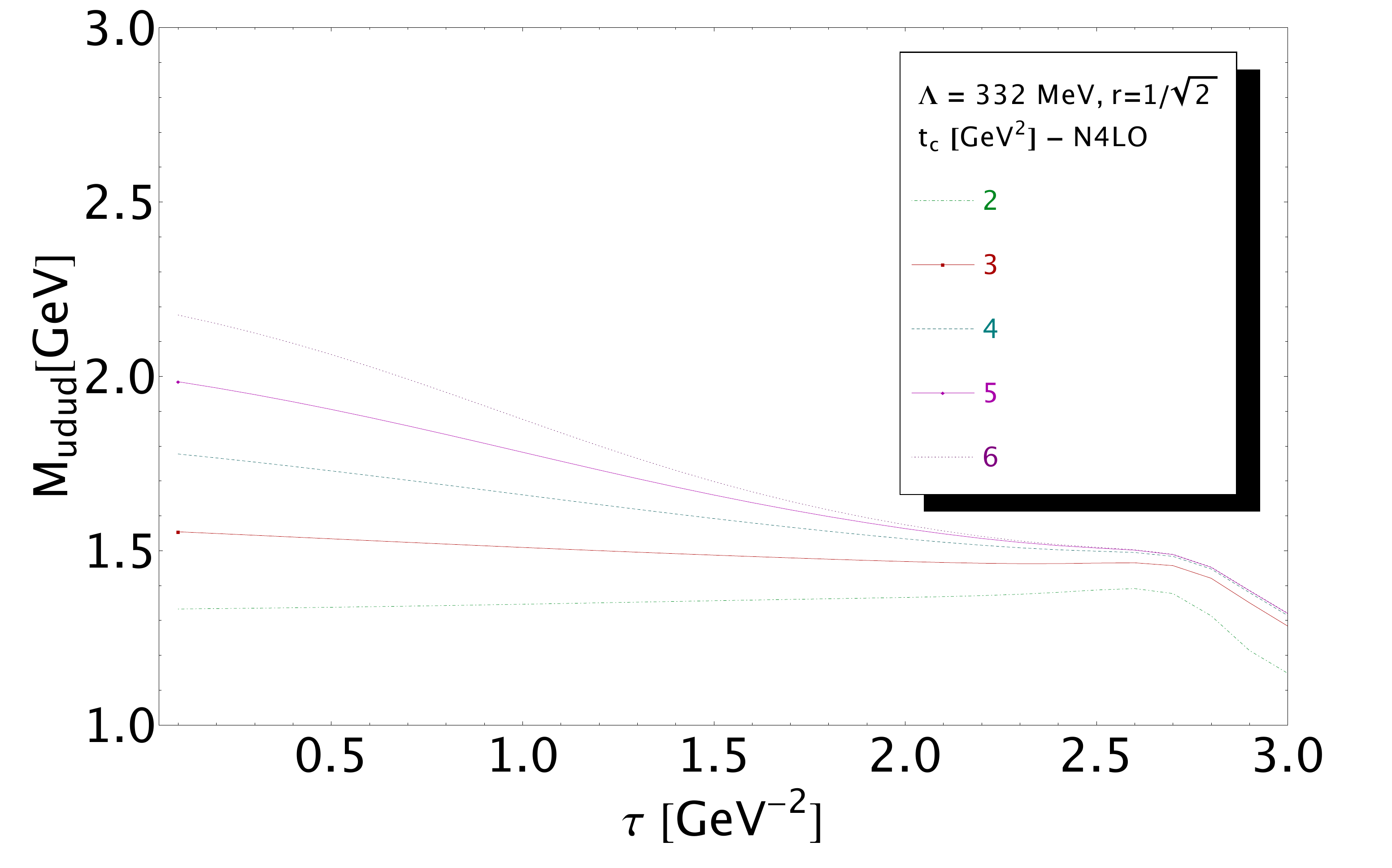}
\includegraphics[width=8.cm]{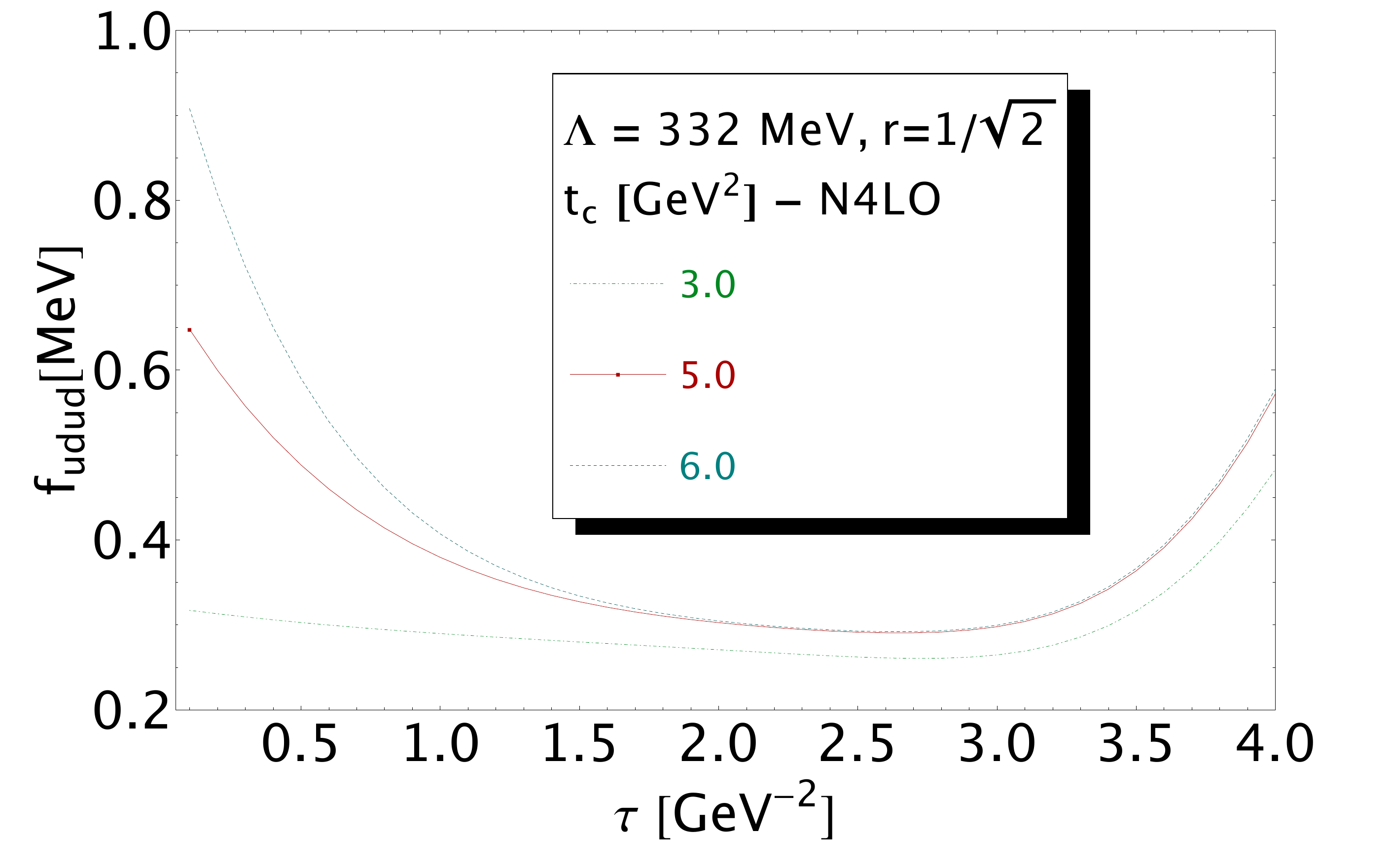} 
\vspace*{-0.5cm}
\caption{\footnotesize   Behaviour of the optimal values in $\tau$ for the  a) mass and  b) coupling of the 1st radial excitation of  the $\bar u\bar d ud$ state versus $\tau$ for different values of $t_c$ for $r=1/\sqrt{2}$ for the (axial) vector currents.} 
\label{fig:4q-rad-VA-2}
\end{center}
\end{figure}

We do a similar analysis for the (axial) vector currents. The analysis is shown in Figs.\,\ref{fig:4q-rad-VA-1} and \ref{fig:4q-rad-VA-2}. 
We obtain:
\bea
M^{V/A(1)}_{\bar u\bar dud}= &&931(193) ~{\rm MeV}  ,\,\,\,\,\,\,\,\,\,\,\,\,\,\,\,\,\,\,\,\,\,\,\,\,\,\,\,\,\,f^{V/A(1)}_{\bar u\bar dud}=\,\,\,\,1137(334)~{\rm keV}.\,\,\,\,\,\,\,\,\,\,\,\,\,\,\,\,\,\,\,\,\,\,\, r=1, \nnb\\
&& 1489(380)~{\rm MeV}  ,\,\,\,\,\,\,\,\,\,\,\,\,\,\,\,\,\,\,\,\,\,\,\,\,\,\,\,\,\,\,\,\,\,\,\,\,\,\,\,\,\,\,\,\,\,\,\,=\,\,\,\,287(99)~{\rm keV}.\,\,\,\,\,\,\,\,\,\,\,\,\,\,\,\,\,\,\,\,\,\,\,\,\, \,\,\,\, \,\,=1/\sqrt{2},
\eea
where the different sources of the errors can be found in Table\,\ref{tab:res-rad}. 
We notice that the $\tau$-stability starts earlier at $t_c=1.5$ GeV$^2$ for the case $r=1$ than the one $r=1/\sqrt{2}$ at 3 GeV$^2$. This feature explains the low mass of the 1st radial excitation in this case in addition to the larger value of $\tau$ where the minimum is obtained (see Fig.\,\ref{fig:4q-rad-VA-1}). The optimal value of the coupling is taken at the inflexion point in $\tau\simeq 2$ GeV$^{-2}$ for $r=1$ and at the minimum $\tau\simeq 2$ GeV$^{-2}$ for $r=1/\sqrt{2}$. 
\subsection*{\b Comments on the radial excitations}
\d Comparing the values of the radial excitation masses with the ones of the ground states, one notice that the mass-splitting for the $\bar qq$ state of about 132 MeV is relatively low compared to the case of the $\rho$-meson of 680 MeV. On the contrary, the ones of the molecule and four-quark states are in the range of 450 to 840\,MeV.  

\d We expect that the masses of the radial excitations with strange quarks are almost degenerated with the non-strange one due to the small $SU(3)$ breakings found for the lowest ground states. This range of mass values is comparable with the one obtained from light front holographic approach\,\cite{BRODSKY}. 

\d The coupling of the 1st radial excitation is comparable with the one of the ground state for the $\bar qq$ state and $r=1$ four-quark state but much smaller for the molecule and $r=1/\sqrt{2},0$ four-quark states. Therefore, one may wonder if the one resonance parametrization done for estimating the ground state mass can be affected by the presence of the nearby 1st radial excitation.

\subsection*{\b Nearby radial excitations effects  on the  $\bar ud$ and $r=1 ~\bar u\bar d ud$ ground states}
\d We re-iterate the determination of the lowest ground state mass and coupling by  including now into the spectral function the effect of the 1st radial excitation having the parameters given in Table\,\ref{tab:res-rad} $\oplus$ the QCD continuum. In this case, we shall only retain the value of $t_c\simeq 2.5$ GeV$^2$ above the 1st radial excitation mass.  

{\it \d  $\bar qq$ ground state}

\begin{figure}[hbt]
\begin{center}
\hspace*{-7cm} {\bf a) \hspace*{8.cm} \bf b)} \\
\includegraphics[width=8.2cm]{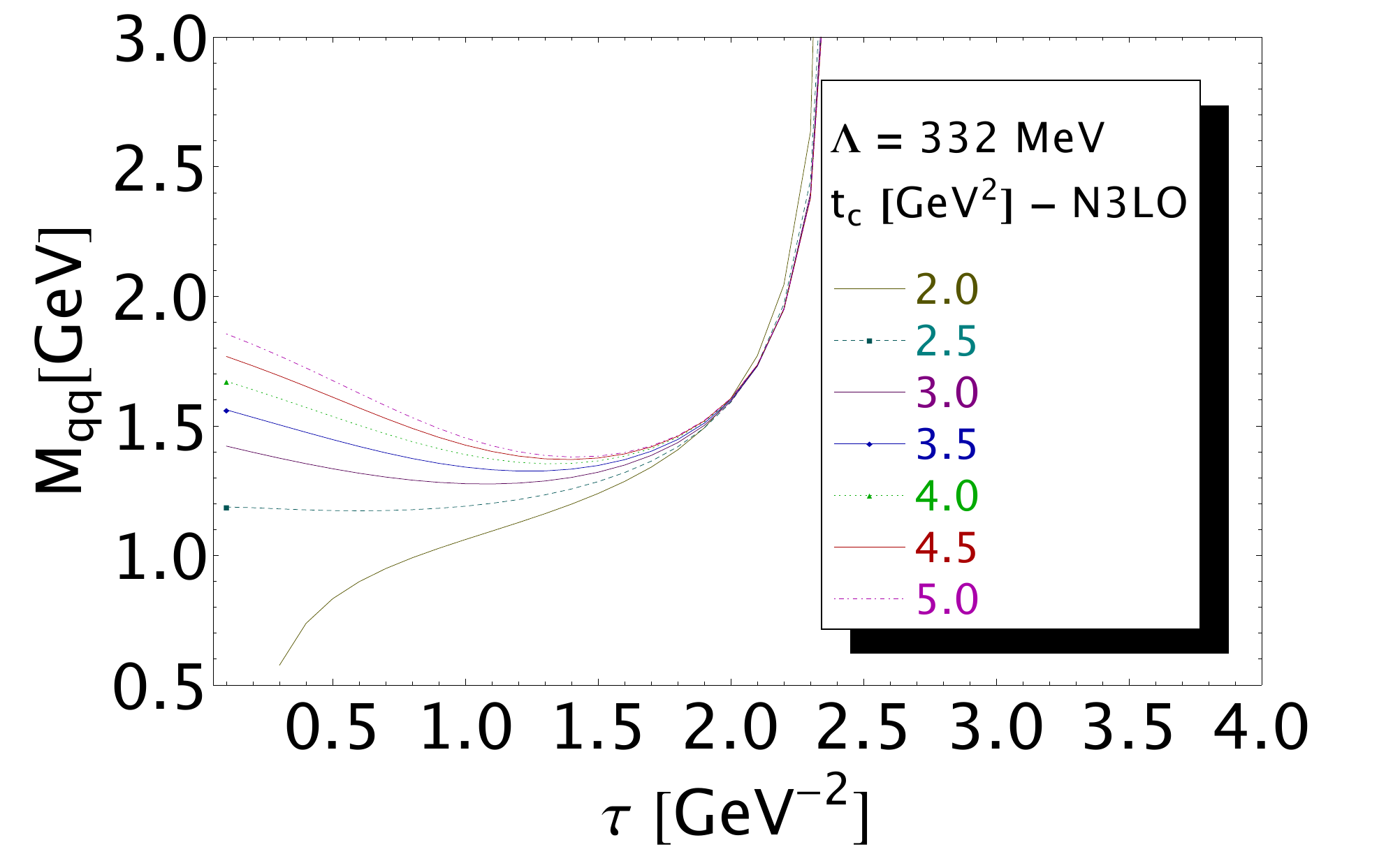}
\includegraphics[width=8.cm]{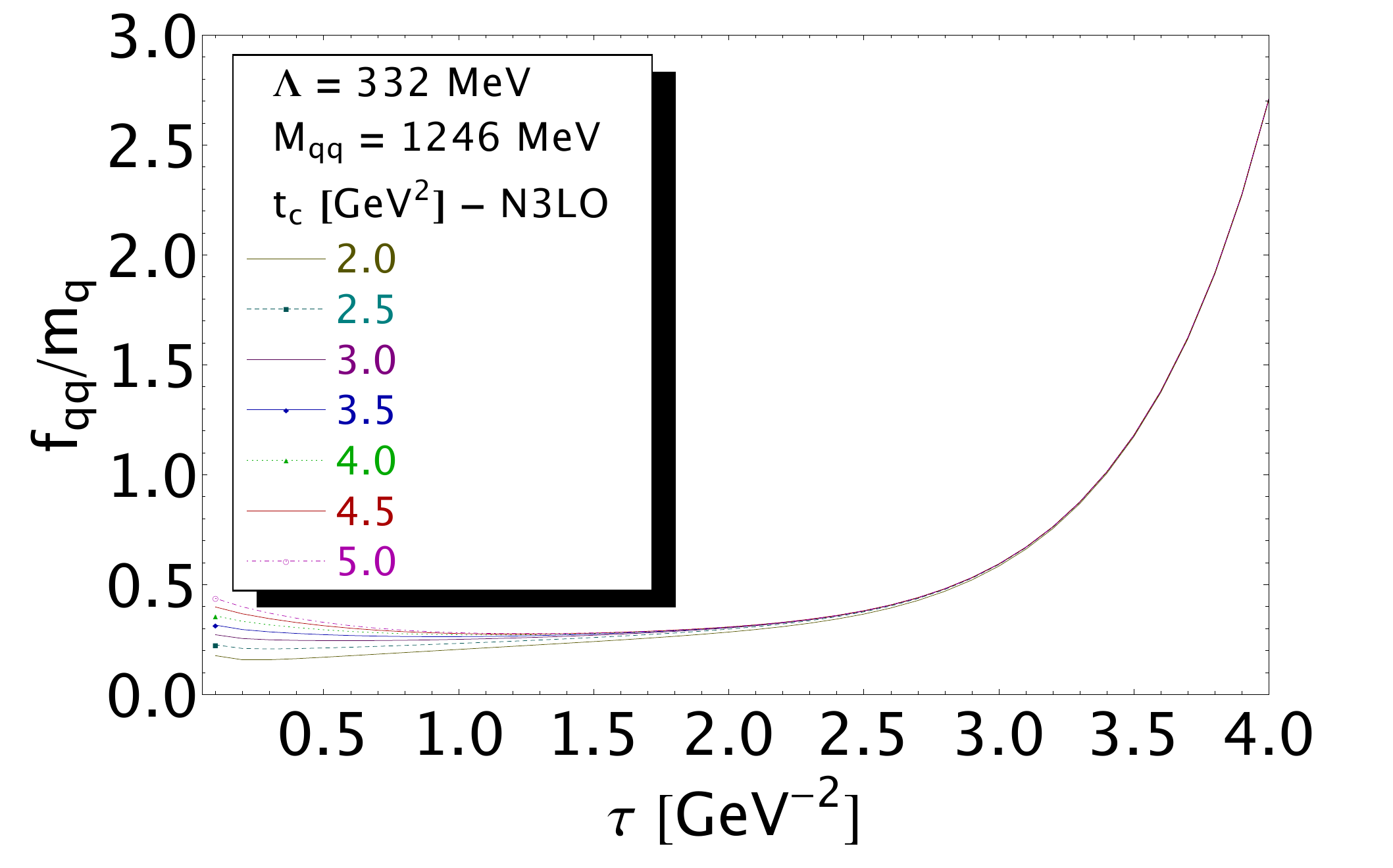} 
\vspace*{-0.5cm}
\caption{\footnotesize   Behaviour of  a) mass and  b) coupling of the  $\bar ud$ ground state within a two-resonances $\oplus$ QCD continuum versus the  LSR variable $\tau$ and for different values of the continuum threshold $t_c$.} 
\label{fig:qq-iter}
\end{center}
\end{figure}

The result of the analysis is shown in Fig.\,\ref{fig:qq-iter} from which we deduce within a NWA:
\beq 
M_{\bar u d} = 1271(124)~{\rm MeV},~~~~~~~f_{\bar qq}/\bar m_q(\tau)=243(43)\times 10^{-3}.
\eeq
This value is about the same (within the errors)  as the one in Eq.\,\ref{eq:nwa}. 

\begin{figure}[hbt]
\begin{center}
\hspace*{-7cm} {\bf a) \hspace*{8.cm} \bf b)} \\
\includegraphics[width=8.2cm]{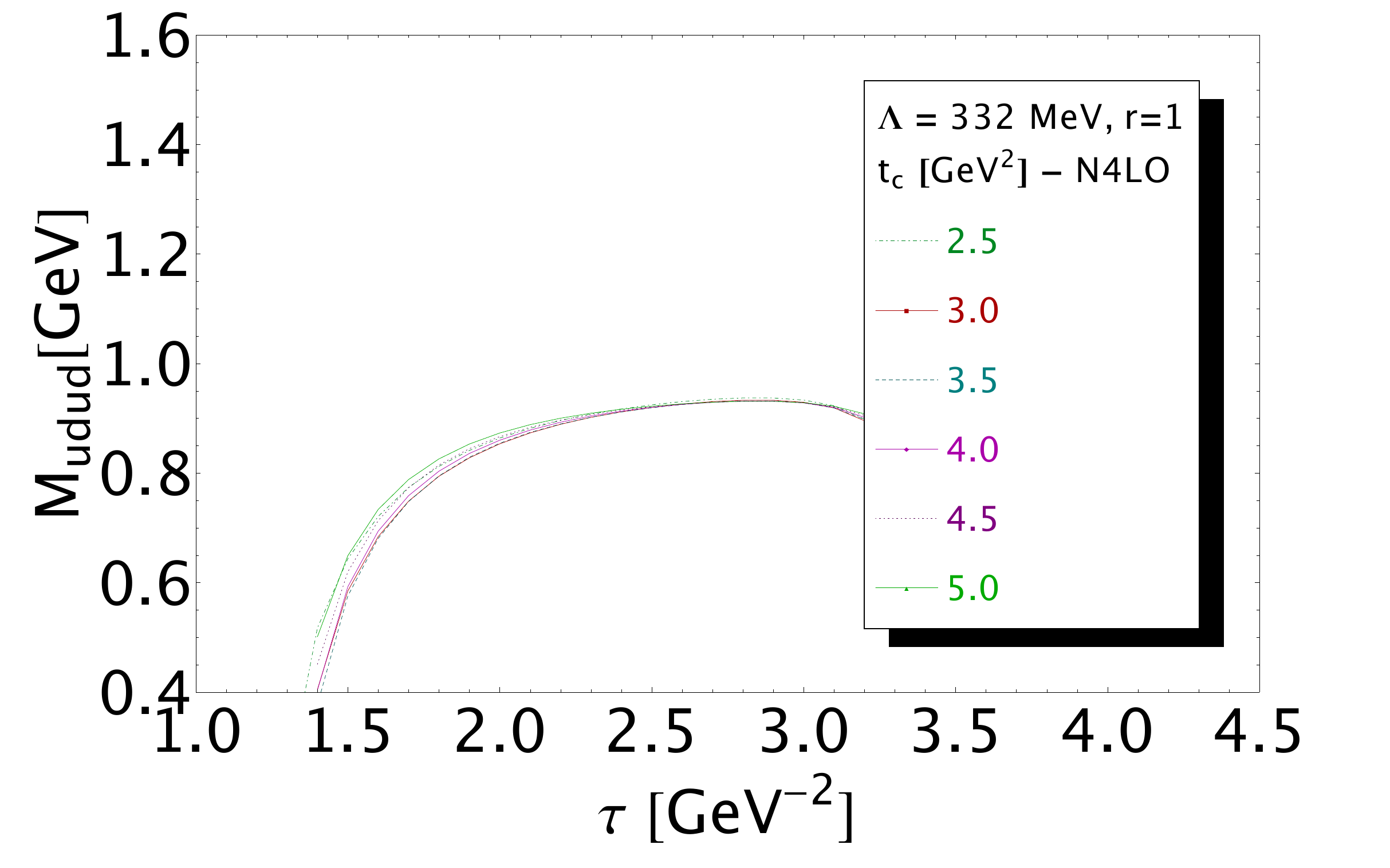}
\includegraphics[width=8.cm]{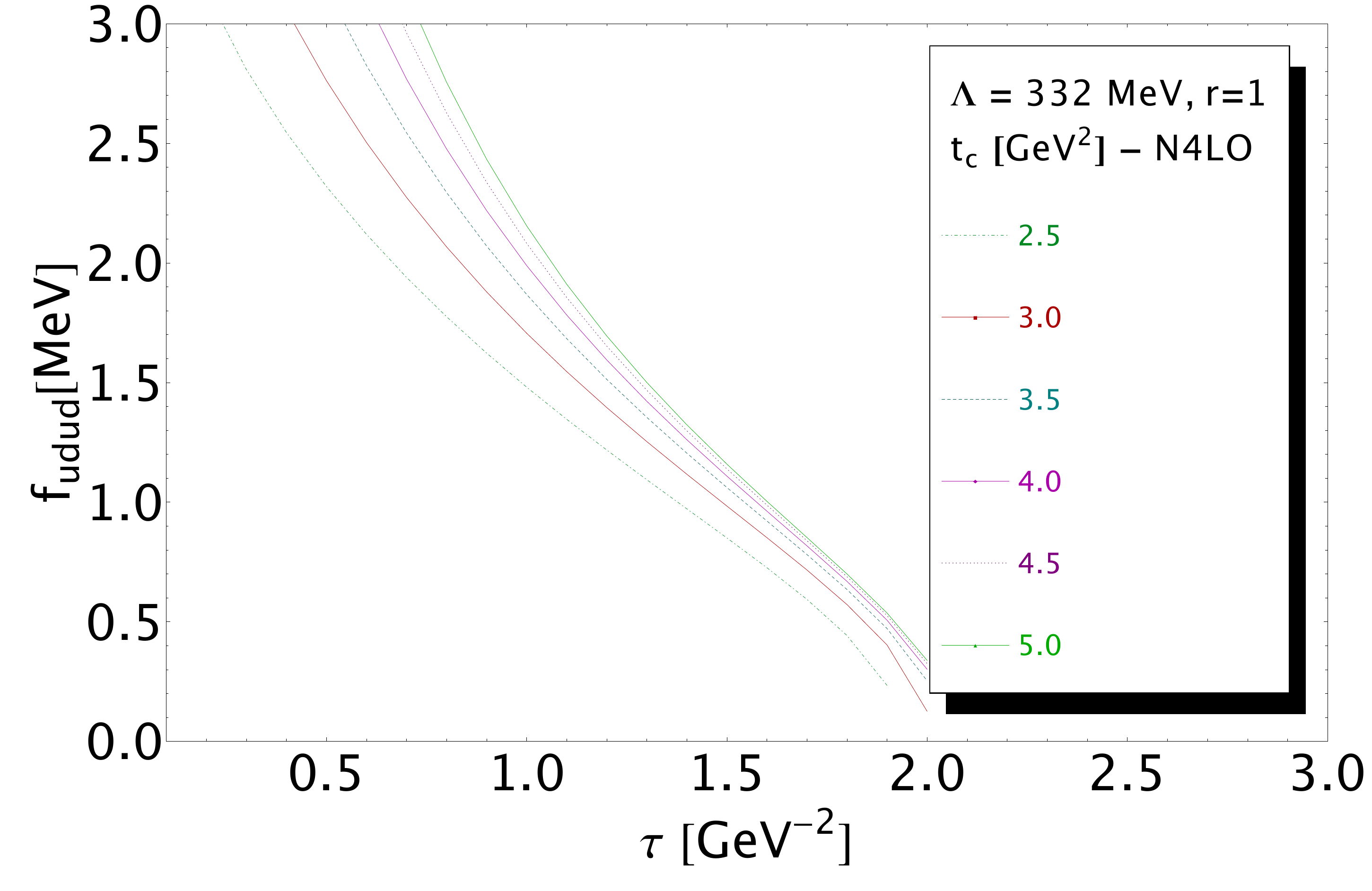} 
\vspace*{-0.5cm}
\caption{\footnotesize   Behaviour of  a) mass and  b) coupling of the  $\bar u\bar d ud$ ground state within a two-resonances $\oplus$ QCD continuum versus the  LSR variable $\tau$ and for different values of the continuum threshold $t_c$.} 
\label{fig:udud-iter}
\end{center}
\end{figure} 

{\it \d  $\bar u\bar d ud$ four-quark ground state (case $r=1$)}

We do an analysis similar to the case of the $\bar ud$ current which we show in Fig.\,\ref{fig:udud-iter}.  One can notice that compared to the case of one resonance (see Fig.\,\ref{fig:ud-1}), the coupling does not present a $\tau$-minimum. It appears a slight inflexion point around $\tau\simeq 1.5$ GeV$^{-2}$. We obtain :
\beq 
M_{\bar u \bar d ud} = 920(205)~{\rm MeV},~~~~~~~f_{\bar u \bar d ud}\approx (850\sim 1139)~{\rm keV}.
\eeq
One can notice that these values are comparable with the ones from MDA obtained in Eq.\,\ref{eq:4ud}.
\subsection*{\d  Comments on nearby radial excitations effects}
We have shown in the previous analysis that the nearby first radial excitation only affects slightly the determination of the lowest ground state mass obtained from  the minimal duality ansatz (MDA) one resonance $\oplus$ QCD continuum parametrization of the spectral function. The errors on the determination of the 1st radial excitations induce slightly larger errors on the ground state mass. 
This feature is essentially due to the exponential damping factor appearing in the LSR and to the smaller coupling of the radial excitation than the ground state one to the corresponding quark currents. In the following, we shall consider as a final result the one obtained from MDA in Tables\,\ref{tab:res} and \ref{tab:res-4q-VA}.

\begin{table}[hbt]
\setlength{\tabcolsep}{0.35pc}
{\footnotesize{
\begin{tabular}{ll ll  ll  ll ll ll ll ll l l c}
\hline
\hline
                Currents  
                    &\multicolumn{1}{c}{$\Delta t_c$}
					&\multicolumn{1}{c}{$\Delta \tau$}
					&\multicolumn{1}{c}{$\Delta \Lambda$}
					&\multicolumn{1}{c}{$\Delta PT$}
					&\multicolumn{1}{c}{$\Delta m_q$}
					&\multicolumn{1}{c}{$\Delta \bar{q}q$}
					&\multicolumn{1}{c}{$\Delta G^2$}
					&\multicolumn{1}{c}{$\Delta \bar q Gq$}
					&\multicolumn{1}{c}{$\Delta G^3$}
					&\multicolumn{1}{c}{$\Delta \bar{q}q^2$}
					&\multicolumn{1}{c}{$\Delta OPE$}
					&\multicolumn{1}{c}{$\Delta M_{S}$}
					&\multicolumn{1}{c}{$\Delta f_{S}$}
					&\multicolumn{1}{c}{Value}
\\
					
\hline
\boldmath $\frac{1}{\sqrt{2}}(\bar uu +\bar dd)$& \\
\it Masses [MeV]
&33&7&36&28&0&0.5&4&1.5&0&104&88&101&51&1378(186)\\
\it Couplings [keV]
&38&1.9&1.8&1.2&0&0.1&0.2&0.1&0&0.9&4&0.1&0.1&212(38)\\
\\
\hline
\boldmath $\pi^+\pi^-$ molecule& \\
\it Masses [MeV]
&167&9&21.5&9&0&113&1.2&1.2&0&109&126&322&303&1621(514)\\
\it Couplings [keV]
&193&28&11.5&14&0&80&3.5&1.7&0&128&62&190&113&665(338)\\
\\
\hline
\boldmath$ 
{\cal O}_{\bar u\bar dud}^{S/P}$ four-quark& \\
\it Masses [MeV]\\
1&128&107&22.5&18.5&0&0.3&4&0.5&0&0.3&0.4&45&98&1670(201)\\
$1/\sqrt{2}$&115&4.3&9&17.5&0&67&2.3&0.8&0&69&53&237&136&1920(317)\\

0&155&7&31.5&0.2&0&110&1.25&1.25&0&153&108&242&361&1588(511) \\
\it Couplings [keV]\\
1&93&35&46&54&0&0.05&2&0.15&0&0.05&0.05&35&16&381(128)\\

$1/\sqrt{2}$&51&4&2.3&5.8&0&21.4&1.25&0.15&0&25.3&10&69&28&498(97)\\

0&41&53&9.5&3.9&0&15&1&0.8&0&61&15&144&77&777(189)\\
\\
\hline
\boldmath$ 
{\cal O}_{\bar u\bar dud}^{V/A}$& \\
\it Masses [MeV]\\
1&102&29&30.5&154&0&15.5&2.4&1.7&0&6.5&33.4&45.5&34.4&931(193)\\
$1/\sqrt{2}$&25&95&9&57&0&4.5&53&14.5&0&0.4&15.5&329&141&1489(380)\\
\it Couplings [keV]\\
1&149&105&40.5&6.5&0&88&13&2.5&0&86.5&125&170.5&129.5&1137(334)\\

$1/\sqrt{2}$&16&6.4&0.5&17&0&0.7&18&3.7&0&0&4.5&92.5&17&287(99)\\


\hline
\hline
\end{tabular}
}}
 \caption{\footnotesize The same caption as for Table\,\ref{tab:res} but for the 1st radial excitations for different assignements. $r=1,1/\sqrt{2},0$ are typical values of the four-quark mixing of currents.}
\label{tab:res-rad}
\end{table}
\subsection*{\b Final predictions for the four-quark states}
One can notice from the previous analysis that the predictions of the four-quark states are a MESS and are not  conclusive as there are too many freedom for fixing the masses and couplings  (choice of the currents and of the mixing parameter). For definiteness, we take the mean of the results from different currents and mixing parameter inside the range $r=0$ to 1. We deduce, as final predictions for four-quark ground states\,:
\bea
\bar M_{\bar u\bar dud}= &&1009(114) ~{\rm MeV}  ,\,\,\,\,\,\,\,\,\,\,\,\,\,\,\,\,\,\,\,\,\,\,\,\,\,\,\,\,\bar f_{\bar u\bar dud}=\,\,\,\,690(66)~{\rm keV},\nnb\\
\bar M_{\bar u\bar sud}= &&991(123) ~{\rm MeV}  ,\,\,\,\,\,\,\,\,\,\,\,\,\,\,\,\,\,\,\,\,\,\,\,\,\,\,\,\,\,\,\,\bar f_{\bar u\bar sud}=\,\,\,\,444(55)~{\rm keV},\nnb\\
\bar M_{\bar u\bar sds}= &&1045(112) ~{\rm MeV}  ,\,\,\,\,\,\,\,\,\,\,\,\,\,\,\,\,\,\,\,\,\,\,\,\,\,\,\,\,\bar f_{\bar u\bar sds}=\,\,\,\,457(55)~{\rm keV},
\label{eq:4q}
\eea
and for the 1st radial excitation:
\bea
\bar M^{(1)}_{\bar u\bar dud}= &&1409(112) ~{\rm MeV}  ,\,\,\,\,\,\,\,\,\,\,\,\,\,\,\,\,\,\,\,\,\,\,\,\bar f^{(1)}_{\bar u\bar dud}=\,\,\,\,449(69)~{\rm keV},
\label{eq:4q-rad}
\eea
where we have taken the error from the most precise determinations. Finite width corrections increase slightly these NWA masses by about (67-80) MeV which is inside the error bars of the determinations The mass of the radial excitations including a strange quark is expected to be (almost) degenerated with the non-strange one due to the small $SU(3)$ breakings. 
\section{Summary and conclusions}
 We have systematically revisited the existing estimates of the masses and couplings of the scalar quarkonia (ordinary $\bar qq$ and four-quark) states and presented new results for $\pi^+\pi^-$-like ($\pi^+\pi^-, \bar KK, \bar K\pi$ and $\eta\pi$) molecule states. Within our choice of the currents for a given configuration, these states can be analyzed separately like in any QCD spectral sum rules approach. The parameters of these unmixed states can be  used  for a further analysis of their mixings which is beyond the scope of this paper. The results are compiled in Table\,\ref{tab:res} to \ref{tab:res-rad}.
 \subsection*{\b QCD expressions}
We have checked some existing expressions of the four-quark and molecules currents given in the literature.
\subsection*{\d  Molecule currents}
We found several errors in the expression given in the pioneer work of Ref.\,\cite{LATORRE} ($\la\alpha_s G^2\ra, \la\bar sGs\ra$ and $\la \bar uu\ra^2 +\la\bar ss\ra^2$) using the pseudoscalar $\oplus$ scalar currents. 

\subsection*{\d Four-quark  currents}

-- We agree with the expression given by Ref.\,\cite{MARINA} for the scalar four-quark current.

-- We recover the different results given in Ref.\cite{ZHU} by putting to zero the  $\la \bar sGs\ra$ mixed condensate contributions absent in their QCD expressions !  Then, we suscpect that the strange quark may have been treated as an heavy quark along the calculation of Ref.\,\cite{ZHU}. 

 \subsection*{\b $SU(3)$ breakings}
One can notice from Table\,\ref{tab:res} to \ref{tab:res-4q-VA} that the $SU(3)$ breakings are tiny (about some few tens of MeV). Howwever, one can observe that for the S/P four-quark states, the central values go in the reverse direction. 

\subsection*{\b $\sigma/f_0(500)$}
The $R_{P/C}$ condition in Eq.\,\ref{eq:rpc} excludes the values $M_{\sigma} \simeq (0.5-0.6)$ GeV obtained in the recent literature\,\cite{ZHU,STEELE} for four-quark state  where its contribution does not exceed 60\% of the QCD continuum one at the $\tau$-stability point but, instead, favours the first estimate of about 1 GeV obtained in Refs.\,\cite{LATORRE,SNB2,SNA0}. This 1 GeV mass obtained in the real axis from LSR  can be identified with the on-shell/ Breit-Wigner mass from fits of the  $\pi\pi$ scattering data given in Eq.\,\ref{eq:sigdata}. 

Therefore, one can see from the previous Tables that the molecules and the mean of the four-quark (Eq.\,\ref{eq:4q}) assignements provide predictions compatible with this on-shell mass definition of the $\sigma$ like also the case of the lightest scalar gluonium\,\cite{VENEZIA,SNG,SNGS}. 

 However, the additional constraint on the $\sigma\bar KK$ coupling from $\pi\pi\to \bar KK$ scattering data\,\cite{KMN,WANG2} quoted in Eq.\,\ref{eq:sig-coupling} does not favour the pure $\bar u\bar d ud$ and $\pi^+\pi^-$ assignement for the $\sigma$ often advocated in the literature.  This is not the case of the light scalar gluonium which is expected to couple universally (up to $SU(3)$ breakings) to pair of pseudoscalar $\bar qq$ states\,\cite{VENEZIA,SNG,SNGS} from the low-energy property of the energy-momentum tensor form factor.  
 
\subsection*{\b $f_0(980)$}
The $\sigma$ and $f_0(980)$ seems to emerge from a maximal meson-gluonium mixing\,\cite{BN} with $(M_{\bar qq},\Gamma_{\pi\pi})= (1229,120)$ MeV from Eq.\,\ref{eq:respi-vertex} and the light scalar gluonium mass $(M_G,\Gamma_G)= (1070,890)$ from\,\cite{SNGS}. 

However, the result in Table\,\ref{tab:res} indicates that a $K^+K^-$ molecule with  $M_{K^+K^-}=1056(214)$ MeV is compatible with the $f_0(980)$, 
while the mean prediction of the different four-quark states leads to $M_{\bar u\bar s ds}$=1045(112) MeV (Eq.\ref{eq:4q}). 

\subsection*{\b $a_0(980)$}
One of the main motivation for introducing the four-quark / molecule assignement for the $a_0(980)$ is its vicinity to the $(\bar uu+\bar dd)$ and its strong coupling to $\bar KK$ states\,\cite{JAFFE}. From our analysis, we found that the $I=1$ isovector $(\bar uu-\bar dd)$ state is too high ($M_{\bar ud}= 1246(94)$ MeV). A $\eta\pi$ molecule with a mass 1040(139) MeV and the mean of four-quark state in Eq.\,\ref{eq:4q} also give the same mass as the one in the case of  the $f_0(980)$ because we have neglected the $SU2$ breakings. 

\subsection*{\b $f_0(1370)$}
This state is reproduced  by the 1st radial excitation of the $\bar qq$ state with a mass 1378(186) MeV (Table\,\ref{tab:res-rad}) which can mix with the scalar gluonium $M_{\sigma'}=1110(117)$ MeV \cite{SNGS} to give the observed  large $\pi\pi$ width. In this picture, the $\sigma$ and $f_0(1370)$ can also be the dragon proposed by Refs.\,\cite{OCHS,MIN}.  However, it can also mix with the 1st radial excitation of the four-quark state with a mean mass given in Eq.\,\ref{eq:4q-rad}. 
\subsection*{\b $a_0(1450)$}
This state coincides with the 1st radial excitation of the $\bar u\bar s ud$ state which should be almost degenerated to the one of $\bar u\bar d ud$ state given in Eq.\,\ref{eq:4q-rad} due to the small $SU(3)$ breakings .

\subsection*{\b $f_0(1500)$}
The $f_0(1500)$ is expected to be a gluonium state from its mass $M_{G'_1}$=1563(141) MeV and from its $U(1)$-like decays ($\eta'\eta,\eta\eta$)\,\cite{VENEZIA,SNG,SNGS}. From the present analysis, one expects to have in this region the 1st radial excitation of the four-quark state having a mass $M^{(1)}_{\bar u \bar d ud}=1409(112)$ MeV (Eq.\,\ref{eq:4q-rad})  which may mix with the previous gluonium state. 

\subsection*{\b $f_0(1710)$}
From the 1st radial excitation masses obtained in Table\,\ref{tab:res-rad},  the $f_0(1700)$
can be likely the radial excitation of the four-quark or/and molecule states. The next radial excitation of a gluonium is predicted to be higher : $M_{G_2}=2992(221)$ MeV. 

\subsection*{\b $K^*_0(700)$}
The four-quark and the $K\pi$ molecule assignements lead to a mass compatible with the Breit-Wigner mass of $(845 \pm 17)$ MeV\,\cite{PDG} (see Eqs.\,\ref{eq:4q} and Table\,\ref{tab:res}). The $\bar us$ assignement leads to a mass 
 $M_{\bar us}=1276(58)$ MeV which is relatively too high.

\subsection*{\b $K^*_0(1430)$}
This state is better fitted by  $M_{\bar us}=1276(58)$ MeV and/or  its radial excitation expected to be around 1400 MeV (Eq.\,\ref{eq:ud-rad}) as the $SU(3)$ breakings are expected to be small. 

\vspace*{0.25cm}
{\it \large We expect that the systematic analysis done in this paper  can help to clarifiy the complex spectra of the light scalar mesons.}




\begin{thebibliography}{999}
 \bibitem{PDG} R.L. Workman et al. (Particle Data Group), Prog. Theor. Exp. Phys. 2022, 083C01 (2022).
 
 \bibitem{AMSLER} Review of C. Amsler et al.  and S. Eidelman et al., in Ref.\cite{PDG}. 
 

\bibitem{LEUT}I. Caprini, G. Colangelo and H. Leutwyler,  {\it Phys. Rev. Lett.} {\bf 96} (2006) 132001.

\bibitem{YND}F.J Yndurain, R. Garcia-Martin, J. R. Pelaez,  {\it Phys. Rev. } {\bf D 76} (2007) 074034.

\bibitem{BES05}  The BES III collaboration : S. Fang, {\it Nucl. Phys. Proc. Suppl.} {\bf 164} (2007) 135.

\bibitem{E741}E.M. Aitala et al., {\it Phys. Rev. Lett.} {\bf 86} (2001) 770.

\bibitem{WANG1} G. Mennessier, S. Narison,  X.G Wang,  {\it Phys. Lett.} {\bf B 688} (2010) 59.

 \bibitem{MNO} G. Mennessier, S. Narison, W. Ochs, {\it Phys. Lett.} {\bf B 665} (2008) 205; {\it Nucl. Phys. Proc. Suppl.} {\bf 238} (2008) 181.

\bibitem{KMN}R. Kaminski, G. Mennessier, S. Narison, {\it Phys. Lett.} {\bf B 680} (2009) 148.

\bibitem{HANHART} M.~Hoferichter, D.~R.~Phillips,  C.~Schat,
Eur. Phys. J. C \textbf{71} (2011), 1743.

\bibitem{WANG2} G. Mennessier, S. Narison,  X.G Wang,    {\it Phys. Lett.} {\bf B 696} (2011) 40.

\bibitem{DOSCH03} H.G. Dosch and S. Narison, {\it Nucl. Phys. Proc. Suppl.} {\bf 121} (2003) 114.

\bibitem{VENEZIA}. S. Narison and G. Veneziano, {\it Int. J. Mod. Phys.} {\bf A4, 11} (1989) 2751.
\bibitem{SNG} S. Narison, {\it Nucl. Phys.} {\bf B 509} (1998) 312/; ibid, 
{\it Nucl. Phys. Proc. Suppl.} {\bf 64} (1998) 210. 

 \bibitem{SNGS} S. Narison, {\it Nucl.Phys.}{ \bf A1017} (2022) 122337. 

\bibitem{JAFFE} R. L. Jaffe, {\it Phys. Rev.} {\bf D15} (1977) 267; {\it Phys. Rev.} {\bf  D15} (1977) 281; {\it Phys.Rept.} {\bf 409} (2005) 1.
\bibitem{ACHASOV}N.N. Achasov, S.A. Devyanin, G.N. Shestakov, {\it Z. Phys.} {\bf  C16} (1984) 55. 
\bibitem{ISGUR} N. Isgur, J. Weinstein, {\it Phys. Rev.} {\bf D41} (1990) 2236.

\bibitem{ROSSI} G. C. Rossi, G. Veneziano, {\it Nucl.Phys.}{ \bf B123} (1977) 507; {\it Nucl. Part. Phys. Proc.}{\bf  312-317} (2021) 140.

\bibitem{OCHS} W. Ochs, {\it J. Phys.} {\bf G 40} (2013) 043001.
 \bibitem{GASTALDI}U. Gastaldi, {\it Nucl. Phys. Proc. Suppl.} {\bf 96} (2000) 234;   {\it Nucl. and Part. Phys. Proc. } {\bf 300-302} (2018) 113.



 \bibitem{SNSCAL} S. Narison, {\it Nucl. Phys. B  Proc. Suppl.} {\bf 186} (2009) 306.

 \bibitem{SNPRD} S. Narison, {\it Phys. Rev.} {\bf D 73} (2006) 114024.

 \bibitem{SNGMIX} G. Mennessier,, S. Narison, N. Paver, {\it Phys.Lett.}{ \bf B158} (1985) 153-157. 

\bibitem{STEELES} T.G. Steele, D. Harnett, R.T. Kleiv, K. Moats, {\it Nucl. Phys. Proc. Suppl.} {\bf B234} (2013) 257.

\bibitem{RICHARD}J.-M. Richard, {\it Few Body Syst} {\bf 57} (2016) 12, 1185.

\bibitem{KLEMPT}  E. Klempt, A. Zaitsev, {\it Phys. Rept.} {\bf 454} (2007) 1.

 
 

 \bibitem{SCAL} S. Narison, N. Paver, E. de Rafael, D. Treleani,  {\bf Nucl.Phys.}{ \bf B212}(1983) 365. 
\bibitem{RRY} L.J. Reinders, H. Rubinstein and S. Yazaki, {\it Phys. Rept. }
{\bf 127}  (1985) 1. 
 \bibitem{SNB2} S. Narison, {\it QCD spectral sum rules, World Sci. Lect. Notes Phys.} {\bf 26} (1989) 1, ISBN 9780521037310. 
 \bibitem{SNA0}S. Narison, {\it Phys. Lett.} {\bf B175} (1986) 88.

 \bibitem{BN}A. Bramon, S. Narison, {\it Mod. Phys. Lett.} {\bf A4} (1989) 1113.

\bibitem{LATORRE} J. I. Latorre and P. Pascual, {\it Jour. Phys.} {\bf G11} (1985) L231. 
\bibitem{MARINA} T.V. Brito, F.S. Navarra, M.Nielsen, M.E. Bracco,, {\it Phys. Lett.} {\bf B608} (2005) 69.
\bibitem{ZHU} H. X. Chen, A. Hosaka and S. L. Zhu, {\it Phys. Rev.} {\bf D76} (2007) 094025. 
\bibitem{STEELE} B. A. Cid-Mora and T. G. Steele, {\it Nucl. Phys.} {\bf A1028} (2022) 122538. 
\bibitem{JAFFE2}M. Alford and R.L. Jaffe, {\it Nucl. Phys.} {\bf B578}, (2000) 367;
 M. Wakayama et al. (the scalar collaboration), {\it Phys. Rev} {\bf D91} (2015) 094508;
S. Prelovsek et al.,  {\it Phys. Rev} {\bf D82} (2010) 094507;
N. Mathur et al.,  {\it Phys. Rev} {\bf D76} (2007) 114505. 

\bibitem{THOOFT}  G. 't Hooft,  G. Isidori,  L. Maiani, A.D. Polosa, V. Riquer, {\it Phys. Lett.} {\bf B662} (2008) 424.

\bibitem{BRODSKY} L.P. Zou, H.G. Dosch, G.F. de Teramond, S. J. Brodsky, {\it Phys. Rev.} {\bf D99} (2019) 114024.
 \bibitem{KNIEHL} B. Kniehl and A. Sirlin, {\it Phys. Rev.} {\bf D77} (2008) 116012.

\bibitem{SVZa}M.A. Shifman, A.I. Vainshtein and V.I. Zakharov, {\it Nucl. Phys.} {\bf B147} (1979) 385; 
\bibitem{SVZb}M.A. Shifman, A.I. Vainshtein and V.I. Zakharov, 
{\it Nucl. Phys.} {\bf B147} (1979) 448.
\bibitem{ZAKA}V.I. Zakharov, Sakurai's Price, {\it Int. J. Mod .Phys.} {\bf  A14} (1999) 4865.
\bibitem{BELLa}J.S. Bell and R.A. Bertlmann, {\it Nucl. Phys.} {\bf B177} (1981) 218; 
\bibitem{BELLb}J.S. Bell and R.A. Bertlmann,  {\it Nucl. Phys.} {\bf B187} (1981) 285.
\bibitem{BERT}R.A. Bertlmann, {\it Acta Phys. Austriaca} {\bf 53}, (1981) 305.  
\bibitem{SNR}S. Narison and E. de Rafael,  {\it Phys. Lett.} {\bf B 103} (1981) 57.
 \bibitem{SNB1}S. Narison, {\it QCD as a theory of hadrons, Cambridge Monogr. Part. Phys. Nucl. Phys. Cosmol.} {\bf 17} (2004) 1-778 [hep-ph/0205006]. 
 
\bibitem{SNB3}S. Narison, {\it Techniques of dimensional regularization and the two-point functions of QCD and QED, Phys. Reports} {\bf 84, $n^0$ 4} (1982), 263.

\bibitem{FESR}R.A. Bertlmann, G. Launer, E. de Rafael, {\it Nucl. Phys.} {\bf B250} (1985) 61.


\bibitem{TQQ} R. Albuquerque, S. Narison, D. Rabetiarivony, {\it Nucl. Phys.} {\bf A1034} (2023) 122637.







\bibitem{BECCHI}C. Becchi, S. Narison, E. de Rafael and F.J. Yndur\`ain,
{\it Z. Phys.} {\bf C8} (1981) 335.


\bibitem{JM95} M. Jamin and M. Munz, {\it Z. Phys.} {\bf C 66} (1995) 633.

\bibitem{SNp15}S. Narison, {\it Phys. Lett.}  {\bf B738} (2014)  346.





\bibitem{SNparam}S. Narison, {\it  Int. J. Mod. Phys.} {\bf A33} (2018) no.10, 1850045, Addendum: {\it Int. J. Mod. Phys.} {\bf A33} (2018) no.10, 1850045 and references therein.

 \bibitem{SNREV1}S. Narison, {\it Nucl. Part. Phys. Proc.}{\bf  312-317} (2021) 87;  
 ibid, {\bf 258-259} (2015) 189. 


\bibitem{SNcb1}S. Narison,  {\it Phys. Lett.} {\bf B693}  (2010) 559, erratum {\it ibid}, 
{\bf B705} (2011) 544; 
ibid, {\bf B706}  (2012) 412; 
ibid, {\bf B707}  (2012) 259.
\bibitem{SNLIGHT}S. Narison, {\it Phys.Lett.} {\bf B738} (2014) 346.

 \bibitem{DOSCHSN}H.G. Dosch, S. Narison, {\it Phys. Lett.} {\bf B417} (1998) 173. 

\bibitem{HBARYON2}R.M. Albuquerque, S. Narison, M. Nielsen, {\it Phys. Lett.} {\bf B684} (2010) 236.


\bibitem{IOFFE}B.L. Ioffe, {\it Nucl. Phys.} {\bf B188} (1981)  317; {\it Nucl. Phys.} {\bf B191} (1981) 591.

 \bibitem{DOSCH}Y. Chung, H. G. Dosch, M. Kremmer, D. Schall, {\it Z. Phys.} {\bf C25} (1984) 151; H.G. Dosch, M. Jamin, S. Narison, {\it Phys. Lett.} {\bf B220} (1989)  251.

 \bibitem{PIVOm}A.A.Ovchinnikov and A.A.Pivovarov, {\it Yad.\ Fiz.}  {\bf 48} (1988) 1135.
\bibitem{SNhl}S. Narison, {\it Phys. Lett.} {\bf B605} (2005) 319.
\bibitem{SNH10}S. Narison,  {\it Phys. Lett.} {\bf B693} (2010)  559; Erratum ibid 705 (2011) 544.
\bibitem{SNH11}S. Narison,{\it Phys. Lett.} {\bf B706} (2011)  412
\bibitem{SNH12}S. Narison,{\it Phys. Lett.} {\bf B707} (2012)  259. 
\bibitem{LNT}G. Launer, S. Narison and R. Tarrach, {\it Z. Phys.} {\bf C26} (1984) 433.
\bibitem{FESR2} R.A. Bertlmann, C.A. Dominguez, M. Loewe, M. Perrottet and E. de Rafael, {\it Z. Phys.} {\bf C39} (1988) 231.
\bibitem{SNTAU}S. Narison, {\it Phys. Lett.} {\bf B673} (2009) 30.
\bibitem{SN95}S. Narison, {\it Phys.Lett.} {\bf B 358} (1995) 113.
\bibitem{SOLA} C.A. Dominguez, J. Sola, {\it  Z. Phys.} {\bf C40} (1988) 63.
\bibitem{SNe23} S. Narison, arXiv: 2306.14639 [hep-ph] (2023).
\bibitem{SN22} S. Narison, talk given at QCD22, Montpellier-FR, {\it Nucl. Part. Phys. Proceed.}{\bf 324-329} (2023) 94 (arXiv 2211.14536 [hep-ph] ).

%

\bibitem{BROAD1}D.J. Broadhurst,  {\it Phys. Lett.} {\bf B101} (1981) 423.

\bibitem{LARIN2}S. G. Gorishny, A. L. Kataev, S. A. Larin, and L. R. Surguladze, {\it Mod.
Phys. Lett.} {\bf A5} (1990) 2703.

\bibitem{CHET3}K. G. Chetyrkin, {\it Phys. Lett.} {\bf B390} (1997) 309.

\bibitem{CHET4}P. A. Baikov, K. G. Chetyrkin and J. H. Kuhn, {\it Phys. Rev. Lett.} {\bf 96} (2006) 012003. 

\bibitem{CNZb}S. Narison and V.I. Zakharov, {\it  Phys. Lett.} {\bf B522} (2001) 266.
\bibitem{ZAKa} V.I. Zakharov, {\it Nucl. Phys. Proc. Suppl.} 
{\bf 164} (2007) 240.
\bibitem{ZAKb}S. Narison,  {\it Nucl. Phys. Proc. Suppl.} {\bf 164} (2007) 225. 
\bibitem{ADS1}	O. Andreev, {\it Phys. Rev.} {\bf D73} (2006) 107901.
\bibitem{ADS2}	O. Andreev and V.I. Zakharov,  {\it Phys. Rev.} {\bf D74} (2006) 025023; {\it ibid},{\bf D76} (2007)047705.
\bibitem{ADS3}	 F. Jugeau, S. Narison, H. Ratsimbarison,  {\it Phys. Lett.}  {\bf  B722} (2013) 111.
\bibitem{CNZa} K.G. Chetyrkin, S. Narison and V.I. Zakharov, {\it Nucl. Phys.} 
{\bf B550} (1999)  353.
\bibitem{SND2}S. Narison, {\it Phys. Lett.} {\bf B300} (1993)  293;
{\it ibid,} {\it Phys. Lett.} {\bf B361} (1995)  121.
\bibitem{TERAYEV}M. Kozhevnikova, A. Oganesian and O. Terayev, {\it EPJ Web of Conferences} {\bf 204} (2019) 02005. 
 

 \bibitem{SNT} S. Narison and R. Tarrach, {\it Phys. Lett.}  {\bf B125} (1983)  217.





\bibitem{XTZ} R. Albuquerque, S. Narison, D. Rabetiarivony,
{\it Nucl. Phys.} {\bf A1023} (2022) 122451;  {\it Phys.Rev.} {\bf D103} (2021) 7, 074015

\bibitem{DK}R. Albuquerque, S. Narison, D. Rabetiarivony, G. Randriamanatrika,
{\it Nucl. Phys.} {\bf A1007} (2021) 122113;

\bibitem{4Q} 
R. Albuquerque, S. Narison, A. Rabemananjara, D. Rabetiarivony, G. Randriamanatrika, {\it Phys. Rev.} {\bf D102} (2020) 9, 094001.

\bibitem{XYZ} R. Albuquerque, S. Narison, F. Fanomezana, A. Rabemananjara, D. Rabetiarivony, G. Randriamanatrika, {\it Int. J. Mod. Phys.} {\bf A31} (2016) 36, 1650196;  {\it Int. J. Mod. Phys.} {\bf A33} (2018) 16, 1850082.


  \bibitem{PICH}A. Pich and E. de Rafael, {\it Phys. Lett.}  {\bf B158} (1985)  477.
  
  \bibitem{SNPIVO}S. Narison and A. Pivovarov,  {\it Phys. Lett.}  {\bf B327} (1994)  341.
  
  \bibitem{SNrho} S. Narison, invited review to appear in the book ``Laplace transform and its applications", edited by Nova Science Publishers, New York, USA.

\bibitem{CHETV}P. A. Baikov, K. G. Chetyrkin, J. H. Kuhn, arXiv:1501.06739 [hep-ph] (2015).

\bibitem{MIN} W. Ochs and P. Minkowski, {\it Eur. Phys. J.}  {\bf C9} (1999) 283.

\end{thebibliography}
\end{document}